\documentclass[12pt,preprint]{aastex}

\usepackage{graphicx}

\usepackage{enumerate}

\usepackage{epstopdf}

\usepackage{rotating}
\usepackage{tablefootnote}
\usepackage[flushleft]{threeparttable}

\begin{document}

\title{The Structure and Stellar Content of the Outer Disks of Galaxies: A New View from the Pan-STARRS1 Medium Deep Survey}

\author{
Zheng Zheng\altaffilmark{1,2}, 
David A. Thilker\altaffilmark{1}, 
Timothy M. Heckman\altaffilmark{1},  
Gerhardt R.\ Meurer\altaffilmark{3}, 
W. S. Burgett\altaffilmark{4}, 
K. C. Chambers\altaffilmark{5}, 
M. E. Huber\altaffilmark{4}, 
N. Kaiser\altaffilmark{4}, 
E. A. Magnier\altaffilmark{4}, 
N. Metcalfe\altaffilmark{5}, 
P. A. Price\altaffilmark{6}, 
J. L. Tonry\altaffilmark{4}, 
R. J. Wainscoat\altaffilmark{4},
C. Waters\altaffilmark{4}
}


\altaffiltext{1}{Department of Physics and Astronomy, Johns Hopkins University, 3701 San Martin Drive, Baltimore, MD 21218, USA}
\altaffiltext{2}{National Astronomical Observatories, Chinese Academy of Sciences, A20 Datun Road, Chaoyang District, Beijing 100012, China}
\altaffiltext{3}{International Center for Radio Astronomy Research, The University of Western Australia, M468, 35 StirlingHighway, Crawley, WA 6009, Australia}
\altaffiltext{4}{Institute for Astronomy, University of Hawaii at Manoa, Honolulu, HI 96822, USA}
\altaffiltext{5}{Department of Physics, Durham University, South Road, Durham DH1 3LE, UK}
\altaffiltext{6}{Department of Astrophysical Sciences, Princeton University, Princeton, NJ 08544, USA}

\begin{abstract}
We present the results of an analysis of Pan-STARRS1 Medium Deep Survey multi-band ($g\,r\,i\,z\,y$) images of a sample of 698 low-redshift disk galaxies that span broad ranges in stellar mass, star-formation rate, and bulge/disk ratio. We use population synthesis SED fitting techniques to explore the radial distribution of the light, color, surface mass density, mass/light ratio, and age of the stellar populations. We characterize the structure and stellar content of the galaxy disks out to radii of about twice Petrosian $r_{90}$, beyond which the halo light becomes significant. We measure normalized radial profiles for sub-samples of galaxies in three bins each of stellar mass and concentration. We also fit radial profiles to each galaxy.  The majority of galaxies have down-bending radial surface brightness profiles in the bluer bands with a break radius at roughly $r_{90}$.  However, they typically show single unbroken exponentials in the reddest bands and in the stellar surface mass density. We find that the mass/light ratio and stellar age radial profiles have a characteristic ÔU-shapeÕ.  There is a good correlation between the amplitude of the down-bend in the surface brightness profile and the rate of the increase in the M/L ratio in the outer disk.  As we move from late- to early-type galaxies, the amplitude of the down-bend and the radial gradient in M/L both decrease. Our results imply a combination of stellar radial migration and suppression of recent star formation can account for the stellar populations of the outer disk.

\end{abstract}

\section{Introduction}
\label{introduction}

The stellar disks of galaxies have been recognized as having exponential radial surface-brightness (SB) profiles 
since the seminal paper published by \citet{fre70}. 
About a decade  later  \citet{vdk79} studied several edge-on galaxies and  found that the surface brightnesses 
of galactic disks do not decline exponentially forever, but can appear truncated after several radial scale-lengths.
More recently, many studies using deeper modern imaging data \cite[e.g.][]{poh02,erw05,erw08} showed that a variety of SB profiles exist. Notably, \citet[][PT06 hereafter]{pt06} systematically studied 90 nearby late-type spiral galaxies using SDSS DR2 imaging data and found that there is no apparent truncation 
in the SB radial profiles of these face-on galaxies.
Instead, they found the radial profiles can be divided into three categories:
Type I, classical single exponential (10\%); Type II, down-bending joint exponential (60\%); 
and Type III, up-bending joint exponential (30\%).
\footnote{ 
  Truncations and Type II breaks are two fundamentally different phenomena. The former is called `truncation' because its outer profile drops much faster than exponential but the outer part of the Type II disk can still be well described by an exponential function. The truncations are only seen in edge-on galaxies and the truncation radius is at about 4.5\,$h$ while the down-bending profiles are seen in all inclinations and the break radii usually appear at 2.5\,$h$, where $h$ is the disk scale-length. \citep{mn14}}
This disk profile classification scheme has been widely adopted by the community 
and a large number of models have been created to explain the formation of these different types of SB radial profiles:

Type I (single exponential,   probably with an outer truncation) galaxies can be formed directly from the collapse of a uniformly rotating protogalactic cloud \citep{fre70,fal80}. Alternatively, they can also be formed through viscous redistribution \citep[e.g.][]{fer01}.
\citet{vdk87} argued that a uniformly rotating protogalactic cloud could collapse into an exponential disk with a sharp cut-off based on angular momentum conservation. 
The model naturally explains the exponential nature of galactic disks and suggests the cut-off radius, which is about 4.5$\,h$, is related to the maximum angular momentum of the cloud.  

This collapsing model is also one of the two traditional scenarios which attempts to explain the formation of Type II (down-bending) disk.  However, as argued by \citet{elm06}, the collapsing model usually forms inner exponential disks with relatively sharp outer cutoffs. 
  Also, the collapsing model would lead to a break radius at about 4.5\,$h$ instead of the observed 2.5\,$h$.
Therefore, the other Type II disk formation scenario, which attributes the formation of breaks to a star formation threshold, was proposed \citep[e.g.][]{ken89,sch04,elm06,mn12}. However, the star formation threshold scenario fails to produce a single exponential profile out to large radii \citep{sch04}.
Instead, secular evolution, especially the outward radial migration of stars is believed to have significant effects on the breaks \citep{deb06,ros08}. The interplay between a star formation cut-off and radial stellar migration produces stellar radial profiles in good agreement with observed Type II disks  \citep{ros08}. Moreover, the predicted sharp change in the radial age profile is indirectly supported by the observational results of \citet{bak08}. We will show that this scenario is supported by the observations we present here. 

An explanation of the Type III (up-bending) disks is more difficult. One might consider that the Type III disk is formed by the enhanced star formation seen in galaxies with extended ultraviolet disk components \citep[XUV disk,][]{t07}. However, there is no obvious correlation between these two kinds of objects \citep{t07}. Radial stellar migration might play an important role in forming Type III disks \citep{you07}, however, it is crucial to include intermittent \citep[e.g. minor mergers,][]{you07} or persistent \citep[e.g. constant gas accretion,][]{min12} interaction with the intergalactic/circumgalactic environment.   A more recent study by \citet{bak12} using much deeper images from the SDSS Stripe82 on 7 nearby galaxies show that the profiles of  previously identified as Type III are actually contaminated by the stellar halo and they claim that there are no true Type III disks. 

  The three types of disks are not only different in their SB profiles but also look distinct in their radial color profiles. \citet{bak08} studied radial $g-r$ color profiles of the 90 PT06 sample galaxies and found that: The $g-r$ color profiles of Type II galaxies typically show a `U'-shape with a minimum at the break radius and a minimum color of $g-r=0.47\pm0.02$. The $g-r$ color profiles of Type I galaxies are almost flat in the outer disk and the average color profile has a relative constant value of $g-r=0.46$. The  color profiles of Type III galaxies drop steeply within the break radius and then rise to form  a plateau around the break radius with a color of $g-r =0.57\pm0.02$. The `U' shape color profile of Type II galaxy implies a `U' shape age profile, which is a signature of the stellar radial migration effect shown by \citet{ros08}. These different behaviors of color profiles provide us more information about the formation history of the outer disk. However, the \citet{bak08} sample of Type I (9 galaxies) and Type III (21 galaxies) is too small to yield fully robust results and the radial color profiles alone are not easy to compare with theoretical models or numerical simulations.

It is important to note that the models are mostly focused on calculating the radial stellar {\it mass} distribution, however, the previous studies generally only present the distribution of the {\it light} in a single band. The light profiles trace a combination of stellar mass and age, with the relative importance of these two factors depending upon the pass-band of the observations.  By observing in multiple bands these two effects can be disentangled, allowing the determination of the radial distribution of both the stellar mass and the characteristics of the stellar population. 
More specifically, the observed spectral energy distribution of the galaxy light can be used to estimate the M/L profiles to first order. This technique, based on the the M/L prescription of \citet{bel03}, was used by \citet{bak08} to derive stellar mass profiles for the PT06 sample.  
  The increase in the $g-r$ color profiles galaxies in the outer disk of Type II and Type III galaxies implies that the M/L also rises in the outer disk. They also showed that the breaks in the Type II SB profiles almost disappear in the stellar surface mass profiles.    The color profiles of the 7 galaxies studied by \citet{bak12} also show a `U'- shape with a slightly increasing value at large radii. Similar results were also obtained using deep stacked SDSS images for late-type galaxies \citep{dsou14}.

The recent studies above suggest that there might be multiple formation mechanisms for the observed outer stellar disks.  Building a more comprehensive galaxy sample was therefore a primary motivation for our study. Motivated by these considerations, we have undertaken an investigation that improves upon these earlier studies in several ways. First, we use a more sophisticated and robust method to derive the stellar mass-to-light and mass profiles by employing multi-band SED fitting (thereby jointly interpreting all the multi-band imaging data rather than just a single color). Second, we will use data that enable us to investigate the faint outer disks in a much large sample of galaxies than in previous studies. Third, our sample is more representative of the full population of disk galaxies than the PT06 sample: our sample is much larger and we have broader ranges in stellar mass and surface mass density.

In this paper, we report the results from our analysis of the images of about 700 galaxies in the Medium Deep Survey (MDS) fields of the Pan-STARRS1 (PS1) project. We measure five-band ($grizy$) SB profiles and analyze them with the  SED fitting program MAGPHYS to derive radial profiles of various parameters, such as stellar M/L, mass surface density, age, and specific star formation rate. In section \ref{data} we describe the PS1MD data and our sample selection; in section \ref{method} we describe the data processing pipeline and briefly introduce the MAGPHYS software.   We then present the results in three tiers. In section \ref{stacked_2d} we show the stacked images of the whole sample and some sub-samples to give an overview of the whole sample in 2D. We use these data to define the radial range of the disk region we are working on. In section \ref{stacked_1d} we present the composite radial profiles to show the 1D structures of generic disk galaxies. In section \ref{detailed_1d} we present a detailed analysis of radial profiles of all the individual galaxies. In particular, we analyze the breaks of SB profiles and compare our results with the PT06 SB profile classification scheme. We finally discuss our results in section \ref{discussion} and summarize the major conclusions in section \ref{summary} The images and radial profiles of each individual galaxy are in the appendix and can be downloaded online. The cosmological parameters we adopt are $h=0.7$, $\Omega_m=0.3$, and $\Omega_{\Lambda}=0.7$.

\section{Description of Pan-STARRS1 data and sample selection}
\label{data}

The imaging data is taken from the Medium Deep Survey (MDS) of the Pan-STARRS1 (PS1) telescope and camera.
The PS1 is the prototype telescope of the Panoramic Survey Telescope \& Rapid Response System \citep[Pan-STARRS;][]{kai02}, located on Mount Haleakala, Hawaii. There are 10 MDS fields, each with 7 square degrees field of view, observed in 5 wide bands \citep[$g$,$r$,$i$,$z$,$y$;][]{ton12}. We use the reference stack images, which have typical exposure times of $\sim 10\,\rm ks$ in the $g$ and $r$ bands and $\sim 20\,\rm ks$ in the $i$,$z$ and $y$ bands. There are also deep stack images, which have typical exposure times $\sim 40\,\rm ks$ in the $g$, $r$, and $y$ bands and $\sim 80\,\rm ks$ in the $i$ and $z$ bands. These are only available in two of the MDS fields (MD04 and MD09).  We use the reference stacks in this work to retain approximate exposure consistency between MDS fields. Further, as we demonstrate in Section \ref{stacked_1d}, the reference stacks are already sufficiently deep to meet our goal of probing out to the disk/halo transition.   Detailed descriptions of the PS1 MDS and the photometric calibration can be found in \citet{ton12}, \citet{res13} and \citet{sch12}.

\begin{figure}[htbp]
\begin{center}
\includegraphics[scale=0.4]{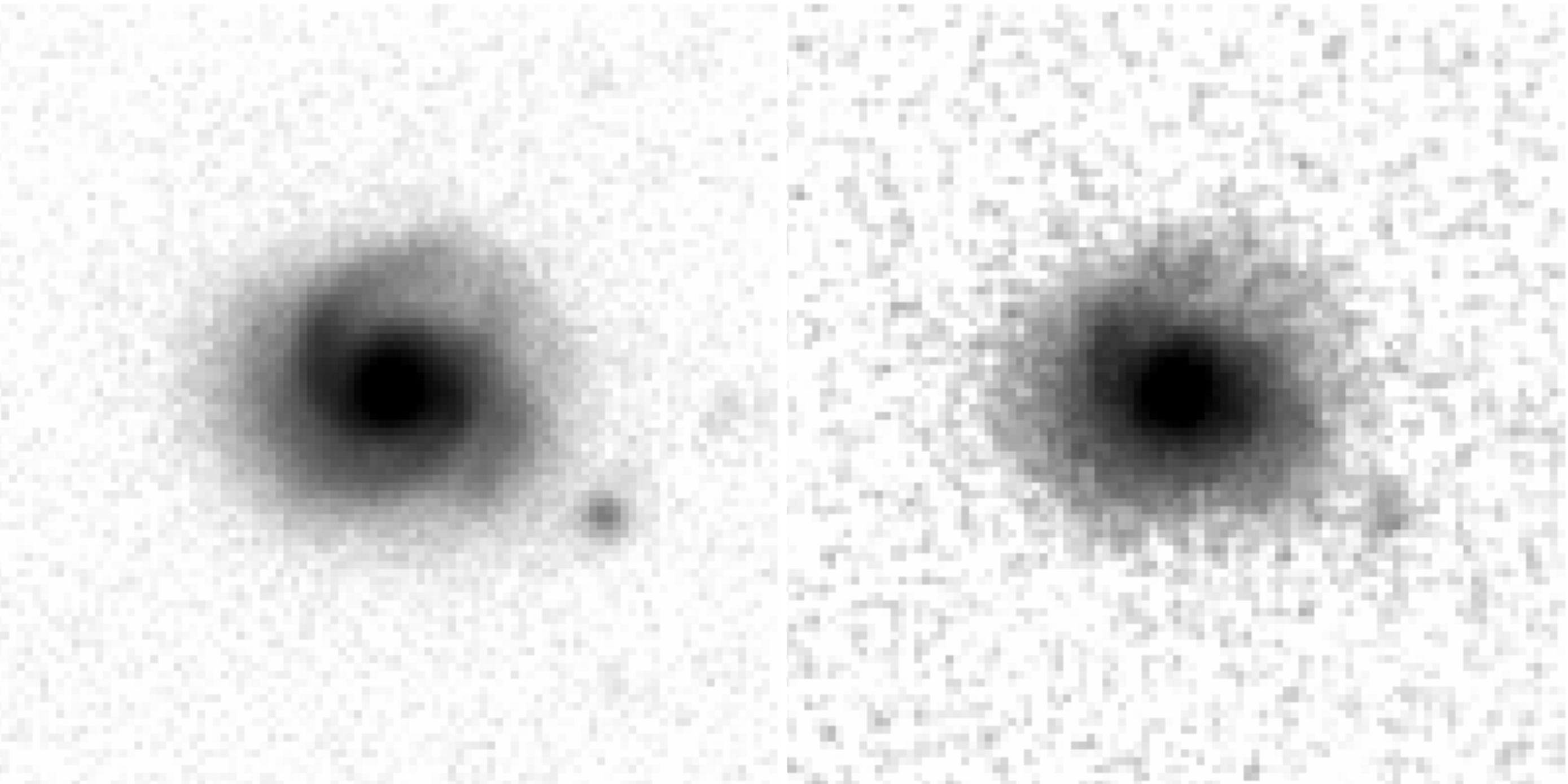}  
\includegraphics[scale=0.4,angle=90]{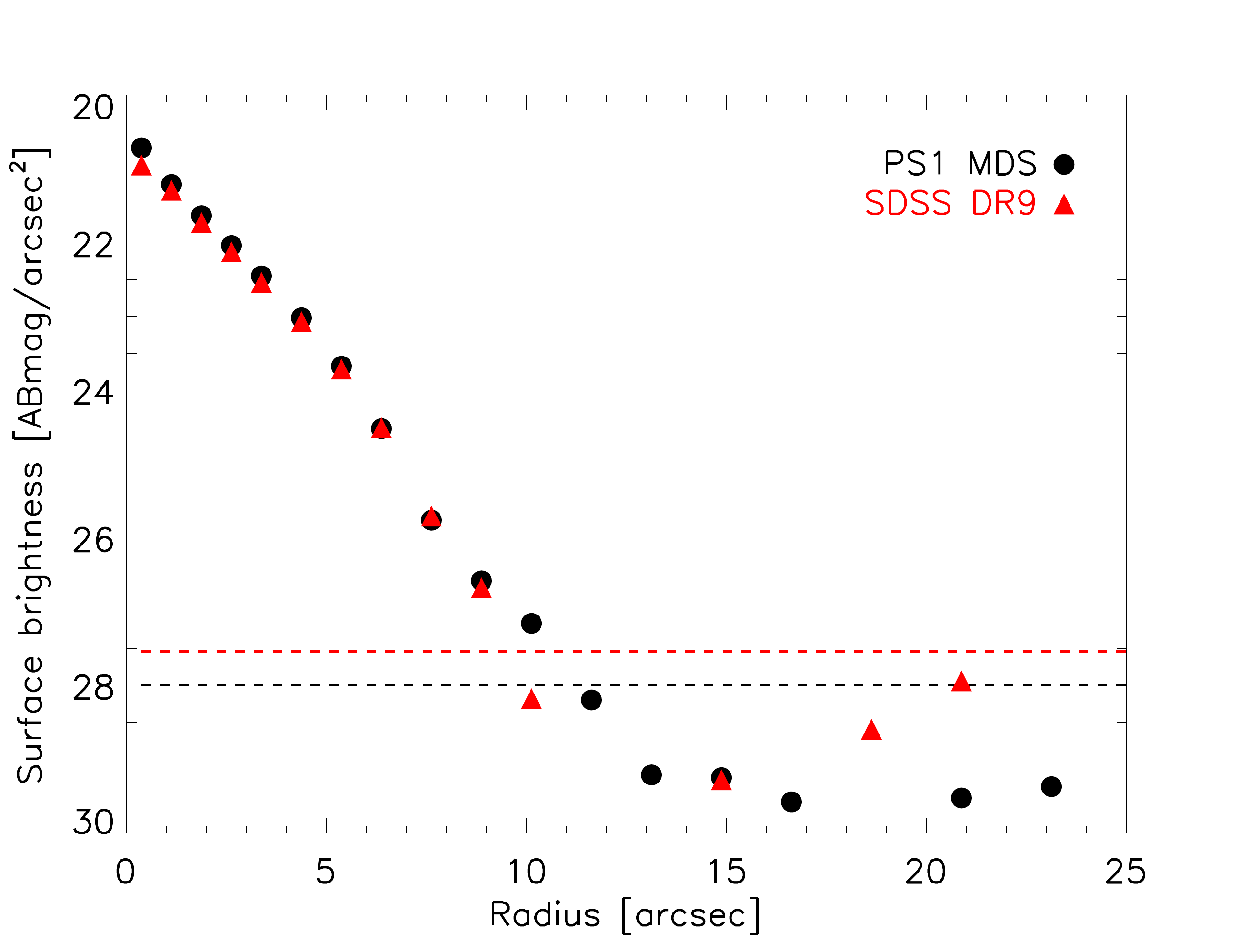}  
\caption{Comparison between PS1 MDS reference stack images and SDSS DR9 mosaic images. The upper panels are $r$-band PS1 image (left) and SDSS image (right). The images are background subtracted and the fluxes are scale to the central area (with radius of 4 pixel). The images are produced in ds9 using logarithmic scale. The lower panel is the $r$-band SB profiles produced using PS1 and SDSS data. The black dots and red triangles are measured SB using PS1 and SDSS data respectively. The black and red dash line show the sky rms of PS1 and SDSS images respectively. The PS1 image has a sharper PSF and lower sky rms than the SDSS image.
The galaxy is centered at RA= 130.95206, DEC=43.800864 (Galaxy 6 in our sample, cf. Table 1). The field of view of the images above are 25'' by 25''.  The equivalent exposure times are 10458s and 1294s for PS1 and SDSS images. Note the SDSS mosaic image is a stack image made through the SDSS DR9 image server using all exposures which has overlap with the target area and it has a much longer equivalent exposure time than single SDSS images. }
\label{sdss_compare}
\end{center}
\end{figure}

The reference stack images can be used to measure SB profiles as faint as $\sim 28-30\,mag/arcsec^2$ in the $r$-band, much deeper than the SDSS images ($\sim 27\,mag/arcsec^2$, e.g. Bakos et al 2008.; PT06). 
For comparison, we show $r$-band images and SB radial profiles of an example galaxy from the PS1 MDS  and SDSS DR9 mosaic  images. in Fig. \ref{sdss_compare}.   The flux of the images are scaled to the average central bulge (within 4 pixel) flux. The pixel size of the PS1 image is 0.25''/pix and the typical FWHM of the PSF is about 1''. The images of Fig. \ref{sdss_compare} are scaled to the same angular size. It is obvious that the PS1 MDS reference stack images are deeper than the SDSS images and the sky background rms of the PS1 image is lower than that of the SDSS image. Therefore the PS1 MDS images are much better for our outer disk study.  

Our sample galaxies are selected from the SDSS-III database using the CASjobs interface\footnote{http://skyserver.sdss3.org/casjobs/}. 
The reason that we use this catalog is because SDSS has many complementary parameters
such as redshifts, galaxy types, properties of neighboring galaxies, etc. 
There are 8 PS1 MDS fields  (MD03 - MD10) observed by the SDSS. 
  In order to choose disk galaxies, we require the fraction attributed to the $r$-band de Vaucouleurs component (fracDeV\_r) in the SDSS composite model to be less than 0.7 \footnote{The galaxy is modeled using a deVaucouleurs profile plus an exponential disk profile. A pure elliptical galaxy should have fracDev=1 and a bulge-less pure exponential disk should have fracDev=0.  According to \citet{mas10}, a galaxy with fracDeV\,$=0.7$ corresponds to a Hubble type of Sa. }  or the galaxy zoo fraction of vote for disk galaxy (p\_cs) to be greater than 50\%.
Similar to PT06, we also require the galaxies to be relatively face-on (  $\rm expAB\_r > 0.5$,  i.e. the ellipticity \footnote{$The ellipticity is defined as e=1-b/a$, where $a$ and $b$ are major and minor semi-axes of the ellipse respectively} is less than 0.5).
In this work, we limit the galaxies by angular size instead of by apparent magnitude. The galaxies should be large
enough to make it possible to measure well-resolved radial profiles ($\rm petroRad\_r >= 5 \, arcsec$).
We also require that the galaxies have a measured spectroscopic redshift by selecting galaxies from the `SpecObj' Casjobs catalog only.
The resulting total number of galaxies using these selection criteria is $782$. The final sample has 698 galaxies after taking out galaxies with an incomplete set of multi-band images.   Since the mother sample of our galaxy sample, the SDSS spectroscopic galaxy sample, is highly complete \citep{str02}, we expect that our sample also has a high completeness. The only source of incompleteness is the size cut, which excludes physically small and distant galaxies.

\begin{figure}[htbp]
\begin{center}
\includegraphics[scale=0.7]{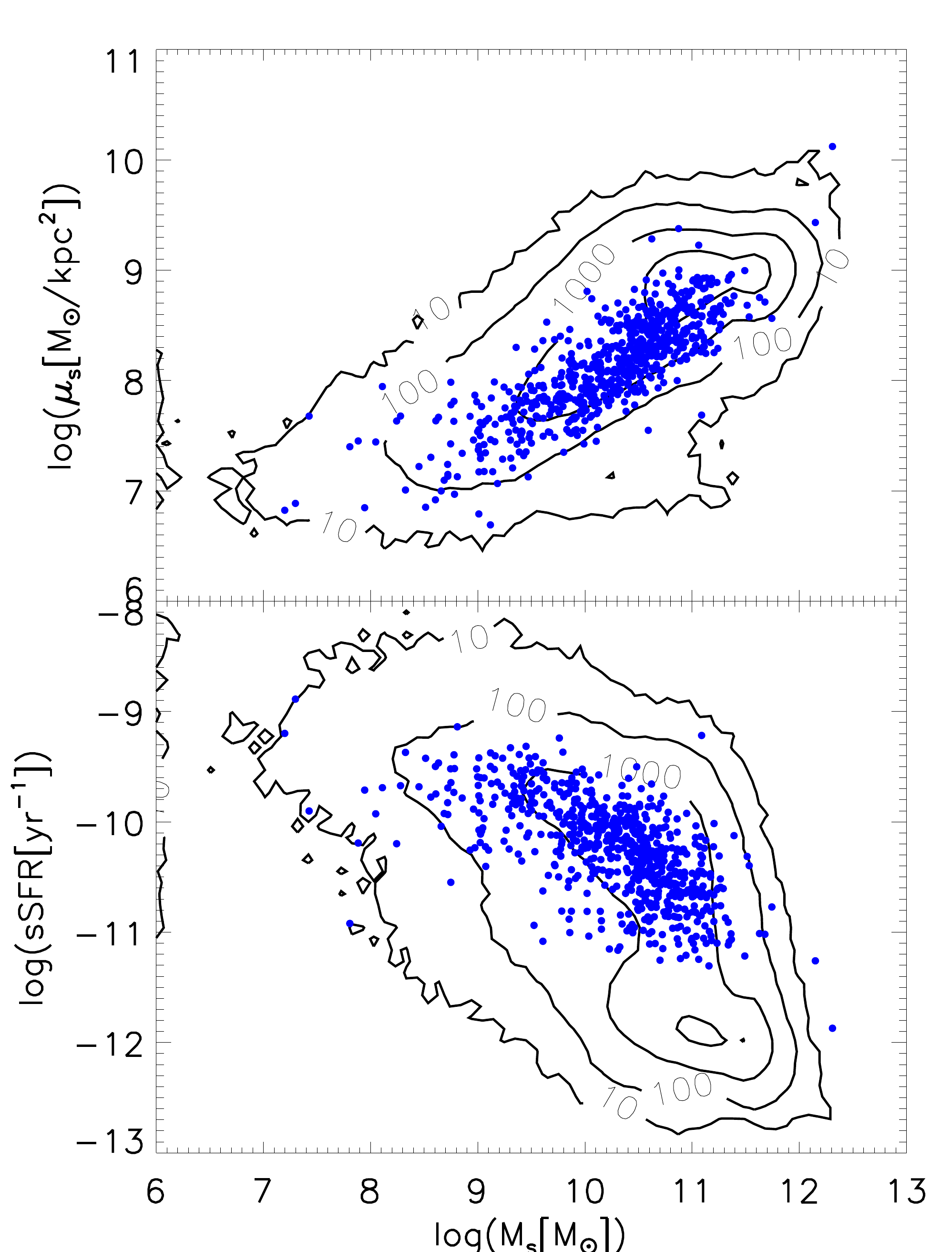}  
\caption{Global properties of our galaxy sample in comparison to SDSS overall.  Upper: Stellar surface mass density versus stellar mass. Lower: specific star formation rate versus stellar mass. Blue dots are our sample galaxies and the background contours show the distribution of the entire MPA-JHU SDSS DR7 sample. Values of the blue points are calculated using PS1 photometry as described in this paper.
}
\label{galaxy_distribution}
\end{center}
\end{figure}

Following \citet{kau03} and \citet{sch07},
Fig. \ref{galaxy_distribution} shows the distribution of our sample using two plots. The first one is the `effective' stellar mass surface  density, $\mu_s$, versus stellar mass, $M_*$. The value for $\mu_s$  is defined as 
\begin{equation}
\mu_s=\frac{M_*}{2\pi r^{2}_{50,z}},
\label{mus_def}
\end{equation} 
where the $r_{50,z}$ is the Petrosian half-light radius in the z band. The other plot is specific star formation rate, $\rm SFR/M_*$, versus $M_*$. Values for $M_*$ and SFR are taken from the MPA-JHU SDSS DR7 release 
\footnote{http://home.strw.leidenuniv.nl/\textasciitilde jarle/SDSS/}. 
The value for $r_{50,z}$ is taken from the SDSS DR8 database.  
It is obvious that our sample is representative of the population of disk galaxies in SDSS over ranges of more than two orders-of-magnitude in stellar mass, mass density, and specific SFR. Our selection of disk dominated galaxies means we have only a sparse sampling of galaxies on the red sequence (sSFR $< 10^{-11} yr^{-1}$ - which are predominantly bulge-dominated galaxies).

\begin{figure}[htbp]
\begin{center}
\includegraphics[scale=0.7]{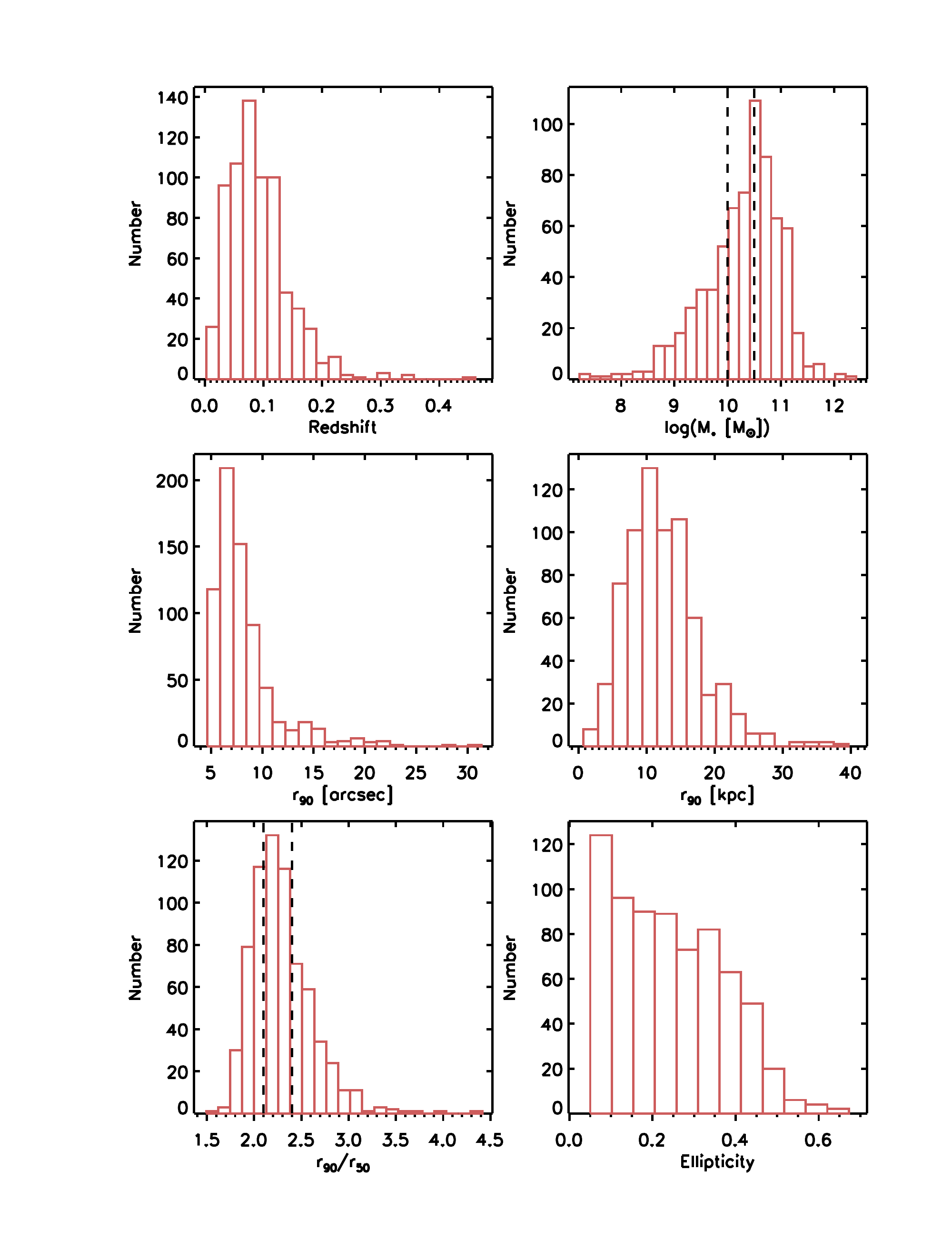}  
\caption{Histograms of redshift (upper left), total stellar masses (upper right), $r$-band r$_{90}$ in arcsec (middle left) and in kpc (middle right), concentration parameter (lower left) and characteristic ellipticity (lower right) of our sample galaxies. The vertical black dash lines indicate the three mass bins (upper right panel) and three concentration bins (lower left panel) we used in this paper.}
\label{rz_mass_histo}
\end{center}
\end{figure}

The distributions of redshift, total stellar mass, $r$-band $r_{90}$, concentration parameter ($C = r_{90}/r_{50}$) and characteristic ellipticity of our sample galaxies are shown in Fig. \ref{rz_mass_histo}.   The $r$-band $r_{90}$ (hereafter $r_{90}$) and $r$-band $r_{50}$ are scale radii defined as the radii containing 90\% and 50\% of the Petrosian flux in the $r$-band respectively.  For our sample galaxies, the $r_{90}$ is roughly 0.7\,$R_{25}$, which we define as the radius of the galaxy to the 25th ABmagnitude isophote in the $r$-band. The parameter $C$ is correlated with galaxy bulge/disk ratio (Gadotti 2009; Lackner \& Gunn 2012), with the range between log $C$ = 0.25 and 0.45 corresponding roughly to bulge/disk ratios of $\sim$ 0 and 1 respectively.  Thus, $C$ can serve as a proxy for Hubble types with smaller $C$ indicating later type disks and larger $C$ indicating earlier type disks.

Since galaxies with different mass can have systematically different Hubble types (concentrations), assembly histories, companion galaxies, halo properties etc. \citep{kau03}, we divided our sample into three subcategories by stellar mass: high mass ($M > 10^{10.5} M_{\odot}$), intermediate mass ($ 10^{10.5} M_{\odot} > M > 10^{10.0} M_{\odot}$), and low mass ($M < 10^{10.0} M_{\odot}$) galaxies. We have also divided our sample into three bins in concentration: $C > 2.4$, $2.4 > C > 2.1$, and $C < 2.1$. These bins roughly correspond to bulge/disk ratios of $B/D > 0.2$, $0.2 > B/D > 0.1$, and $B/D < 0.1$ (Gadotti 2009; Lackner \& Gunn 2012).

We list our sample galaxies in the appendix. A part of the galaxy list is also listed below in Table \ref{gal_list}.

\begin{sidewaystable}[htdp]
\tiny
\caption{Galaxy list example}
\label{gal_list}
\begin{center}
\begin{tabular}{cccccccccccccccccccccccc}
\hline
\hline
ID & Field &  RA & DEC & z & $r_{90}$ & $r_{50}$ & $r_b$& $e$ & M$_*$ & $\mu_s$ & sSFR& $\mu_{r10}$ & $\mu_{r20}$ & $k_{r1}$ & $k_{r2}$& $\mu_{m10}$ & $\mu_{m20}$ & $k_{m1}$ & $k_{m2}$& $\Upsilon_{10}$ & $\Upsilon_{20}$ & $k_{\Upsilon1}$ & $k_{\Upsilon2}$ \\
 & & & & &  [kpc] & [kpc] & [kpc] &  & [M$_{\odot}$] & [$\frac{\rm M_{\odot}}{kpc^2}$] & [yr$^{-1}$] & \multicolumn{2}{c}{[mag] } &\multicolumn{2}{c}{[$mag/r_{90}$] } 
 & \multicolumn{2}{c}{[$\frac{M_{\odot}}{kpc^2}$]}  & \multicolumn{2}{c}{[$r_{90}^{-1}$]} & 
  \multicolumn{2}{c}{[$M_{\odot}/L_{\odot}$]}  &\multicolumn{2}{c}{[$r_{90}^{-1}$]} \\
\hline

1   & MD03 &       129.289 &       44.3036 & 0.13 & 10.96 & 5.26 & 10.71 & 0.05 & 10.62 & 8.46 & -10.65 & 20.4 & 19.5 & 3.36 & 4.28 & 8.7 & 8.9 & -1.19 & -1.54 & 0.3 & 0.0 & 0.13 & -0.14 \\  
2   & MD03 &       130.200 &       44.2520 & 0.08 & 7.83 & 3.68 & 7.25 & 0.29 & 10.01 & 8.05 & -10.24 & 20.5 & 19.0 & 2.53 & 4.38 & 8.4 & 9.0 & -1.05 & -1.47 & 0.1 & 0.1 & 0.01 & 0.13 \\  
3   & MD03 &       131.078 &       45.0381 & 0.15 & 13.43 & 5.91 & 11.06 & 0.24 & 10.91 & 8.80 & -10.69 & 20.6 & 19.7 & 2.81 & 4.07 & 8.8 & 8.6 & -1.37 & -1.17 & 0.5 & -0.3 & -0.36 & 0.25 \\  
4   & MD03 &       130.901 &       44.0561 & 0.11 & 9.95 & 5.24 & 7.42 & 0.09 & 10.35 & 8.12 & -10.11 & 20.5 & 19.1 & 2.87 & 4.35 & 8.6 & 8.7 & -1.46 & -1.47 & 0.3 & -0.6 & -0.26 & 0.25 \\  
5   & MD03 &       131.361 &       44.9439 & 0.15 & 17.67 & 6.88 & 11.68 & 0.12 & 11.41 & 9.00 & -11.13 & 21.0 & 19.6 & 2.40 & 4.67 & 8.8 & 8.9 & -1.46 & -1.25 & 0.4 & 0.2 & -0.24 & -0.34 \\  
6   & MD03 &       130.952 &       43.8009 & 0.19 & 16.70 & 8.08 & 15.64 & 0.17 & 10.82 & 8.30 & -10.15 & 20.6 & 19.1 & 2.94 & 4.59 & 8.8 & 9.1 & -1.66 & -1.85 & 0.4 & -0.1 & -0.52 & 0.09 \\  
7   & MD03 &       131.782 &       43.6363 & 0.09 & 8.97 & 4.62 & 6.37 & 0.21 & 10.03 & 7.99 & -10.01 & 20.9 & 19.3 & 2.33 & 4.30 & 8.3 & 8.7 & -1.10 & -1.64 & 0.1 & -0.3 & -0.27 & 0.41 \\  
8   & MD03 &       130.372 &       43.2660 & 0.18 & 14.92 & 7.55 & 9.51 & 0.05 & 10.66 & 8.26 & -9.69 & 20.1 & 19.7 & 3.15 & 3.84 & 8.8 & 8.6 & -1.78 & -1.73 & 0.3 & -0.6 & -0.81 & 0.18 \\  
9   & MD03 &       129.000 &       44.2391 & 0.31 & 24.48 & 11.36 & 17.24 & 0.10 & 11.51 & 8.68 & -10.31 & 20.8 & 19.7 & 2.24 & 4.48 & 9.0 & 8.6 & -1.48 & -1.21 & 0.5 & -0.4 & -0.60 & 0.74 \\  
11  & MD03 &       129.123 &       44.1469 & 0.13 & 13.20 & 5.93 & 8.66 & 0.09 & 10.65 & 8.47 & -10.19 & 20.5 & 19.6 & 2.51 & 4.25 & 8.7 & 8.4 & -1.66 & -1.08 & 0.4 & -0.4 & -0.45 & -0.15 \\  
12  & MD03 &       131.171 &       44.3604 & 0.09 & 9.20 & 3.92 & 8.36 & 0.29 & 10.67 & 8.69 & -10.13 & 19.0 & 18.6 & 3.16 & 3.95 & 9.2 & 8.7 & -1.57 & -1.16 & 0.3 & -0.4 & -0.34 & 0.30 \\  
13  & MD03 &       130.159 &       44.1456 & 0.08 & 8.80 & 3.84 & 5.86 & 0.05 & 10.28 & 8.33 & -10.51 & 20.2 & 20.4 & 4.49 & 3.51 & 8.8 & 8.7 & -1.50 & -1.84 & 0.4 & 0.1 & -0.34 & 0.27 \\  
14  & MD03 &       131.413 &       45.0530 & 0.14 & 14.22 & 6.66 & 9.47 & 0.32 & 10.85 & 8.35 & -10.93 & 20.7 & 20.1 & 2.10 & 3.20 & 9.0 & 9.1 & -1.43 & -1.70 & 0.6 & 0.2 & -0.55 & -0.14 \\  
15  & MD03 &       130.280 &       44.0019 & 0.09 & 9.70 & 5.10 & 6.47 & 0.22 & 10.00 & 7.77 & -10.17 & 21.2 & 20.0 & 2.48 & 3.86 & 8.4 & 9.1 & -1.54 & -2.30 & 0.2 & 0.5 & -0.42 & -0.49 \\  
16  & MD03 &       131.864 &       44.2269 & 0.11 & 11.76 & 6.11 & 11.86 & 0.13 & 10.18 & 7.99 & -9.97 & 21.0 & 18.0 & 2.43 & 5.74 & 8.2 & 9.2 & -1.26 & -2.15 & 0.1 & -0.5 & -0.23 & 0.17 \\  
17  & MD03 &       130.336 &       45.3785 & 0.15 & 15.11 & 7.63 & 12.48 & 0.05 & 10.47 & 7.90 & -9.90 & 20.9 & 19.8 & 2.83 & 4.41 & 8.5 & 8.4 & -1.74 & -1.56 & 0.2 & -0.5 & -0.72 & 0.23 \\  
18  & MD03 &       130.371 &       43.7678 & 0.13 & 14.51 & 6.48 & 10.60 & 0.13 & 10.64 & 8.33 & -10.49 & 20.9 & 20.8 & 3.20 & 3.55 & 8.7 & 8.6 & -1.68 & -1.52 & 0.5 & 0.2 & -0.44 & -0.33 \\  
19  & MD03 &       130.212 &       44.7354 & 0.15 & 14.85 & 6.64 & 15.40 & 0.39 & 11.04 & 8.64 & -10.77 & 20.3 & 18.1 & 2.50 & 4.58 & 9.1 & 9.0 & -1.46 & -1.39 & 0.6 & -0.5 & -0.46 & 0.34 \\  
20  & MD03 &       128.884 &       44.8572 & 0.13 & 13.74 & 5.83 & 10.04 & 0.35 & 10.97 & 8.74 & -10.50 & 19.9 & 19.5 & 2.83 & 3.46 & 9.1 & 9.1 & -1.54 & -1.42 & 0.6 & 0.1 & -0.39 & 0.00 \\  
22  & MD03 &       130.514 &       43.0018 & 0.09 & 9.87 & 4.77 & 7.17 & 0.30 & 9.76 & 7.53 & -9.75 & 21.0 & 19.6 & 2.59 & 4.10 & 8.3 & 8.1 & -1.46 & -1.32 & 0.1 & -1.0 & -0.57 & 0.62 \\  
23  & MD03 &       129.013 &       43.8074 & 0.15 & 15.98 & 6.85 & 11.30 & 0.16 & 10.80 & 8.33 & -10.58 & 20.6 & 20.3 & 3.35 & 3.87 & 8.6 & 8.6 & -1.44 & -1.25 & 0.4 & 0.1 & -0.03 & 0.20 \\  
24  & MD03 &       129.303 &       43.3983 & 0.15 & 14.73 & 6.93 & 9.81 & 0.33 & 10.75 & 8.33 & -10.35 & 20.8 & 19.7 & 2.21 & 3.96 & 8.7 & 8.5 & -1.51 & -1.12 & 0.5 & -0.2 & -0.57 & 0.28 \\  
25  & MD03 &       130.742 &       43.6373 & 0.09 & 10.79 & 4.09 & 7.54 & 0.25 & 10.64 & 8.75 & -9.81 & 18.9 & 19.0 & 3.89 & 3.61 & 8.9 & 8.4 & -1.56 & -0.86 & 0.1 & -0.7 & -0.18 & 0.36 \\  
26  & MD03 &       131.239 &       45.0734 & 0.12 & 13.61 & 5.19 & 9.40 & 0.14 & 10.10 & 7.95 & -9.66 & 20.1 & 21.7 & 4.71 & 3.03 & 8.5 & 7.8 & -1.94 & -0.95 & -0.0 & -0.2 & -0.27 & -0.30 \\  
27  & MD03 &       129.171 &       45.0970 & 0.04 & 4.51 & 2.05 & 3.60 & 0.23 & 9.50 & 8.08 & -10.15 & 20.1 & 19.5 & 3.20 & 3.83 & 8.6 & 8.4 & -1.59 & -1.14 & 0.2 & -0.4 & -0.38 & 0.02 \\  

\hline
\hline

\end{tabular}
    \begin{tablenotes}
      \small
      \item Note -- Galaxy list example. This table contains basic parameters of 25 galaxies selected from our 698-galaxy sample. The complete table of the whole sample can be downloaded online in electronic form. The RA, DEC and redshift (z) are from SDSS DR8 catalog. The $r_{90}$, $r_{50}$, break radius $r_b$, $e$,  M$_*$, $\mu_s$ (defined in Eq. \ref{mus_def}), inner and outer $r$-band SB profile slope ($k_{r1}$ and $k_{r2}$), stellar mass surface density profile slope ($k_{m1}$ and $k_{m2}$), $r$-band M/L profile slope ($k_{\Upsilon_r1}$ and $k_{\Upsilon_r2}$) and sSFR are measured using PS1 data. The definition of the slopes can be found in Section \ref{detailed_1d}. The $\rm M_*$, $\mu_s$ sSFR, $\mu_{m10}$, $\mu_{m20}$, $\Upsilon_{10}$ and $\Upsilon_{20}$ are in logarithmic scale.
          \end{tablenotes}
\end{center}
\end{sidewaystable}%

\section{Method}
\label{method}

\subsection{Data processing and radial profile extraction}
\label{data_processing}

  We create 8' by 8' (1920pix by 1920pix) image cutouts for each galaxy. All the images are background subtracted stacked images. The background level was determined by the PS1 Image Processing Pipeline (IPP) using blocks of 400 by 400 pixels (100" by 100"). Since most of our galaxies are smaller than this block size the background subtraction performed by IPP should be appropriate and not cause over-subtraction of low spatial frequency emission. 
However, in order to study the outer disk more accurately, 
we need to do our own custom background subtraction. Here we describe the pipeline used to mask out all the background/foreground galaxies and stars, derive galactic geometric parameters, subtract the background and extract multi-band radial profiles. All the data are processed in IRAF and IDL.

Since we have a fairly large sample of galaxies, it would be a very tedious job to mask all the contaminating sources by hand. 
Therefore we developed a semi-automatic graphic user interface (GUI) software to mask the contaminants such as foreground stars and background galaxies. First, we extract all the sources by using SExtractor applied to a 5-band composite chi-square detection image \citep{sza99}. All sources above 2$\sigma$ local background variation are detected by SExtractor.  We then examine the 3-color image and the mask segments on top of the 3-color image and select the mask segments which correspond to the target galaxy by clicking the mouse.  We can also mask out SExtractor undetected stars and unwanted regions using circular apertures defined within this GUI software.    The masks are then grown using the IDL function DILATE with a square shape operator. The size of the operator matrix is proportional to the size of the mask and has a minimum value of 5 pixel by 5 pixel.

The characteristic center, position angle (PA) and ellipticity of each galaxy are derived using the IRAF \footnote{IRAF is the Image Reduction and Analysis Facility, a general purpose software system for the reduction and analysis of astronomical data. IRAF is distributed by the National Optical Astronomy Observatory, which is operated by the Association of Universities for Research in Astronomy, Inc., under cooperative agreement with the National Science Foundation.} task {\it ellipse} mainly following the two-step procedure provided by \citet{li11}. In the first step, we fit all three parameters (center, PA and ellipticity) using the $i$-band image. We then have radial profiles of the center coordinates ($x_c(r)$ and $y_c(r)$), position angle (PA(r)) and ellipticity (e(r)) for each galaxy. The characteristic center is determined using the origin value corresponding to the annuli at $0.5\,R_{25}$ 
\begin{equation}
\begin{array}{l}
\displaystyle X_c = x_c(r=0.5\,R_{25}) \\
\displaystyle Y_c = y_c(r=0.5\,R_{25}).
\end{array}
\end{equation}
In the second step, we fix the center determined in the first step and re-fit the PA and ellipticity. The final PA and ellipticity are held constant using the value of the annuli at $R_{25}$. We then visually examine every ellipse fit and re-fit the bad ones by hand.

After deriving the characteristic center, PA and ellipticity of the galaxy, we extract the SB profile over the radial range from 5 pixels (1.25'') up to 3 $R_{25}$. The SB profiles are extracted with the characteristic parameters fixed in rings with the widths of the rings increasing logarithmically (in log step of 0.03). The values of the unmasked pixels within each ring are averaged to get the mean flux of each ring. The radius assigned to each ring is the average value of the inner and outer boundaries of each ring along the major axis. At this point we have a SB radial profile including the sky background (after deriving all the mean fluxes of every ring). 

The outer most ring (  the region between $5\,r_{90}$ and $6\,r_{90}$) is used to derive the background. The outer most ring is divided azimuthally into 36 regions (10 degrees for each region). We calculate the mode of each region and then derive the background value by averaging all the 36 mode values. The standard deviation of the 36 values is taken as 1$\sigma$ of the sky background. We then subtract the background value from each ring to get the background subtracted SB radial profile in each band.   The error source for the SB is the combination of the sky background rms and the Poisson noise of the annulus. We did not include the dispersion of pixel values inside the annulus in our S/N calculation because this may be exaggerated by the internal structures such as spiral arms and star-forming regions.

The 5-band background subtracted SB radial profiles are corrected for Galactic reddening using the \citet{cal89} extinction curve assuming foreground E(B-V) values based on the \cite{sfd98} dust maps.
The K-corrections \footnote {Note the input SB profiles for MAGPHYS are not K-corrected. The K-corrected SB profiles are only used for color analysis.} were derived using the K-corrections calculator developed by \citet{chi10} and \citet{chi12}.

\subsection{Deriving the stellar mass: SED fitting}
Previous studies that convert SB into mass surface density usually use a single mass-to-light ratio \citep[e.g.][]{fre70} or simple empirical mass-to-light ratios 
\citep[e.g.][]{bak08}. To improve on these simple methods, 
in this paper, we use the Multi-wavelength Analysis of Galaxy Physical Properties (MAGPHYS) program \citep{dac08,bc03,cf00} to interpret the 5-band SB radial profiles. To do this, we do SED fitting using PS1 {\it grizy} photometry for each annulus of each galaxy and combine the fitted results to get radial profiles of various physical parameters, such as the stellar mass, average age, recent star formation rate (averaged over the last 0.1 Gyr). MAGPHYS supplies a comprehensive library of model SEDs (based on \citet{bc03}) at the same redshift and in the same photometric bands as the observed galaxy including various star formation histories (generally declining smoothly with time but allowing for bursts at late epochs). The model SEDs are made using a wide range of plausible physical parameters. MAGPHYS then compares the observed SED with the model SEDs and builds a marginalized likelihood distribution for each physical parameter. As a result, our estimation of uncertainty for each inferred parameter implicitly includes a contribution from any model parameter degeneracies in addition to the limitations from the photometry.   Nevertheless we took pains to reduce the impact of degeneracy as described below, for the specific case of extinction as a function of galactocentric radius.

\begin{figure}[htbp]
\begin{center}
\includegraphics[scale=0.35,angle=90]{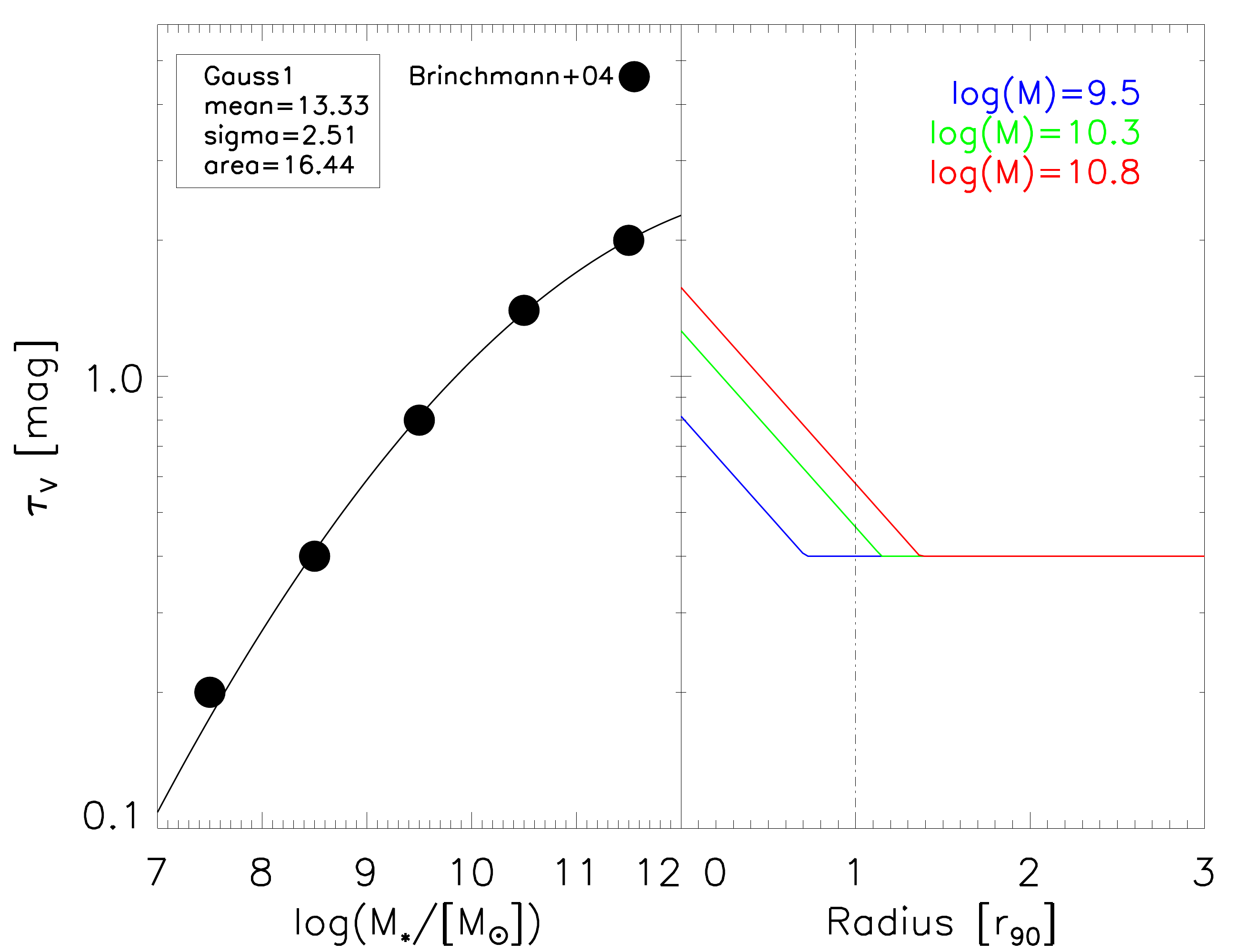}  
\includegraphics[scale=0.35,angle=90]{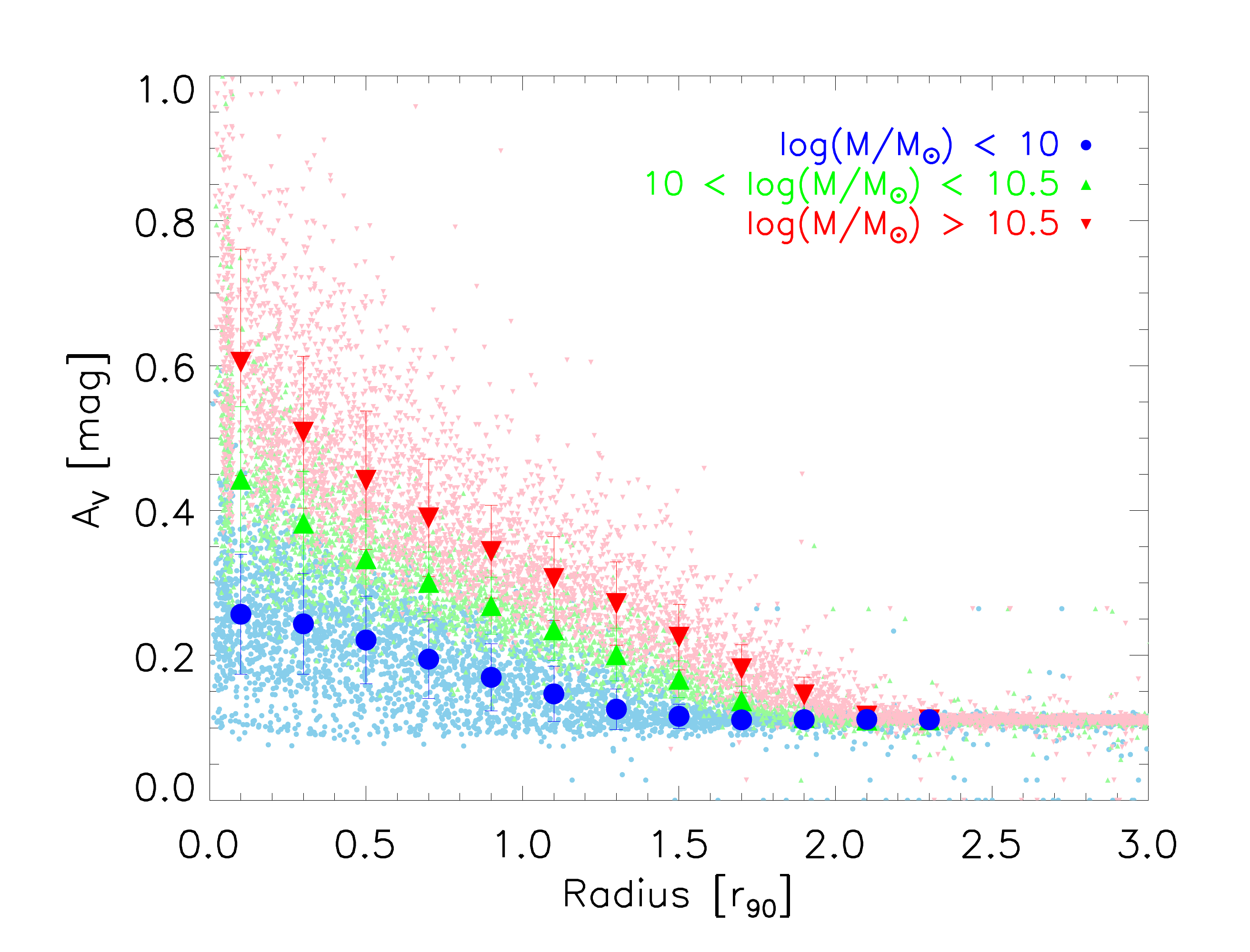}  

\caption{MAGPHYS extinction priors (upper panels) and output Av (lower panel). Upper-left: central total dust extinction of the {\it emission-line gas} in low-redshift star forming galaxies as a function of stellar mass. The black dots are scaled data from \citet{bri04} and the solid line is the Gaussian functional fit to the data. Upper-right: parameterized typical radial variation in the emission-line extinction used for the MAGPHYS fitting; only three example galaxies are shown here: high mass (red), intermediate mass (green) and low mass (blue) galaxies; each parameterization starts with the corresponding central extinction from the left panel and decreases exponentially until hit the lower limit $\tau_V=0.4$. To implement such an extinction prior, the fitted $\tau_V$ values are allowed to range from 0 to 2 times the radial extinction curves shown here.  Lower panel: The MAGPHYS output $\rm A_{V}$ composite radial profile. This refers to the extinction affecting the {\it starlight}. The $\rm A_V$ output for the annuli of all the galaxies are plotted versus radius (normalized in $r_{90}$) The points are color coded in stellar mass as describe in the legend and in Section \ref{data}. }
\label{av_plot}
\end{center}
\end{figure}

The effects of dust reddening and age on the optical SED of galaxies conspire to produce a very similar relationship between color and stellar mass-to-light ratio (M/L) \cite[e.g.][]{bdj01}. This means that the optical SED can yield a robust measurement of the stellar M/L. However, if we wish to make inferences about the age of the stellar population we need some independent way to estimate the extinction. The optimal method would be to use high-resolution ($\sim$arcsec) ultraviolet and infrared images, which we do not have. Instead we use an extinction prior in our model fitting that is based on galaxies that are similar to those in our sample, but for which the radial dust extinction has been independently measured. We show these priors in Fig. \ref{av_plot}. They were calculated as follows. 
  The extinction radial profile is assumed to decay from a central extinction value $\tau_{\rm V0}$ exponentially until reaches a lower limit $\tau_{\rm V,min}$. We set the scale-length of the exponential part of the extinction profile for all galaxies to be $1 r_{90}$ \citep{mm09}. The central extinction value varies from galaxy to galaxy depending on total stellar mass of the galaxy \citep{bri04}. We fit a  Gaussian functional form to the observed $\tau_{\rm V0}$ of low-redshift star forming galaxies provided by \citet{bri04} so that we can determine the $\tau_{\rm V0}$ for every galaxy as a function of its stellar mass. The extinction lower limit is set to be $\tau_{\rm V,min}=0.4$ and it usually meets the exponential profile in the radial range  $r / r_{90} =1 - 2.5$. The Gaussian fitting to the $\tau_{\rm V0}$ and the assumed extinction profiles are shown in the upper panel of Fig. \ref{av_plot}. The actual extinction prior is then set individually for each annulus to be between the 1/2 and 2 times of the parameterized typical extinction described above.
  This choice of $\tau_{\rm V}$ is justified by the MAGPHYS A$_{\rm V}$ output, which is also plotted in Fig. \ref{av_plot}. The A$_{\rm V}$, which is the stellar extinction, is defined as 
\begin{equation}
{\rm A_V}=1.086 \mu \tau_{\rm V},
\end{equation}
where $\mu$ is the fraction of $\tau_V$ contributed by dust in the diffuse ISM. The MAGPHYS output $A_{\rm V}$ for our sample is consistent with previous spectroscopic observations of similar galaxies \citep[CALIFA group in private communication and][]{mm09}. Actually, the MAGPHYS output of stellar mass is not very sensitive to the extinction prior. The stellar mass only varies about a few percent  when we double or reduce by half the extinction prior. 
While the mean age and sSFR are more sensitive to the extinction, they change by only +/-0.05 dex and +/-0.1 dex respectively when we vary the extinction priors in this way.  The most important point is that the shapes of the radial profiles of the stellar surface mass density, M/L ratio, age and sSFR do not change significantly as we vary the extinction priors.

\begin{figure}[htbp]
\begin{center}
\includegraphics[scale=1]{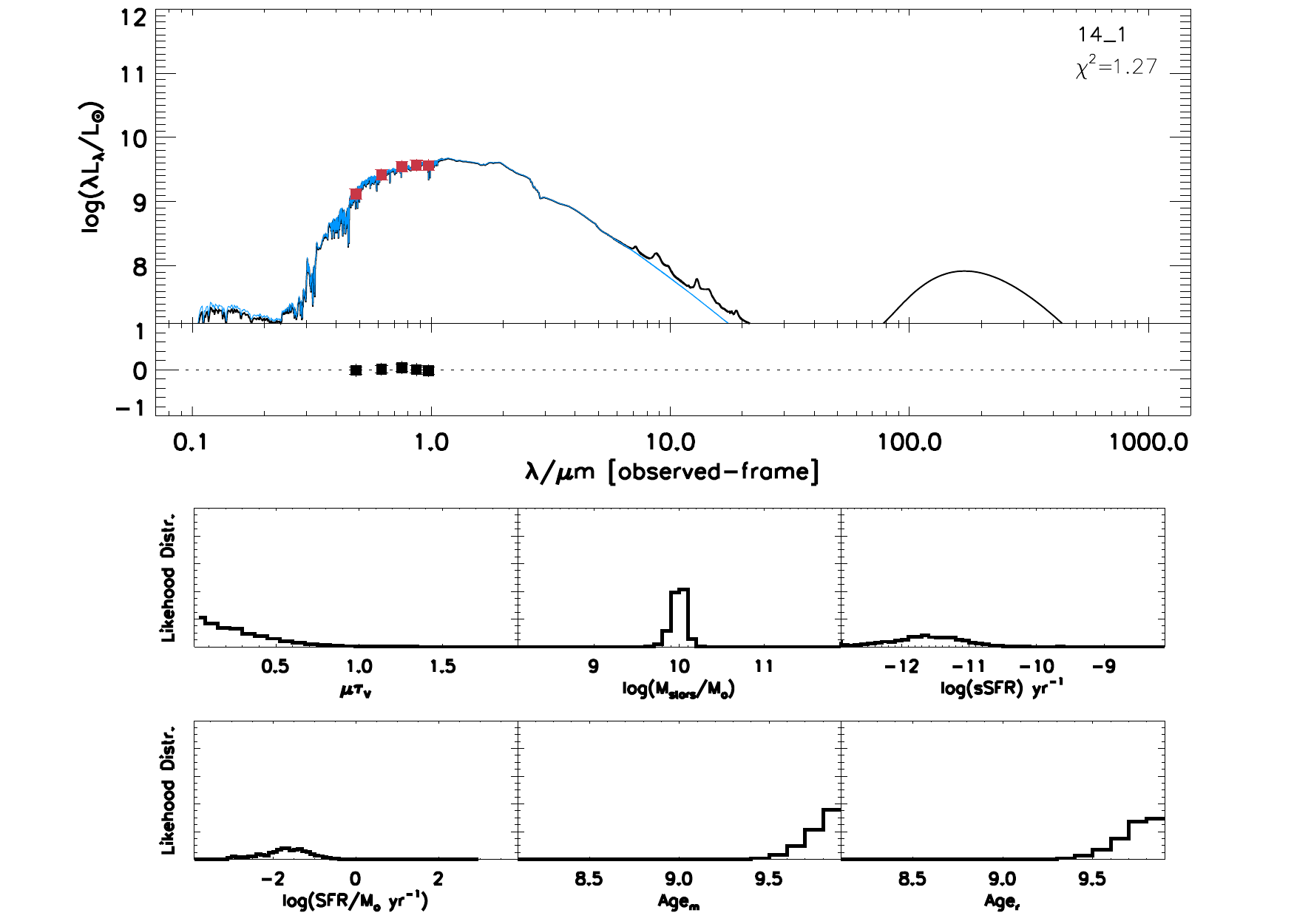}  
\caption{Example of MAGPHYS SED fitting and the distribution of different physical parameters. Upper panel: The red points are the observed 5-band photometry for one annulus of one of our galaxies. The black curve is the fitted attenuated SED and the blue curve is the fitted unattenuated SED. The black dots show the residuals. Lower panel: Marginalized likelihood distributions of various physical parameters: clockwise from upper left, these are the dust attenuation of the starlight, the stellar mass, the specific SFR, the log r-band-weighted mean age in years, the log of the mass-weighted mean age in years, and the SFR.}
\label{magphys1}
\end{center}
\end{figure}

One example of MAGPHYS SED fitting and the distribution of different physical parameters is shown in Fig. \ref{magphys1}. 
We take the 50\% (median) value of the likelihood distribution to be the value of that physical parameter and take  half of the difference between the 16th and 84th percentile values to be the 1$\sigma$ uncertainty.

The SED fitting using annular photometry may lead to biases in our results because each annulus  unavoidably blends contributions from physically unassociated regions, which may have different star formation histories.  For instance, arm and inter-arm locations are jointly measured within an annulus. Hence, we may be somewhat insensitive to very low levels of star formation (such as seen in XUV-disks) if the optical luminosity is dominated by an older stellar population. The best way to derive the stellar mass radial profiles might be to use pixel SED fitting and then to sum up the parameter results. However, this method is much more computational intensive (cf.  POGS project, Vinsen\,\&Thilker 2013; Thilker et al. 2014 in prep) and difficult to apply to the low SB regime we are attempting to probe.

\section{The 2D stack images}
\label{stacked_2d}

\subsection{Motivation}

The $g-r$ color profiles measured by \citet{bak08} have revealed that the outer disk is different from the inner disk by having a redder color and (probably) an older age. The `red' and `old' descriptions also apply to stellar halo stars, which raises the possibility that we could be seeing the stellar halo instead of the outer disk.
\citet{bak12} studied 7 of the PT06 galaxies using deep imaging data and claimed that up-bending disks were caused by the contamination of halo light in the outer disk. It is therefore interesting to investigate where the stellar halo starts to emerge in the outer region and make a clear definition of the outer border of the stellar disk region. 
Previous studies \citep{zib04, dsou14} show that the stellar halo usually has a much rounder shape than the stellar disk, so the best way to study the stellar halo is to examine the 2D stacked images and look for changes in the galaxy shape.

\citet{zib04} stacked 1047 local (z $\sim$ 0.05) edge-on disk galaxies imaged in five bands by SDSS.  They contract or expand each galaxy before adding it to the stack in such way that the scale-length of each galaxy has the same angular size. \citet{dsou14} made stack images using a much larger sample, 45508 galaxies, and the sample contains both early type and late type galaxies. \citet{dsou14} made a further step by rebinning each galaxy to produce an image scaled to the common value of redshift $z=0.1$. They also divided their sample into different mass bins to study the properties of stellar halo as a function of stellar mass. However, they did not rescale the size of the galaxies as done by \citet{zib04}. Assuming the exposure time for each SDSS image is about 100s, the total exposure times of the stack images made by \citet{zib04} and \citet{dsou14} are about 100ks and 4.55Ms. Since typical exposure time of our PS1 MD images is about 10ks, the total exposure time of our sample is about 10Ms. 

\subsection{Stack method and the stack images}

Here we process and stack the galaxy images following a method combining the techniques provided by \citet{zib04} and \citet{dsou14}: Galaxy images and masks are shifted, rotated and contracted/expanded by the IRAF task GEOTRAN so that the transformed image is centered on the center of the galaxy, the major axis of the galaxy is horizontal and  $r_{90}$ is 7'' (28 pixels). The GEOTRAN task was run without flux conservation so that the SB is preserved during the transformation. We then scale the linear SB $I_{\nu}$ to redshift $z=0.1$ according to the equation 
\begin{equation}
I_{\nu,z=0.1} = I_0 (\frac{1+z}{1+0.1})^3,
\end{equation}
where $I_0$ is the original intensity in unit of $\rm erg\,s^{-1}cm^{-2}\,Hz^{-1}\,arcsec^{-2}$ and $z$ is the redshift of the galaxy. This equation corrects both the $(1+z)^4$ cosmological SB dimming and the $(1+z)^{-1}$ frequency compression effect on the monochromatic flux. The SB is then corrected for Milky Way extinction according to \citet{sfd98}. The K-correction is ignored.
Note that \citet{dsou14} did the stacking in a different way: they ran the IRAF GEOTRAN with flux-conservation so that the SB is not preserved. They also did not count for the $(1+z)^{-1}$ frequency compression effect.

We then apply the masks to each image and subtract the sky background calculated using the IDL routine MMM and combine the masked background subtracted images to make the stack image. For each pixel of the stack image, we only use intensity values within the 16th and 84th percentiles of the pixel value distribution at that specific pixel. Following \citet{zib04} we made a mean stack image, a median stack image and a mode stack image.  The mode stack image is calculated using the approximation \citep{zib04}
\begin{equation}
Mode\,=\,3\,\times\,Median\,-\,2\,\times\,Mean.
\end{equation}
We also have the rising background problem in the outer regions of the mean and median stack images \citep{zib04}. Therefore, we use the mode stack images, which are much less affected by this problem. Following \citet{dsou14}, we make stack images using sample galaxies divided in three different mass bins: high mass  ($M_* > 10^{10.5}M_{\odot}$) galaxies, intermediate mass ($10^{10}M_{\odot}  <  M_* < 10^{10.5} M_{\odot}$) galaxies and low mass  ( $M_* < 10^{10}M_{\odot}$) galaxies respectively. We also make a super stack image using all the galaxies in our sample.  

\begin{figure}
\begin{center}
\includegraphics[scale=0.3]{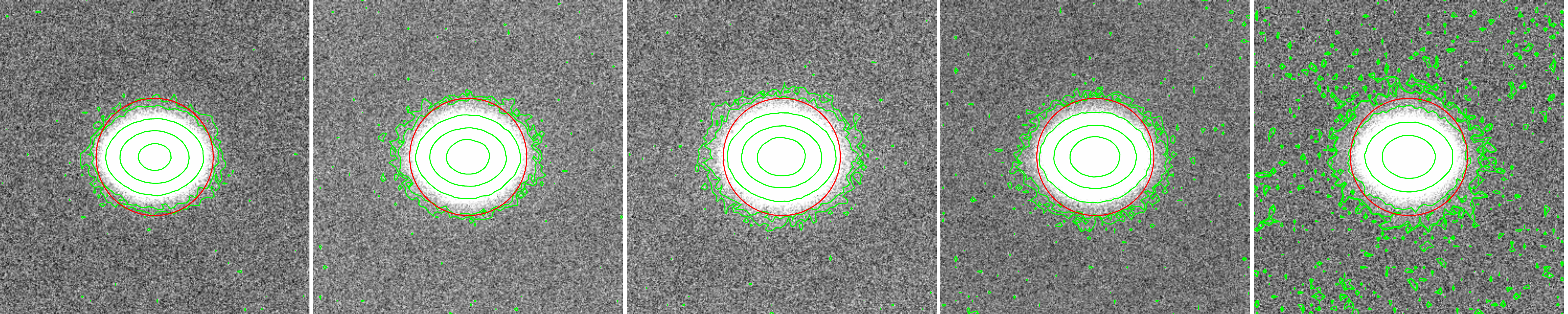}  
\includegraphics[scale=0.3]{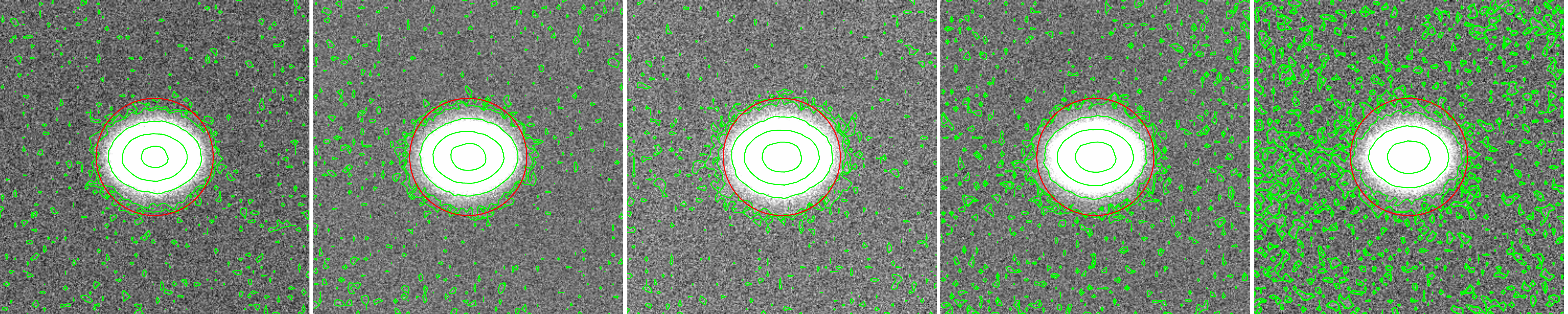}  
\includegraphics[scale=0.3]{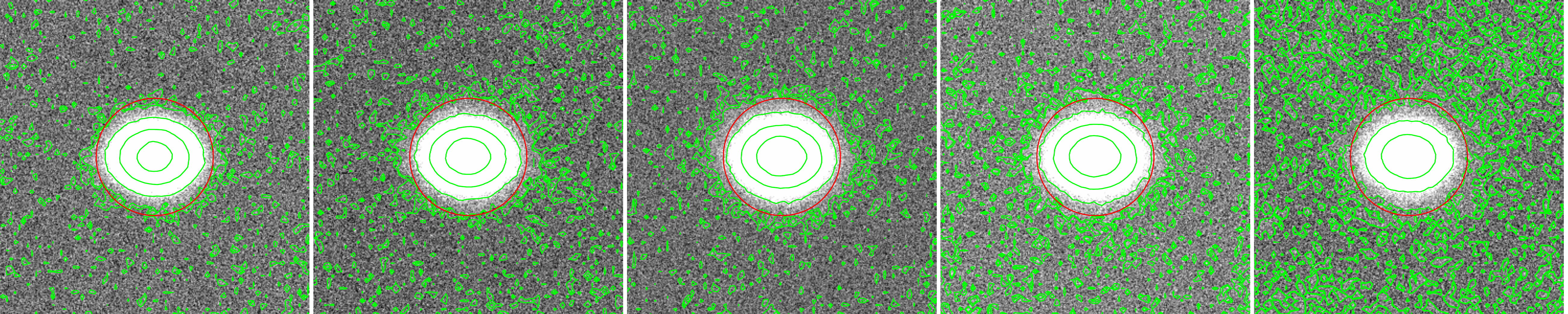}  
\includegraphics[scale=0.3]{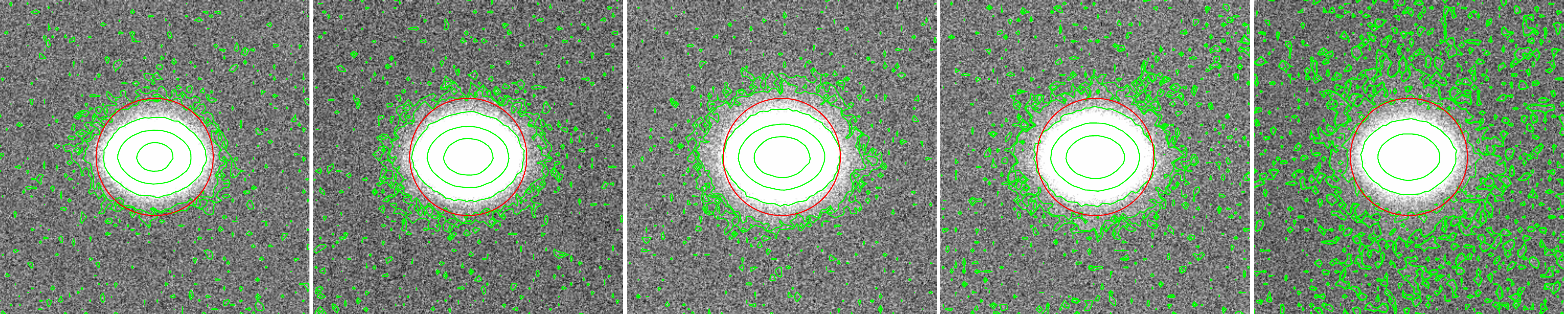}  
\caption{2D stack images. From top to bottom: all galaxies, low-mass galaxies, intermediate-mass galaxies, high-mass galaxies. From left to right: g, r, i, z, y band. The contour levels represent 23, 25, 27, 29, 30 ABmag/arcsec$^2$. The red circles indicate $2\, r_{90}$. The 30 ABmag/arcsec$^2$ contour is close to $3\,r_{90}$.}
\label{ps12_fig_stackimg}
\end{center}
\end{figure}

The stack images are shown in Fig. \ref{ps12_fig_stackimg}. We plot isophote contours (green lines) and 2\,$r_{90}$ (14'', red circles) on top of the images. It is obvious that the emission of the stacked galaxy extends out of $2\,r_{90}$. The isophotes inside the $2\,r_{90}$ all have similar ellipticity, while isophotes outside the $2\,r_{90}$ are significantly rounder than those inside $2\,r_{90}$. The extended PSF wing (up to $3\,r_{90}$) might  cause some of the emission outside $2\,r_{90}$ but should have minimum effect on disk light within $2\,r_{90}$ (cf. appendix \ref{psf} for detailed discussion).

\subsection{Evidence of the stellar halo} 

\begin{figure}
\begin{center}
\includegraphics[scale=0.7, angle=90]{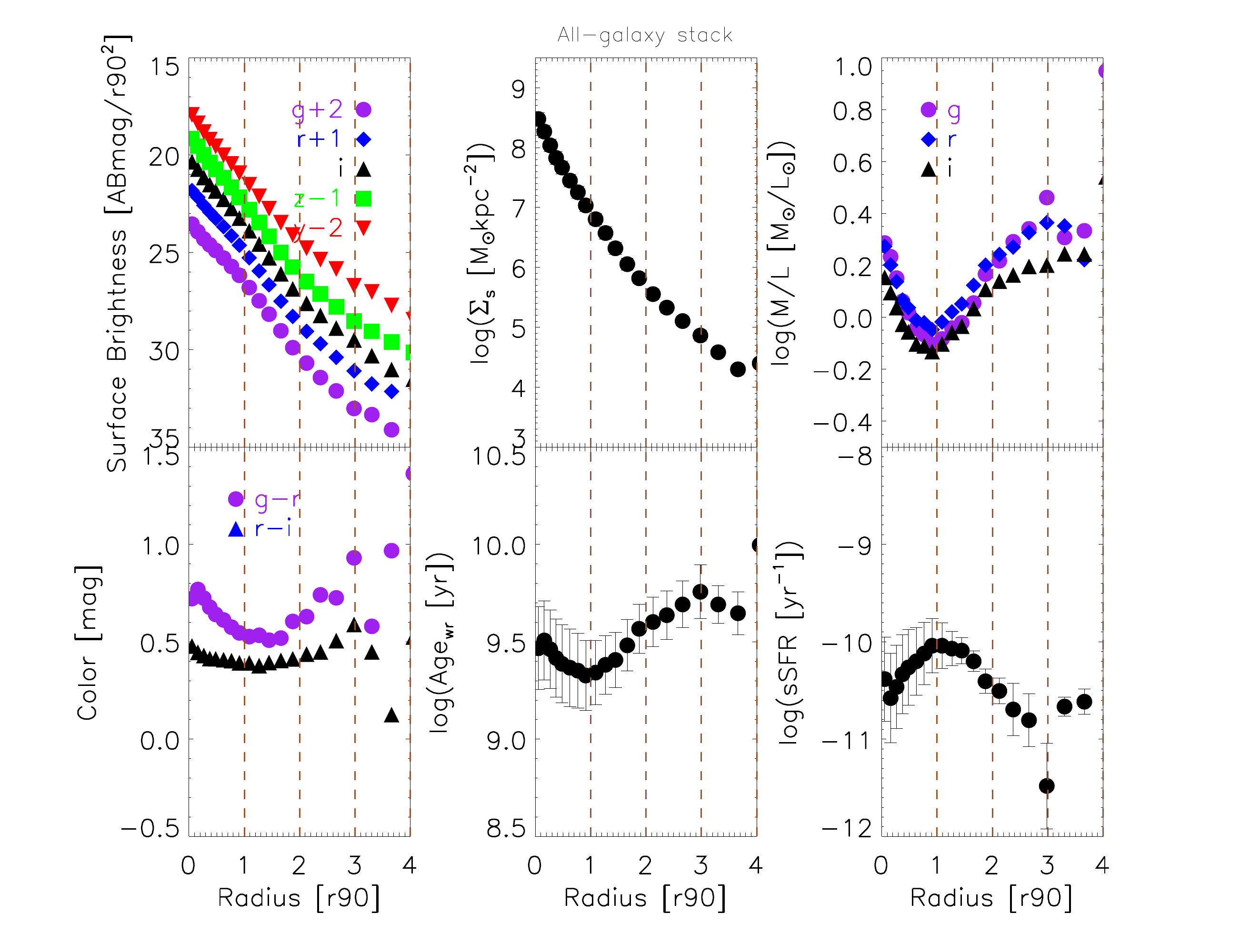} 

\caption{Radial profiles of stack image of all galaxies. Vertical dash lines show 1, 2, 3 times $r_{90}$. The profiles show (clockwise from the upper left) the multi-band surface brightness, the stellar surface mass density, the mass-to-light ratios in three bands, the specific SFR, the log of the r-band-weighted age, and the colors. The up-turn/break in the stellar surface mass density profile at around $2,r_{90}$ and red colors beyond that radius suggest a transition from disk to halo at around this radius.}
\label{ps12_fig_all_magphys}
\end{center}
\end{figure}

\begin{figure}
\begin{center}
\includegraphics[scale=0.7, angle=90]{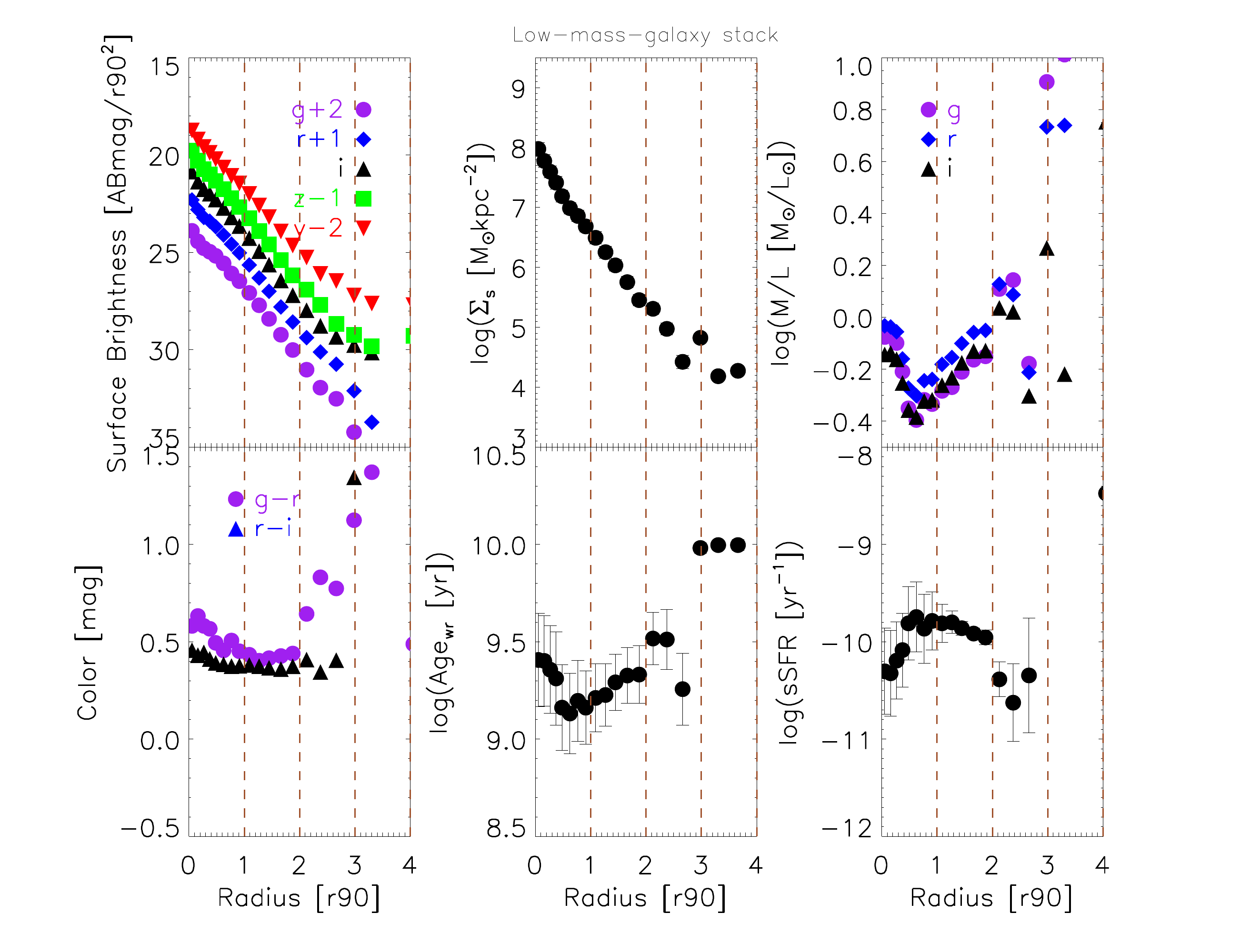} 

\caption{Radial profiles of stack image of low mass galaxies. The profiles inside $3\,r_{90}$ are robust, however, the results outside $3\,r_{90}$ are unreliable because the SB profiles are dominated by the sky background uncertainties in this regime.}
\label{ps12_fig_lowm_magphys}
\end{center}
\end{figure}

\begin{figure}
\begin{center}
\includegraphics[scale=0.7, angle=90]{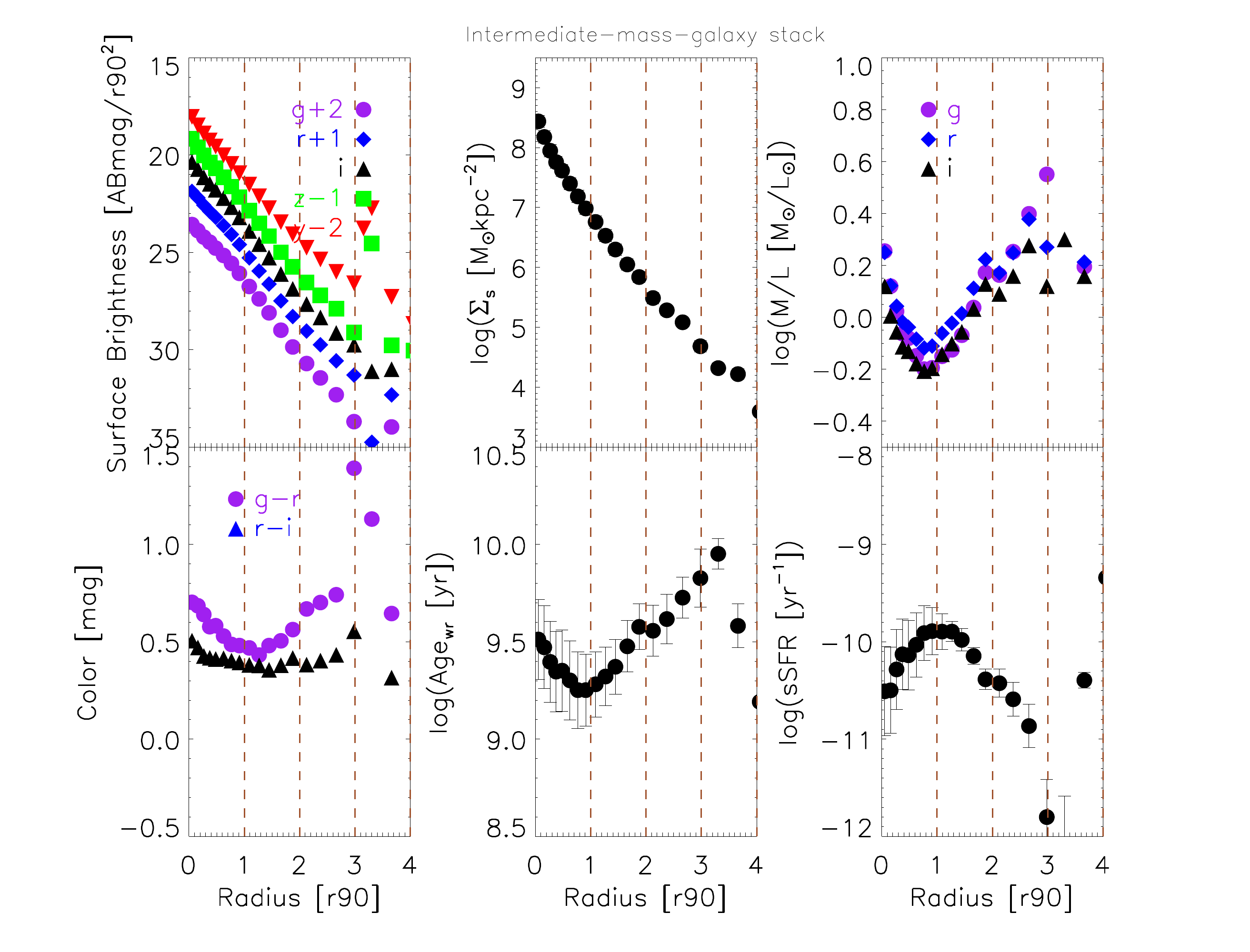} 

\caption{Radial profiles of stack image of intermediate mass galaxies. The profiles inside $3\,r_{90}$ are robust, however, the results outside $3\,r_{90}$ are unrelaible because the SB profiles are dominated by the sky background uncertainties in this regime.}
\label{ps12_fig_medm_magphys}
\end{center}
\end{figure}

\begin{figure}
\begin{center}
\includegraphics[scale=0.7, angle=90]{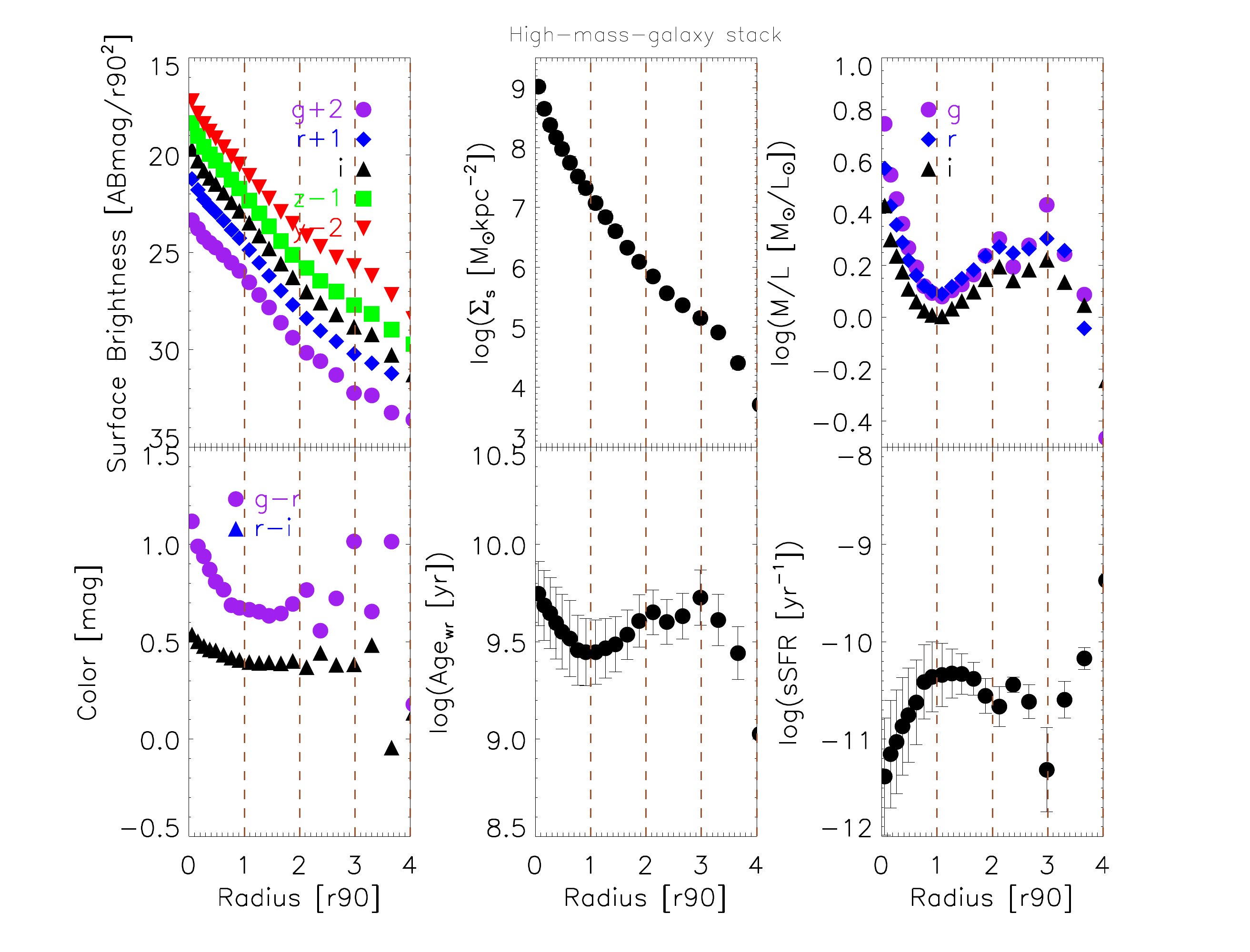} 

\caption{Radial profiles of stack image of high mass galaxies. The profiles inside $3\,r_{90}$ are robust, however, the results outside $3\,r_{90}$ are unreliable because the SB profiles are dominated by the sky background uncertainties in this regime.}
\label{ps12_fig_highm_magphys}
\end{center}
\end{figure}

We extracted SB and color profiles of the stack images using the method described in section \ref{method} and fit the SB profiles of the stack images using MAGPHYS.  The results are shown in Fig. \ref{ps12_fig_all_magphys} - Fig. \ref{ps12_fig_highm_magphys}. The stack images using galaxies in different mass bins have lower signal-to-noise ratio in the very outer region ($> \sim 3\,r_{90}$) and the figures from Fig. \ref{ps12_fig_lowm_magphys} - Fig.\ref{ps12_fig_highm_magphys} are plotted to show the variations of the profiles in different mass bins. 

The stack image made using all the galaxies (Fig. \ref{ps12_fig_all_magphys}) has the best SNR and we are confident about the profiles out to about $3\,r_{90}$. 
The main conclusion from this figure is that the stellar surface density profile shows an obvious change in slope (up-turn) beyond $2\,r_{90}$ and the color and M/L profiles show a relatively constant red color and high M/L. The colors in the region $2\,r_{90}$ and $3\,r_{90}$ are $g-r\approx 0.8$ and $r-i\approx0.6$, which are very close to the values found by \citet{zib04}. The stellar mass measured in the stellar halo of the all-galaxy stack image (within the region $2\,r_{90}$ and $4\,r_{90}$) is about $10^{8.4}M_{\odot}$ and is about 1\% of the total stellar mass of the stack galaxy. We suggest that the upward break in the surface mass density profile at around $2\,r_{90}$ represents the transition from the region dominated by the outer disk and that contaminated by the halo.

\begin{figure}
\begin{center}
\includegraphics[scale=0.7, angle=90]{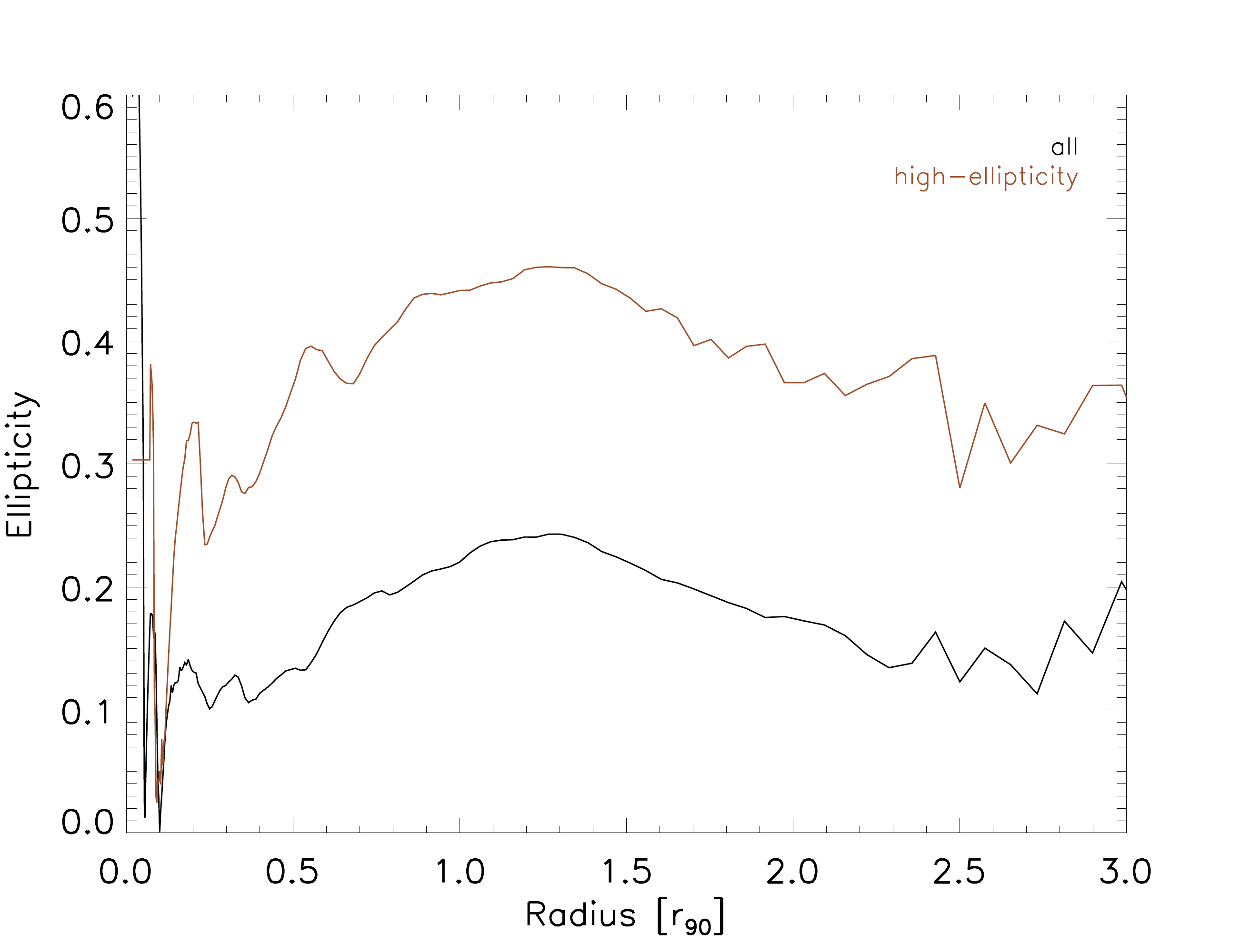} 

\caption{Ellipticity profiles of stacked $i$-band images of all galaxies and galaxies with ellipticity $e>0.4$. The ellipticity remains high in the region between 1 and 2\,$r_{90}$ and we conclude the light in this region is dominated by the disk rather than the stellar halo.}
\label{ps12_fig_ellprof_stack}
\end{center}
\end{figure}

The behavior of the galaxy ellipticity as a function of radius provides additional clues, since the halo is intrinsically much rounder than the disk. 
In order to maximize this effect of the stellar halo on the geometry of the galaxy shape, we made a stack image using galaxies with highest ellipticities, i.e.  $e>0.4$. The ellipticity profile of this high-ellipticity stack image is plotted in Fig. \ref{ps12_fig_ellprof_stack}.
The ellipticity profile rises slowly between $\sim$0.5 and $\sim$1.4\, $r_{90}$ and then slowly declines out to $\sim 3\,r{90}$.

Taken together, we conclude that the images of the outer disk are not significantly contaminated by light from the stellar halo at radii interior to $\sim2\,r_{90}$.   This argument is also consistent with the 1D study of stellar halo by \citet{bak12}, who found the stellar halo starts to affect the disk SB profile at $\mu_r\sim28 \rm mag/arcsec^2$, which is about $2\,r_{90}$.

\section{Composite radial profiles: the generic shapes of 1D profiles}
\label{stacked_1d}

Perhaps the most compact way to visualize the general properties of the radial profiles of the 698 sample galaxies is to plot all the radial profiles onto a single figure with the radius normalized by the galaxy's intrinsic scale radius ($r_{90}$). We will do this by creating two sets of composite profiles: one in which the galaxies are binned in stellar mass and one in which they are binned in the concentration parameter ($C = r_{90}/r_{50}$). 

\begin{figure}[htbp]
\begin{center}
\includegraphics[scale=0.3,angle=90]{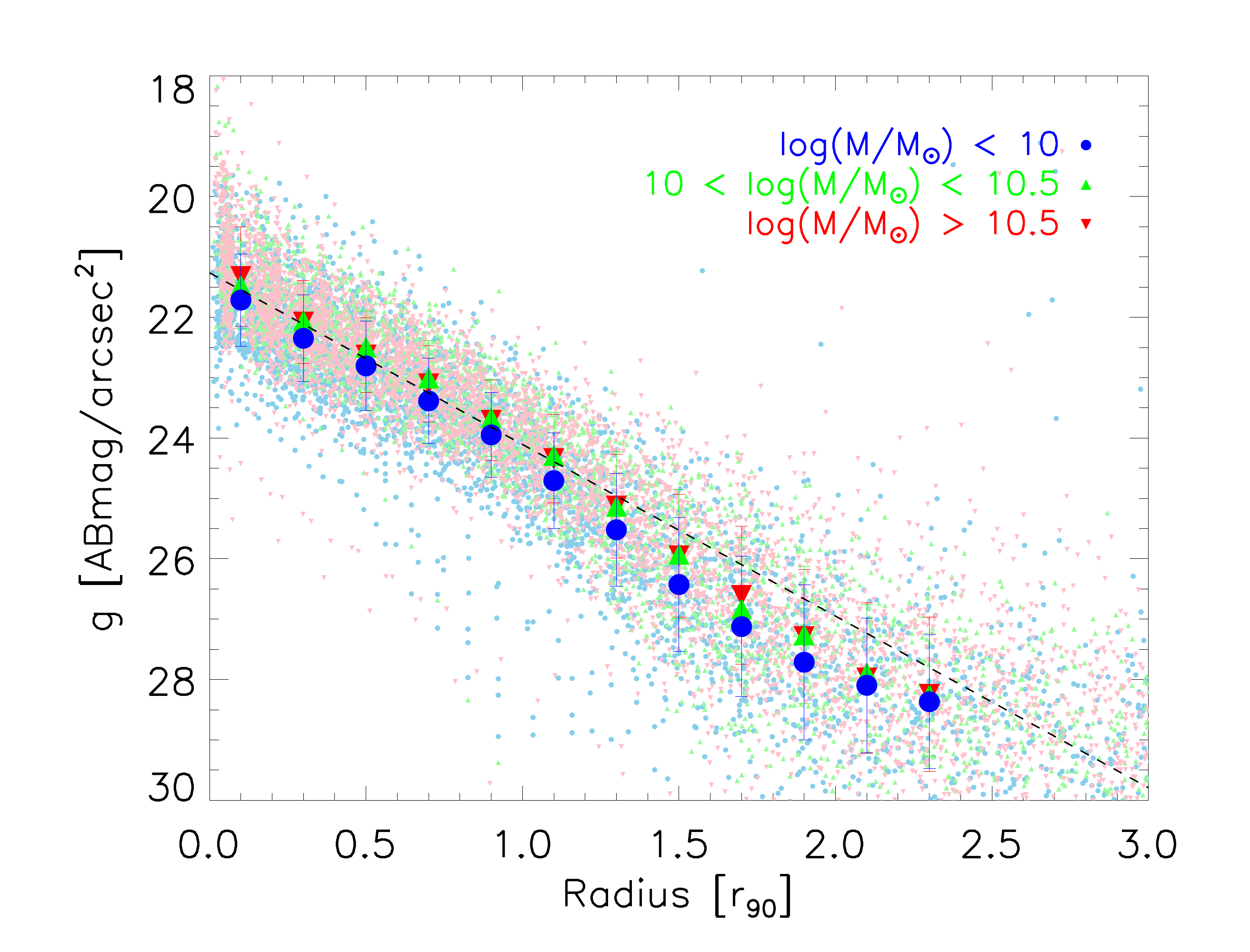}  
\includegraphics[scale=0.3,angle=90]{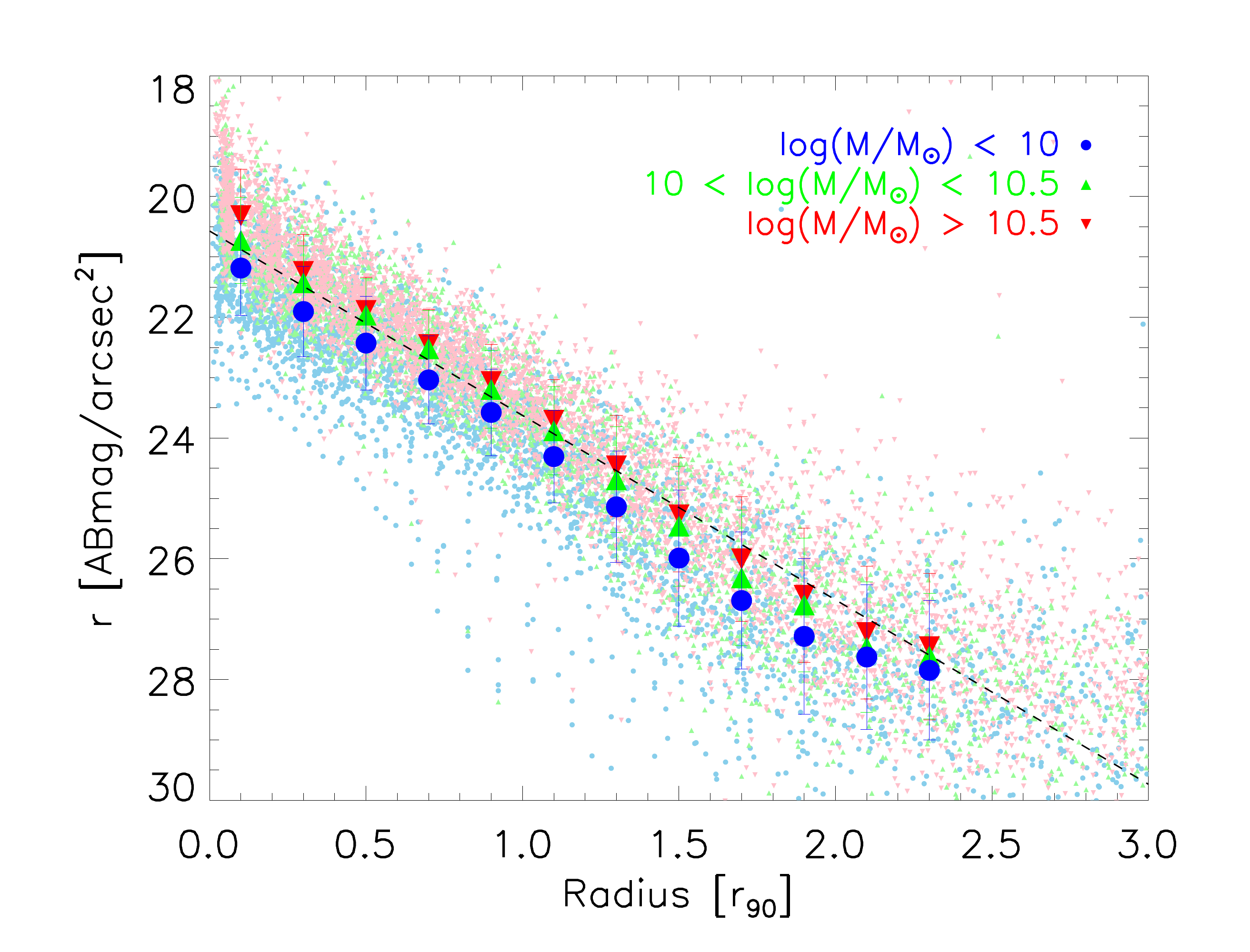}  
\includegraphics[scale=0.3,angle=90]{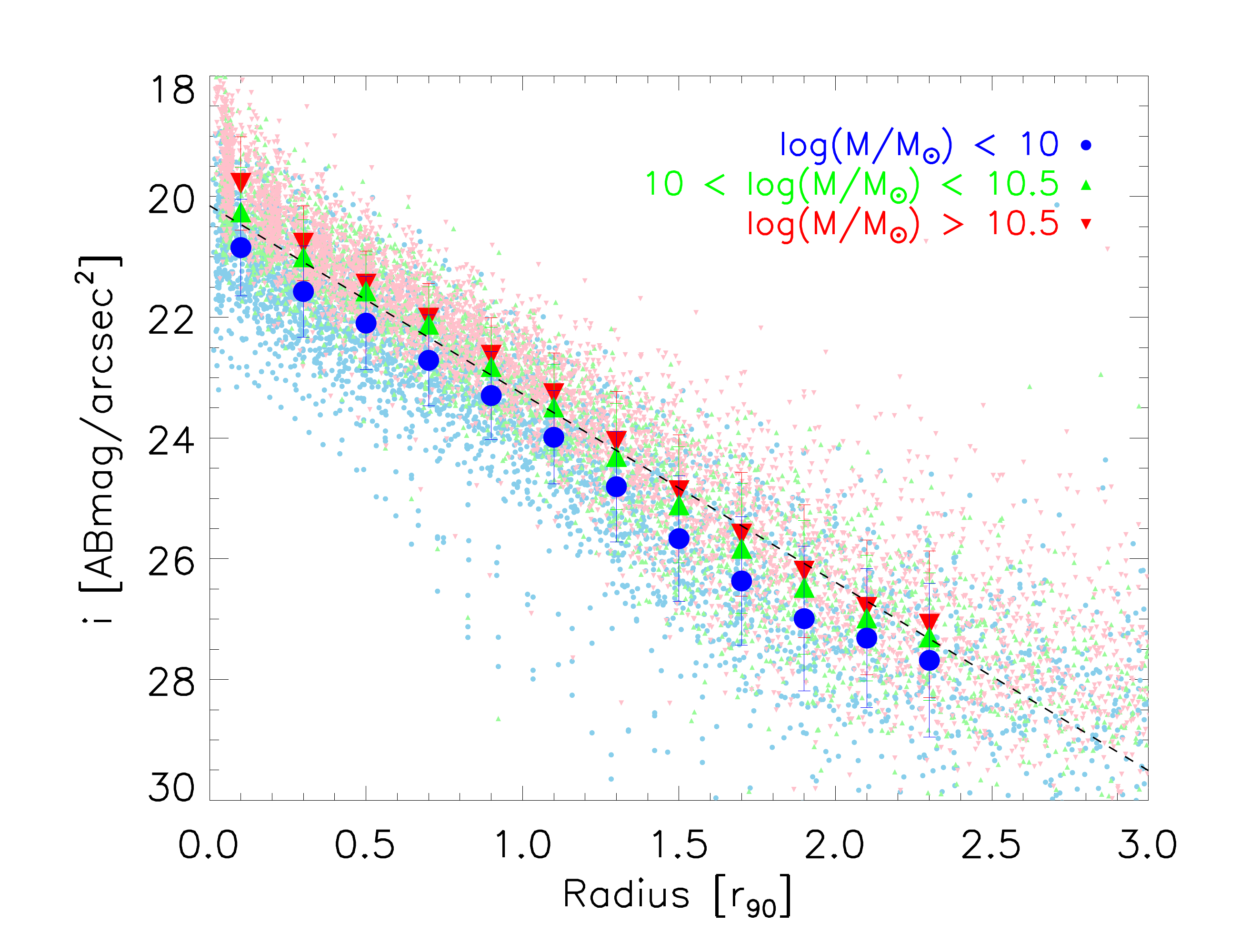}  
\includegraphics[scale=0.3,angle=90]{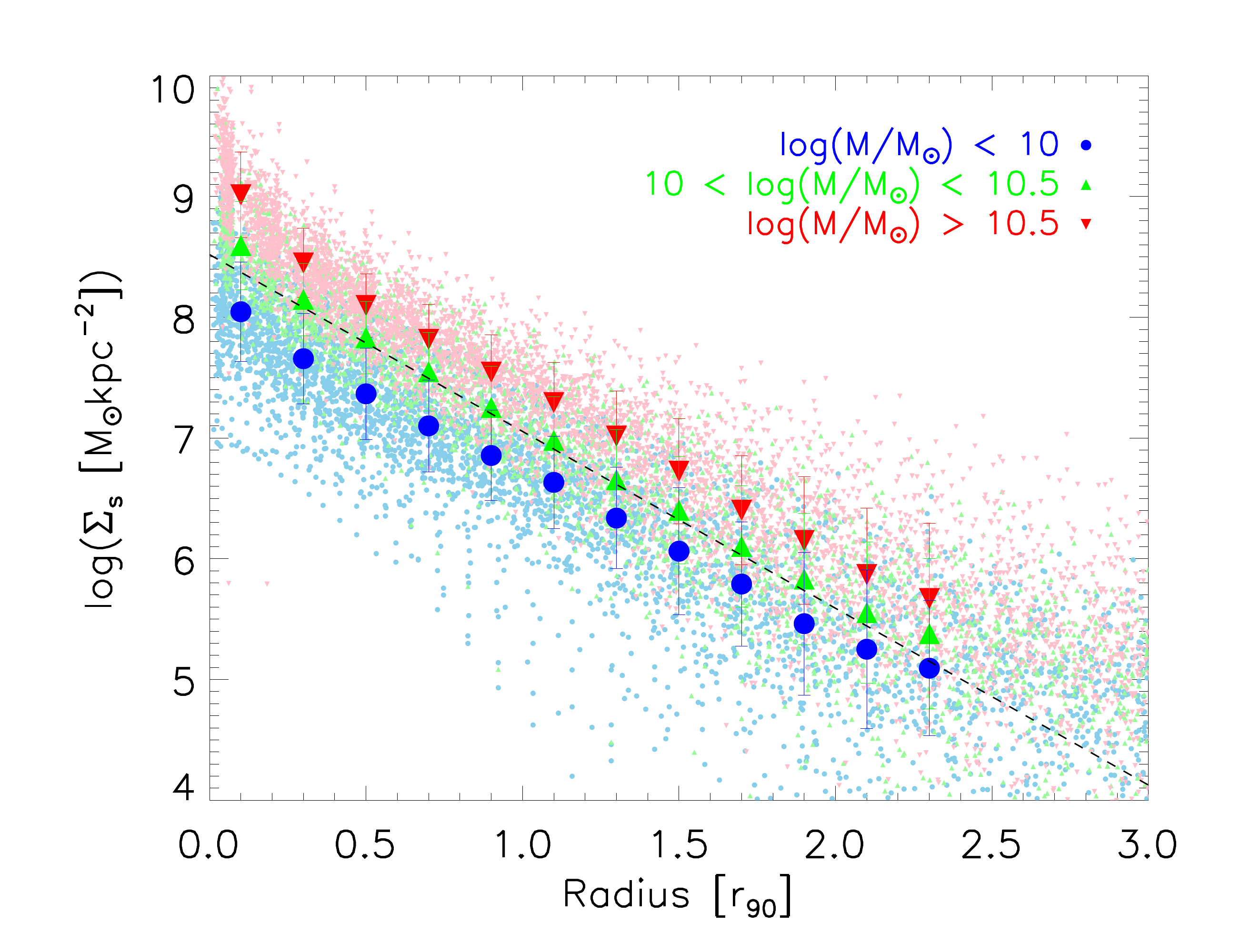}  
\caption{Composite radial profiles of $g$, $r$, $i$ -band SB and stellar mass surface density. Each point corresponds to single annulus in a singe galaxy, and are color-coded according to galaxy mass, as indicated in the legend.  The large symbols show the median value within bins of width 0.2 $r_{90}$.  The vertical bars represent the dispersion in the observed/derived median values and not the typical error on each of the thousands of measurements.  The photometric magnitudes are Milky Way extinction corrected and K-corrected.   The dash line in each panel is a pure exponential fit to the data between 0.3 and 1 $r_{90}$. The bluer bands show down-bending radial profiles starting at about 1.0\,$r_{90}$, while the redder bands and the surface mass density are single exponentials beyond the inner bulge-dominated region inside $\sim0.3\,r_{90}$. This is true in all bins of stellar mass.}
\label{mass_stack}
\end{center}
\end{figure}

\begin{figure}[htbp]
\begin{center}
\includegraphics[scale=0.5,angle=90]{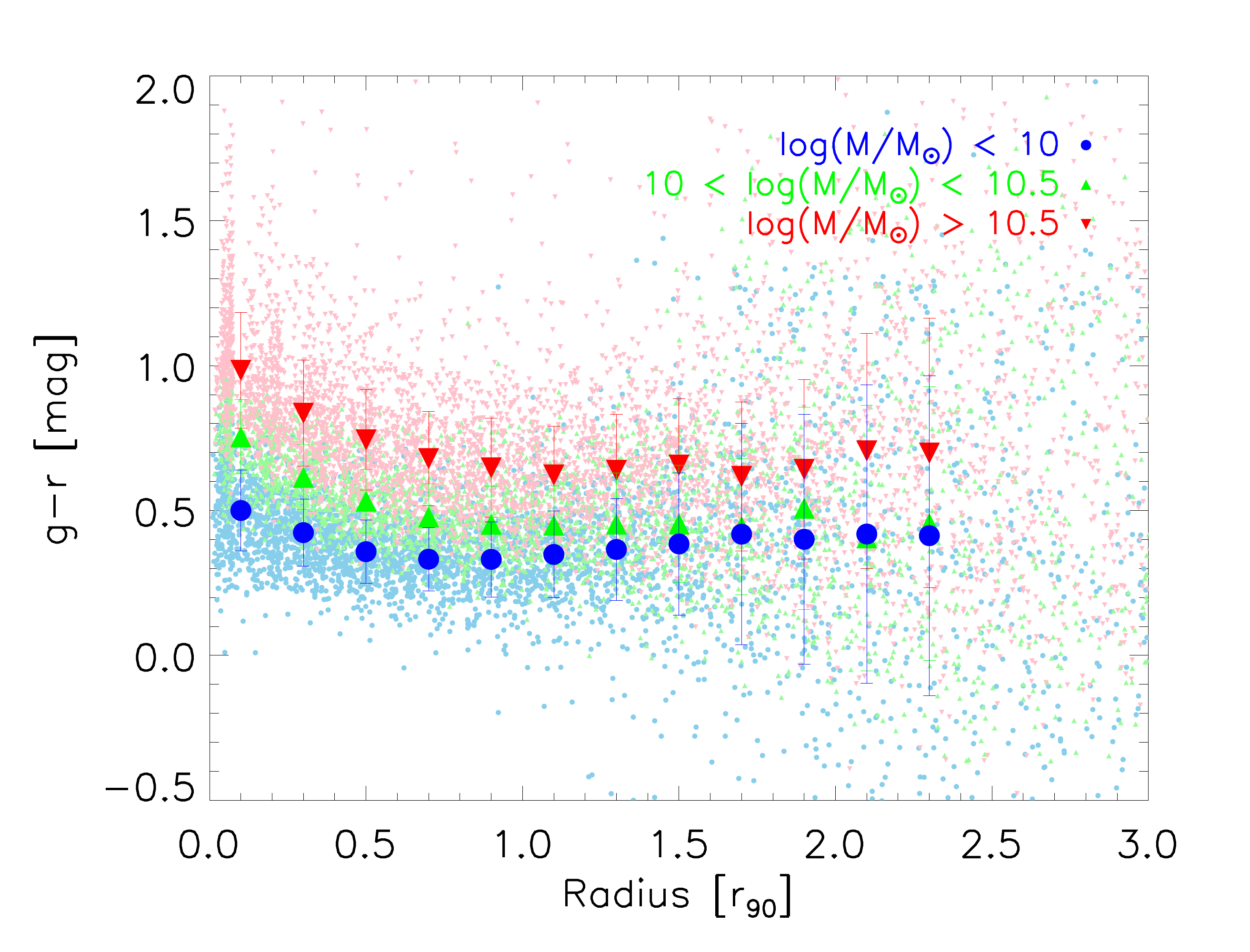}  
\includegraphics[scale=0.5,angle=90]{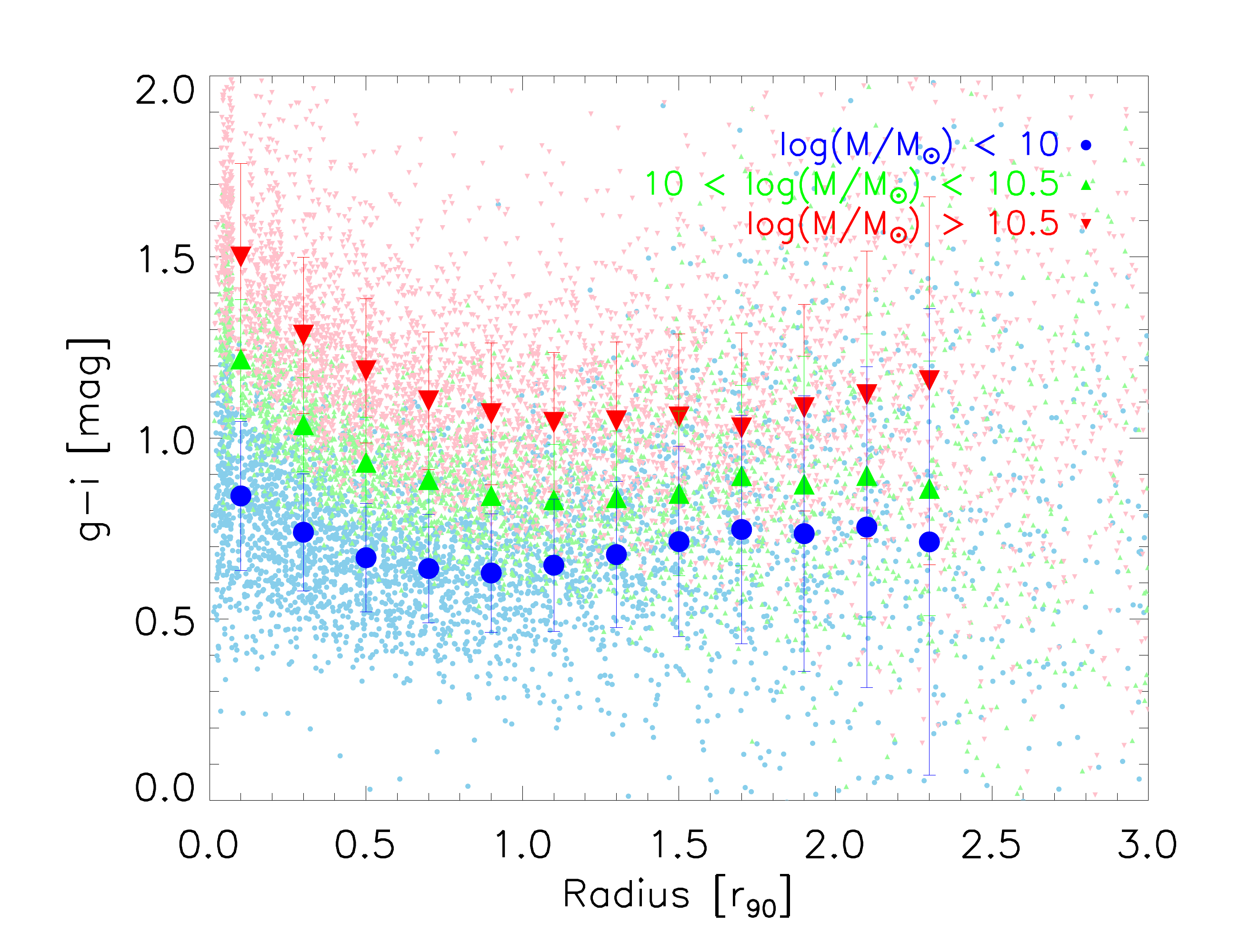}  
\caption{Composite radial profiles of colors. Upper panel: $g - r$; lower panel: $g-i$. The colors are Milky Way extinction corrected and K-corrected.  Symbols and color-coding are as in the previous figure. Both the color profiles show a `U`-shape with minima at around $r_{90}$.}
\label{color_stack}
\end{center}
\end{figure}

\begin{figure}[htbp]
\begin{center}
\includegraphics[scale=0.5,angle=90]{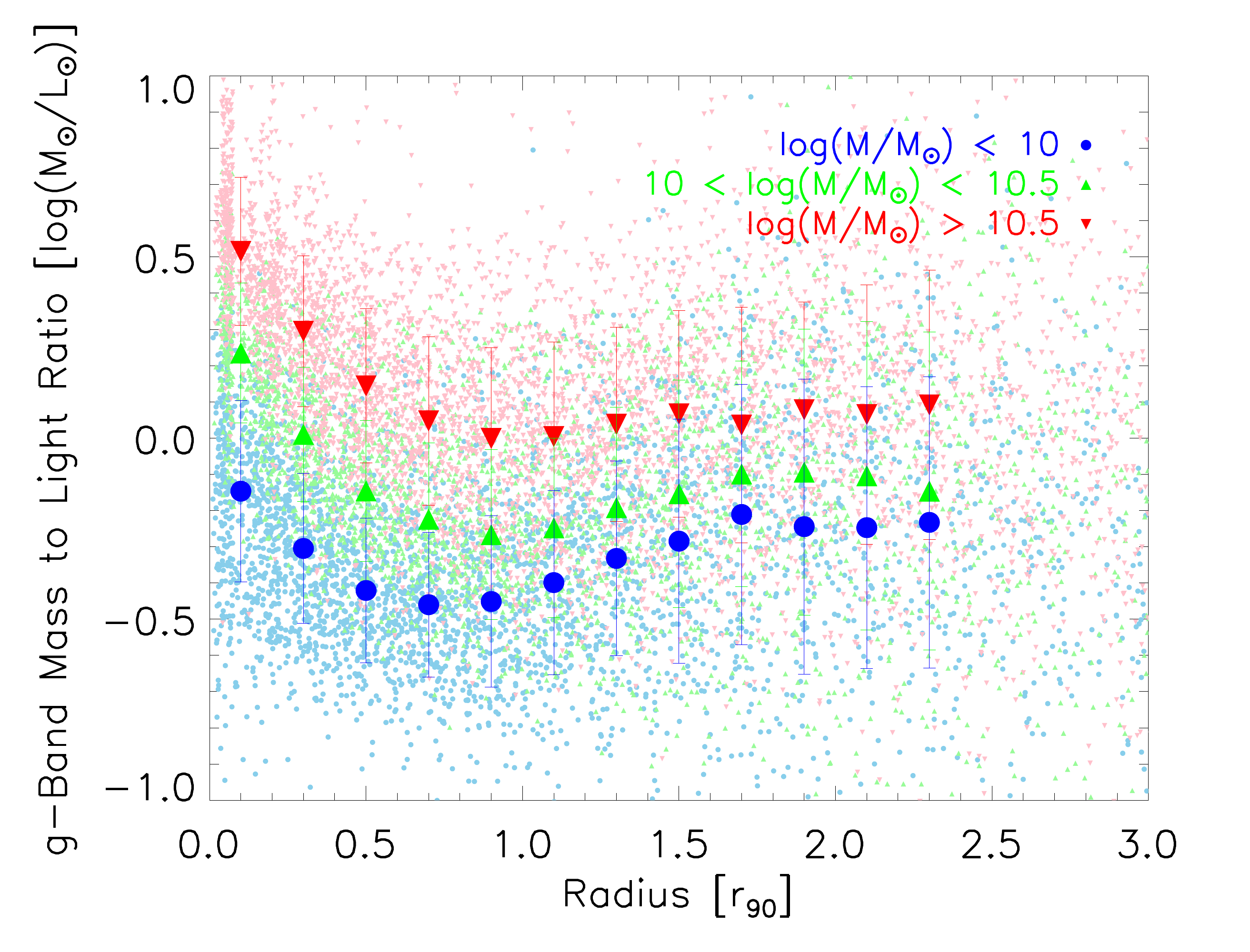}  
\includegraphics[scale=0.5,angle=90]{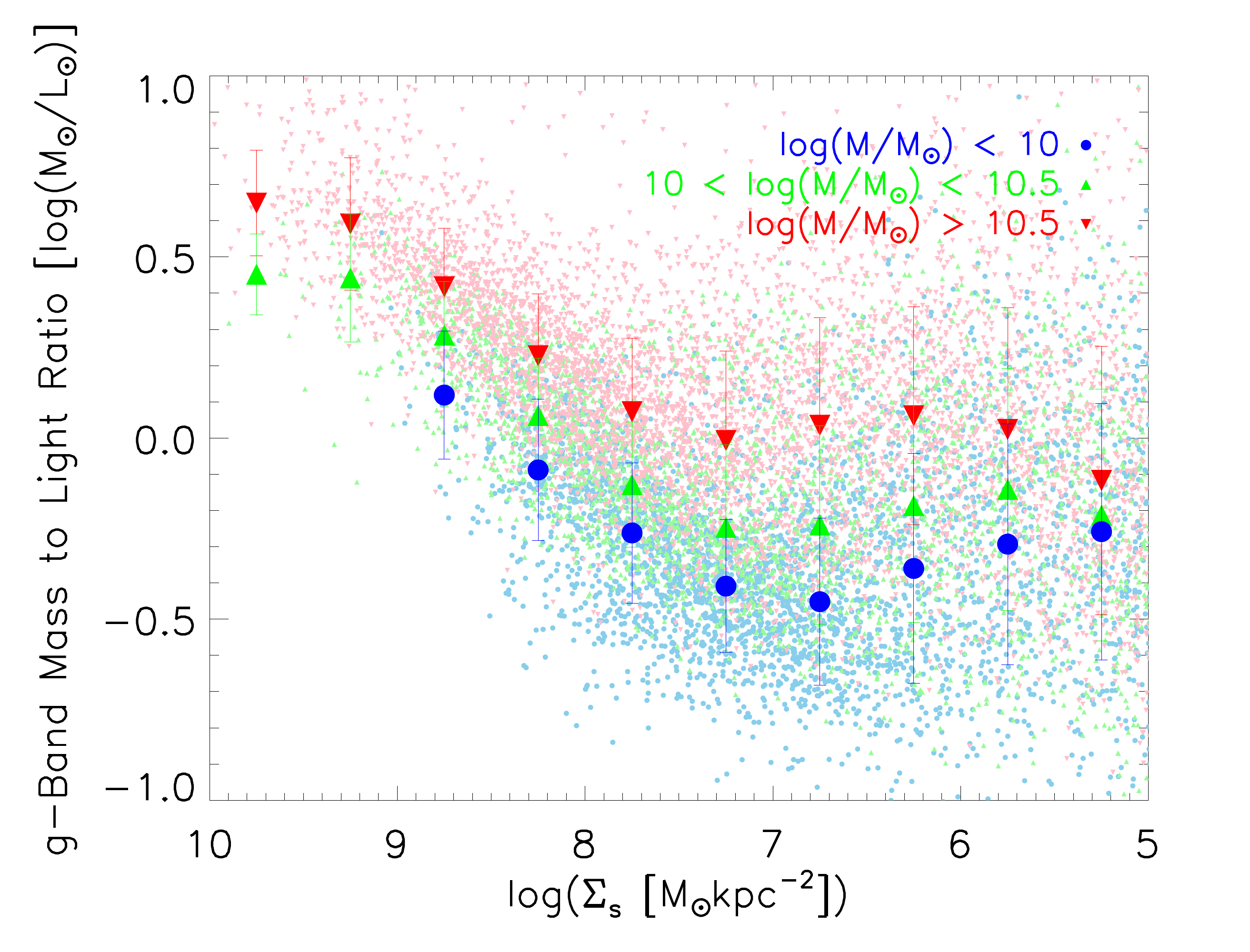}  
\caption{Composite $g$-band stellar M/L profile plotted as a function of radius (top) and local stellar surface mass density (bottom). Both plots show a `U'-shape with a minimum at about $r_{90}$ (top) and $\Sigma_S \sim 10^7$ M$_{\odot}$ kpc$^{-2}$ (bottom).}
\label{gm2l_stack}
\end{center}
\end{figure}

We first divide our sample into three bins in stellar mass, and show the resulting composite SB radial profile plots and stellar surface mass radial profile plot in Fig. \ref{mass_stack}), and then present the corresponding $g-r$ and $g-i$ color radial profiles in Fig. \ref{color_stack}.  Lastly, we show the corresponding $g$-band stellar M/L plots in Fig. \ref{gm2l_stack}. 

The resulting SB profiles have a dynamic range of about 7 magnitudes. 
The composite $i$, $z$, $y$ SB and the stellar mass surface density radial profiles are very close to pure exponentials with a scale-length $\sim 0.3\,r_{90}$. 
The plots in the bluer bands, especially the $g$-band SB radial profile, show a down-bending feature with a break radius at $\sim r_{90}$. This implies that most $g$-band profiles are down-bending and this is consistent with the PT06 results. There are no obvious breaks in the $i$, $z$, $y$ and $\Sigma_s$ radial profiles. This is consistent with the results in 
\citet{bak08} and \citet{bak12}.

The changes from $g$-band to the $y$-band and then to the $\Sigma_s$ radial profiles imply that disk galaxies have a positive color gradient and thus a increasing stellar M/L beyond $\sim 1 r_{90}$.
The color profiles (Fig. \ref{color_stack}) do indeed show that the colors turn up at larger radii. The higher mass galaxies tend to be redder at all radii, but the shapes of the radial color profiles are the same in all three mass bins.  Our sensitive PS1 imaging in the $i$, $z$, and $y$-bands provides improved contrast between the minimum in the color profiles (near $r_{90}$) and the comparatively red outermost disk. 

Irrespective of whether the color gradients are driven purely by the age of the stellar population or whether dust reddening plays a role, the redder colors in the outer disk imply that that the stellar M/L is larger there. This is shown explicitly in the top panel of Fig. \ref{gm2l_stack}. The $g$-band stellar M/L at first decreases with increasing radius as we move outward from the center of the galaxies, reaches a minimum value, and then rises as we move into the outer disk. The normalized radius at which the minimum occurs in stellar M/L increases slightly with increasing stellar mass (from $\sim$ 0.8 to 1.0 $r_{90}$). In the botton panel we plot the
$g$-band stellar M/L at a given location vs. the local stellar mass surface density. This shows the same qualitative behavior as the top panel (an obvious `U'-shape), but the minimum in $M/L$ occurs at the same value ($\Sigma_s \sim 10^7$ M$_{\odot}$ kpc$^{-2}$) in all three mass bins.  The behavior of the stellar mass-to-light ratios for other bands are similar to that of the $g$-band but with diminished amplitude. 

One important clue as to the nature of the breaks in the g and r band composite profiles is that they occur exactly at the location of the bluest point in the color profiles (Fig. \ref{color_stack}), $\sim r_{90}$. This is consistent with results in previous studies \cite[e.g.][]{bak08,bak12}. According to Fig. \ref{mass_stack}, the value of $r_{90}$ is about 3 scale-lengths in the SB profiles. This is consistent with the locations of the PT06 breaks, but conflicts with the \citet{vdk87} model.

\begin{figure}[htbp]
\begin{center}
\includegraphics[scale=0.5,angle=90]{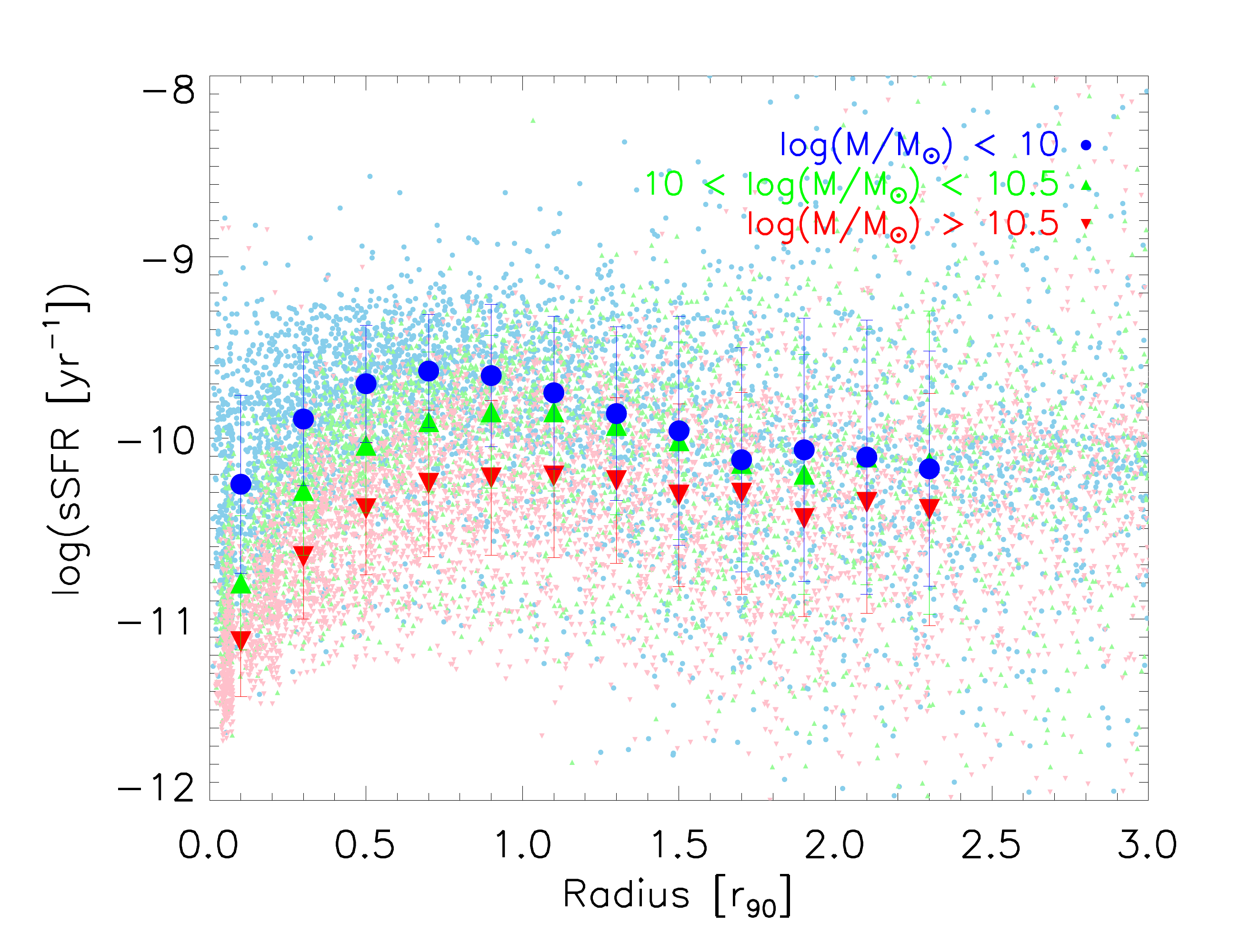}  
\includegraphics[scale=0.5,angle=90]{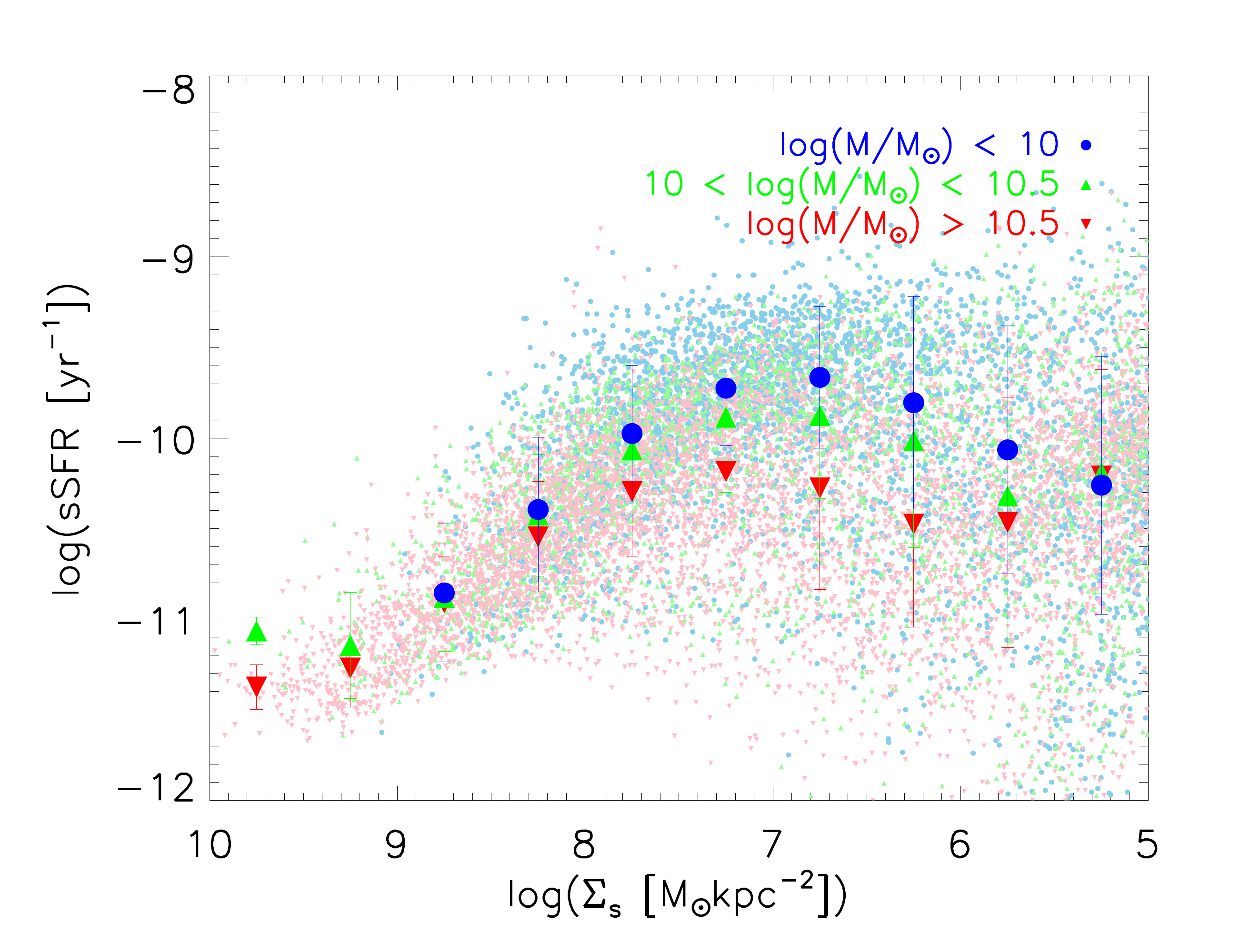}  
\caption{Composite radial profiles of specific star formation rate (sSFR). In all mass bins the specific star-formation rate reaches a maximum value at a radius of $\sim r_{90}$ (top) and $\Sigma_s \sim 10^7$ M$_{\odot}$ kpc$^{-2}$ (bottom).}
\label{ssfr_stack}
\end{center}
\end{figure}

\begin{figure}[htbp]
\begin{center}
\includegraphics[scale=0.5,angle=90]{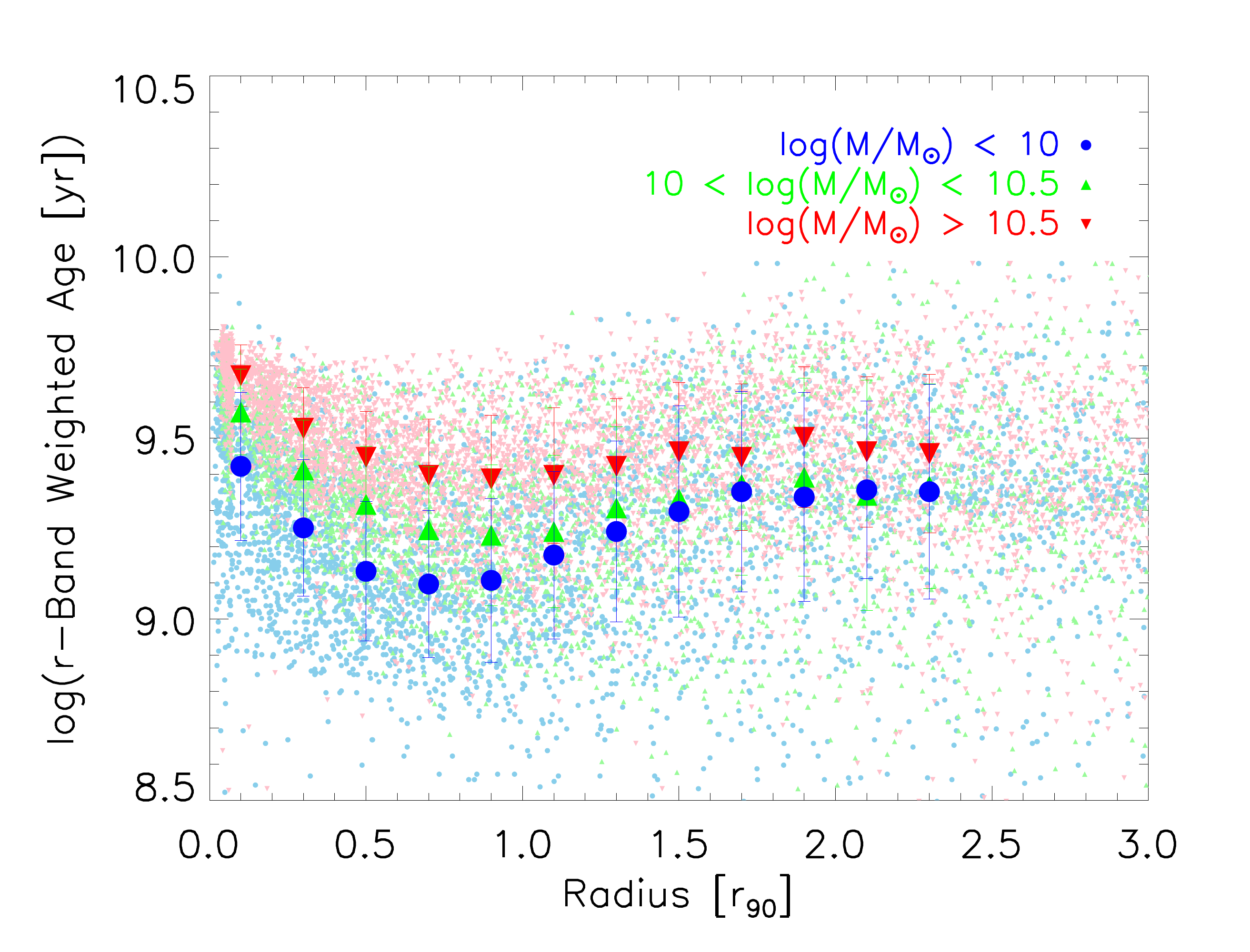}  
\includegraphics[scale=0.5,angle=90]{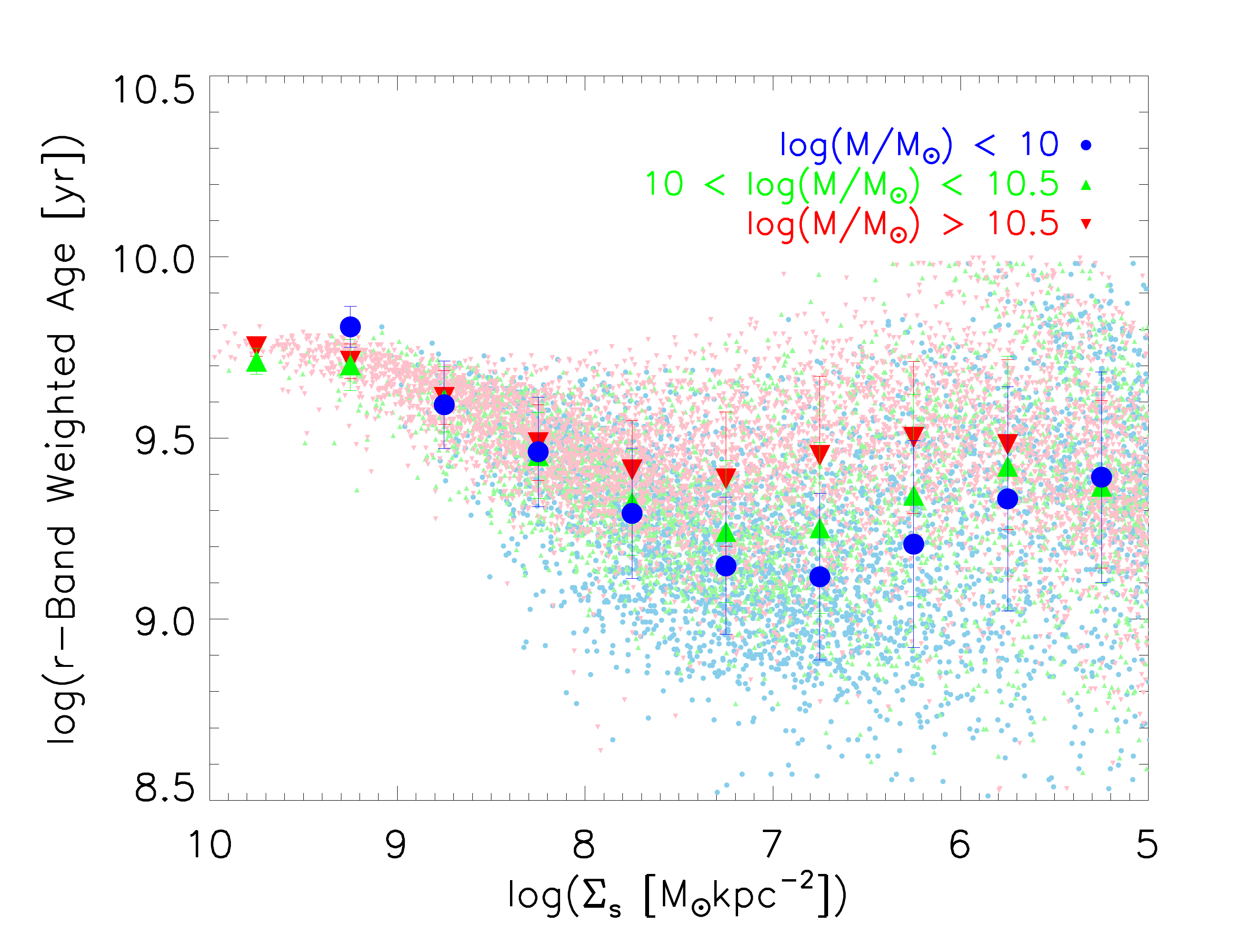}  
\caption{$r$-band weighted age radial profile stack plot. In all bins of stellar mass, the population is youngest at a radius of $\sim r_{90}$ (top) and $\Sigma_s \sim 10^7$ M$_{\odot}$ kpc$^{-2}$ (bottom). }
\label{ager_stack}
\end{center}
\end{figure}

To interpret the nature of the `U'-shaped stellar M/L profile, we plot the composite radial profiles of the specific star formation rate (sSFR) in Fig. \ref{ssfr_stack}. This figure shows that the sSFR drops with increasing normalized radius in the outer disk in all three mass bins. The bottom panel plots sSFR vs. $\Sigma_s$ and shows that sSFR reaches a (mass-dependent) maximum value at $\Sigma_s \sim 10^7$ kpc$^{-2}$. A complementary parameterization of the stellar population is shown in Fig. \ref{ager_stack} which shows the $r$-band luminosity weighted mean age of the stars plotted as a function of normalized radius (top) and local surface mass-density (bottom). 

Note that without UV and IR data we cannot unambiguously disentangle the effects of age and dust reddening on the observed colors (even though we have attempted to do so, as described in section 2). Nevertheless, it is difficult to imagine that the redder colors in the outer disk (or in the regions of low stellar mass surface density) are caused by an increase in the amount of dust reddening. This is completely unphysical at such low surface mass densities, where the corresponding gas and dust column densities will be small.

Finally, as expected, we note that our data show higher mass disk galaxies tend to have higher SB and be redder and lower mass galaxies tend to have lower SB and be bluer. This is also consistent with many previous studies ( e.g. \citet{bak08,bak12, kau03}).

\begin{figure}[htbp]
\begin{center}
\includegraphics[scale=0.35,angle=90]{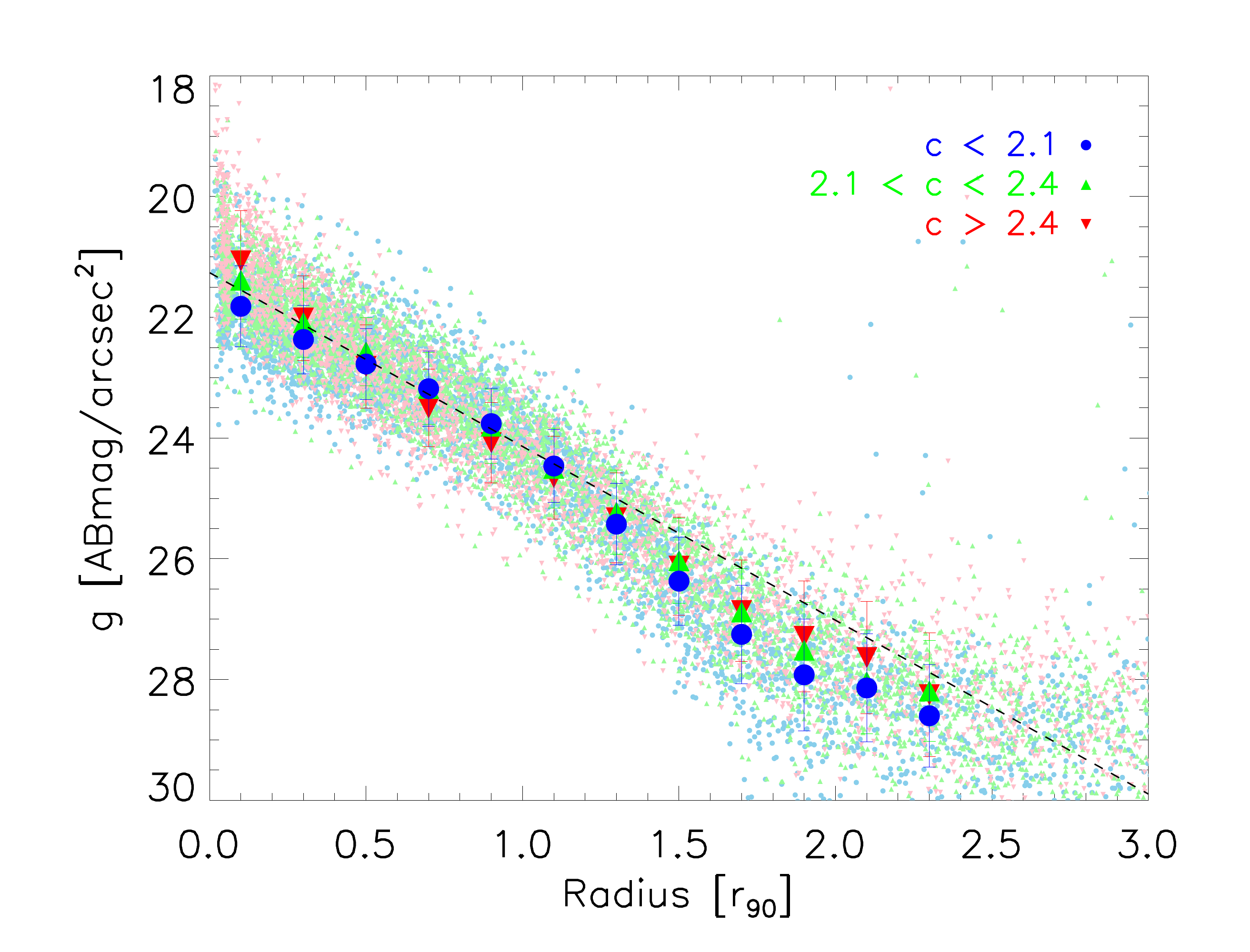}  
\includegraphics[scale=0.35,angle=90]{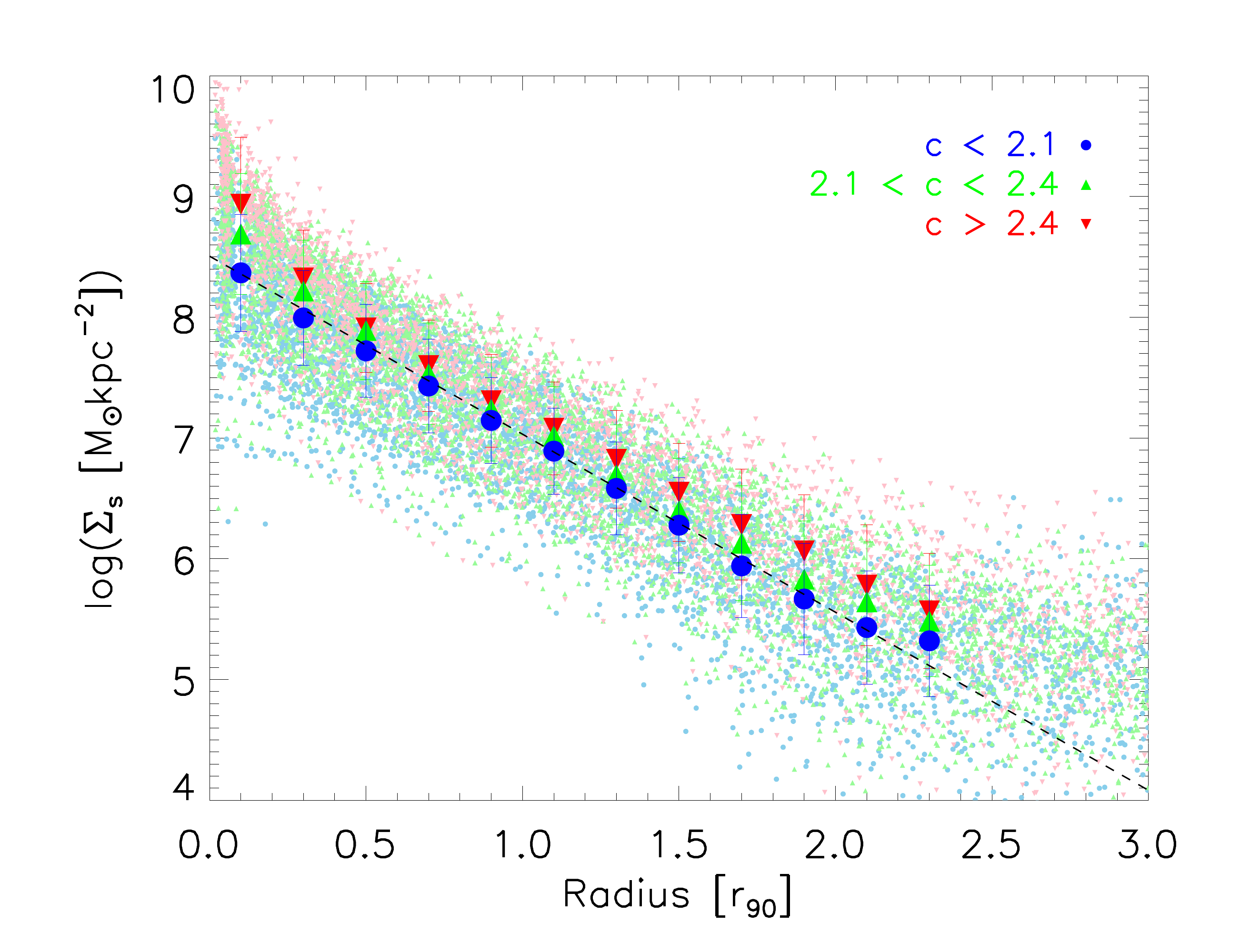}  
\includegraphics[scale=0.35,angle=90]{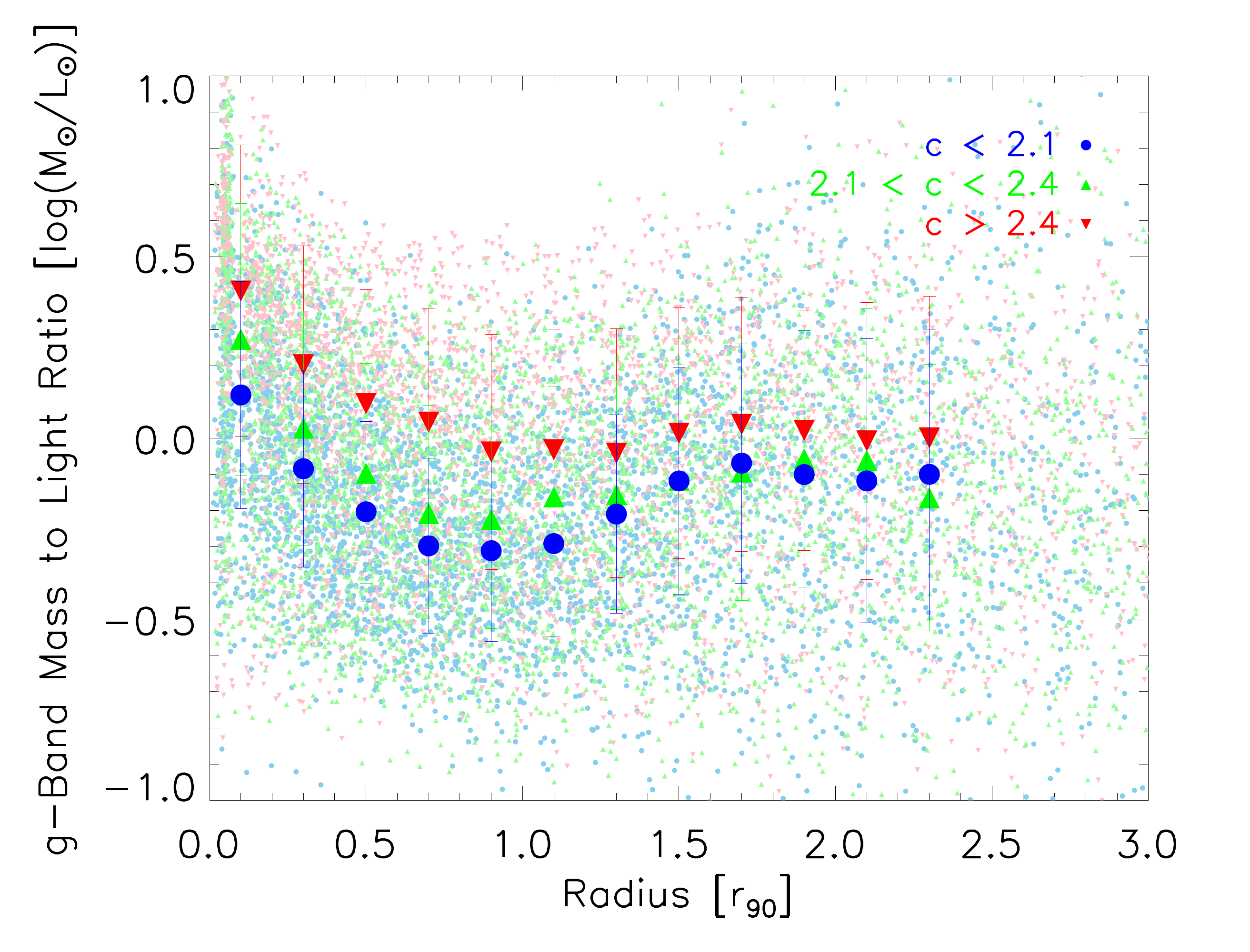}   
\caption{Composite radial profiles of $g$-band SB, stellar mass surface density and $g$-band M/L for our sample binned in three ranges of $C$. Symbols follow the convention of Fig. \ref{mass_stack}}
\label{stack_prof_con}
\end{center}
\end{figure}

In Figure \ref{stack_prof_con} we show composite radial profiles of the g-band SB, the stellar surface mass density, and the g-band M/L for our sample binned in three ranges of $C$.

Three results are evident. First, the amplitude of the down-bending in the SB profile is largest for the lowest values of $C$ and smallest for the largest values of $C$. Second, the amplitude of the `U'-shaped mass-to-light radial profile follows the same pattern. Third, the stellar surface mass density radial profile is well described as a single exponential for all three bins in $C$. We conclude that there is a relationship between $C$ (the bulge/disk ratio; Hubble type) and the radial gradients in the age of the stellar population between the inner and outer disk, even though the relative radial distribution of stellar mass is the same for both later and earlier-type disks (pure exponential). 

\section{Properties of Individual Galaxies}
\label{detailed_1d}
PT06 and several following studies \cite[e.g.][]{bak08,her13} classify the SB radial profiles into three categories. However, these studies classify the profiles more or less subjectively in terms of establishing the division between Type I, II, and III. Also, the breaks are less obvious in redder bands than in bluer bands. Even in the same photometric band the  strength of the break is different from galaxy to galaxy. Thus, a better approach is to develop an automatic algorithm and derive a number which can describe the breaks quantitatively instead of qualitatively. In this section, we develop an objective automatic break finding algorithm and use the inner-outer disk slope ratio $k_1/k_2$ as the break strength indicator, where $k_1$ and $k_2$ are defined as 
\begin{equation}
\label{ps12_eq_k1}
\begin{array}{lr}
\mu(r)=\mu_{10}+k_1\,r  & ({\rm if}\, r<r_b),
\end{array}
\end{equation}
and 
\begin{equation}
\label{ps12_eq_k2}
\begin{array}{lr}
\mu(r)=\mu_{20}+k_2\,r &({\rm if}\, r > r_b),
\end{array}
\end{equation}
where $r$ is the radius of the profile, $\mu$ is the SB profile in $\rm ABmag/arcsec^2$, $\mu_{10}$ and $\mu_{20}$ are the central SBs interpolated using the inner and outer disk SB profiles respectively, and $r_b$ is the break radius. The inner and outer disks are divided by the break radius $r_b$. Here we define the slope ratio as 
\begin{equation}
R=k_1/k_2.
\end{equation}
Therefore, a large slope ratio ($R> 1$) corresponds to the Type III (up-bending) profile; a small slope ratio ($R < 1$) corresponds to the Type II (down-bending) profile; and a unity slope ratio ($R\approx 1$) corresponds to the Type I (single exponential) profile.

\subsection{The break-finding algorithm}

It is very difficult to classify $\sim 3500$ SB radial profiles accurately and fully automatically for a couple of reasons:
A typical SB profile contains three different parts: a S\'{e}rsic profile for the bulge, an exponential profile for the inner disk and another exponential profile for the outer disk. Some galaxies even show a very extended profile for the stellar halo. The outer boundary of the bulge and the inner boundary of the stellar halo can be very different from galaxy to galaxy. Furthermore the radial profile can have wiggles (due to irregularities like spiral arms) and be very noisy in the outer low SB regime. Finally, galaxies can be classified as a combination of Type II and Type III (PT06). Previous studies \cite[e.g. PT06 and][]{her13} use human inspection in determining the type of the SB profiles. 

Here we develop a fully automatic SB profile classification strategy. The strategy is based in part on the method developed by PT06 and several properties of the SB profiles described in section \ref{stacked_1d}. 
According to section \ref{stacked_1d}, breaks are generally more obvious in bluer bands, e.g. $g$ or $r$ band. Furthermore, the SB radial profile break radius will not change from filter to filter, even though the inner/outer disk slopes ratio varies \cite[cf section \ref{stacked_1d} and ][]{bak08}. 
Since the $r$-band data have higher S/N than the $g$-band data, we use the $r$-band SB profile to derive the break radius and perform the profile classification.

One of the key points of the classification is finding the boundaries of the stellar disk and the location of the break radius. The inner boundary of the stellar disk is where the SB of the bulge is surpassed by the SB of the stellar disk and the outer boundary of the stellar disk is where the SB of the stellar disk is surpassed by the SB of the stellar halo or the rms of the sky background, which ever is higher. These boundaries are very hard to determine automatically: PT06 use human examination to determine the location of the inner boundary and set the outer boundary when the SB of the radial profile meets the rms of the sky background. The inner boundaries of PT06 are more accurate but the method is not suitable for our big sample. The outer boundaries of PT06 sometimes include part of the stellar halo \citet{bak12}. Fortunately, for most of the galaxies in our sample, the bulge is significant (if at all) only within $0.3\,r_{90}$ and the stellar halo becomes significant outside $2\,r_{90}$ (cf. section \ref{stacked_2d} and \ref{stacked_1d}).  Therefore, we simply fix the inner and outer boundaries as $0.3\,r_{90}$ and $2\,r_{90}$ in this paper.

The location of the break radius is determined using two different methods:
The first method is given by PT06: We first calculate the derivative of the $r$-band radial profile within $0.3\,r_{90}$ and $2\,r_{90}$. The derivative profile is then smoothed to remove noisy points. The smoothed derivative profile should be a step function in an ideal situation (cf. Fig. 6 of PT06). We then search for the location of the biggest jump in the derivative profile around $r_{90}$. This location should be the break radius in most cases. 
The second method is based on the property of the M/L profile: The M/L profile usually shows a `U' shape and the location of the `U' shape minimum is almost always the location of the break radius. Since the `U' shape is more obvious than the SB profile break it is much easier to fit the $r$-band M/L profile than to fit the $r$-band SB profile. We further simplify the algorithm by fitting the M/L profile using a skewed `V' shape function (two straight lines joined around $r_{90}$). The minimum location of the `V' shape function should be the break radius. 
After deriving the break radii using two independent methods, we fit the $r$-band SB profile using a broken exponential function with the break radius fixed using the values derived above. The break radius which results in the smallest $\chi ^2$ value is then our final best break radius for this specific galaxy.

The method works well for our galaxies but the reader should bear in mind that this classification is not as accurate as the method provided by PT06 because the inner bulge and outer stellar halo regions are assigned fixed values in radius. It is possible that some individual galaxies with an extended bulge and/or brighter stellar halo are misclassified in this strategy. However, our goal here is to draw statistical conclusions based on the results for our full sample of 698 galaxies, as derived by this algorithm. 

\subsection{The slope-ratio $R$ (in light and mass) and its correlation with other physical parameters} 
\label{ps12_sloperatio}
\begin{figure}
\begin{center}
\includegraphics[angle=90,scale=0.6]{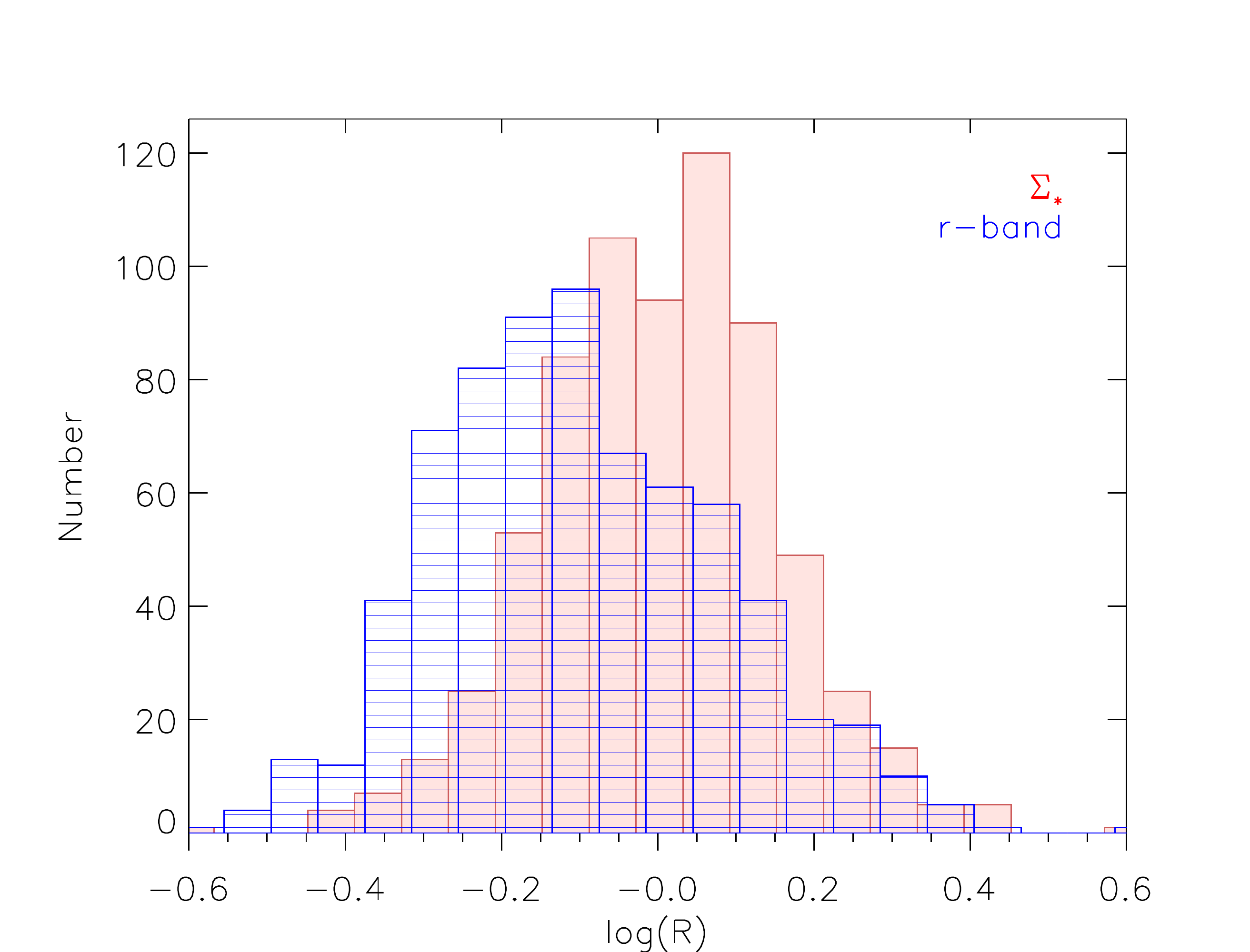}  

\caption{Ratio of inner and outer disk slopes ($R$). The distribution for the r-band implies most disk galaxies have down-bending outer surface brightness profiles ($R < 1$). The distribution for $\Sigma_s$ peaks at $R = 1$, corresponding to a pure exponential profile. The typical errors in both forms of $R$ are about 0.08 (about half the rms dispersion in the distributions).}
\label{ps12_fig_slope_ratio}
\end{center}
\end{figure}

After determining the break radius, we calculate the slopes of the inner ($k_{r1}$) and outer ($k_{r2}$) disk $r$-band SB profiles. Similarly, we calculate the inner and outer disk slopes ($k_{m1}$ and $k_{m2}$) of the stellar surface density profiles. The slopes of the stellar surface density profile are defined in a similar way as Eq. \ref{ps12_eq_k1} and Eq. \ref{ps12_eq_k2} except that the SB $\mu$ is replaced by stellar surface densities $\Sigma$ in a logarithmic scale. 
We plot the distribution of the $r$-band slope ratio ($R_r$) and the stellar surface mass density profile slope ratio ($R_m$) in Fig. \ref{ps12_fig_slope_ratio}. The distribution of log($R_r$) looks like a skewed  Gaussian centered around -0.12 ($k_{r1}/k_{r2}=0.75$), which means most galaxies have down-bending SB profiles. In comparison, the distribution of log($R_m$) is a more symmetric Gaussian but centered around 0.0 ($k_{m1}/k_{m2}=1$), implying a single smooth exponential form.  The statistical parameters of the distributions of log($R_r$) and log($R_m$)  are listed in Table \ref{R_statistics}. The log($R_r$) distribution has a slightly a larger rms and a some skewness, while the log($R_m$) has a somewhat smaller rms and no skewness.

\begin{table}[htdp]
\caption{Statistics of log($R_r$) and log($R_m$) distribution.}
\begin{center}
\begin{tabular}{cccccc}
\hline
\hline
 & Mean & Median &  Mode & rms & skew \\
\hline

log($R_r$)    &    -0.10 &   -0.11   &  -0.14    &  0.18 & 0.04   \\
log($R_m$) & 0.006   & 0.01  &  0.005 &     0.15 & 0.0      \\
\hline
\hline
\end{tabular}
\end{center}
\label{R_statistics}
\end{table}%

\begin{figure}
\begin{center}
\includegraphics[scale=0.6, angle=90]{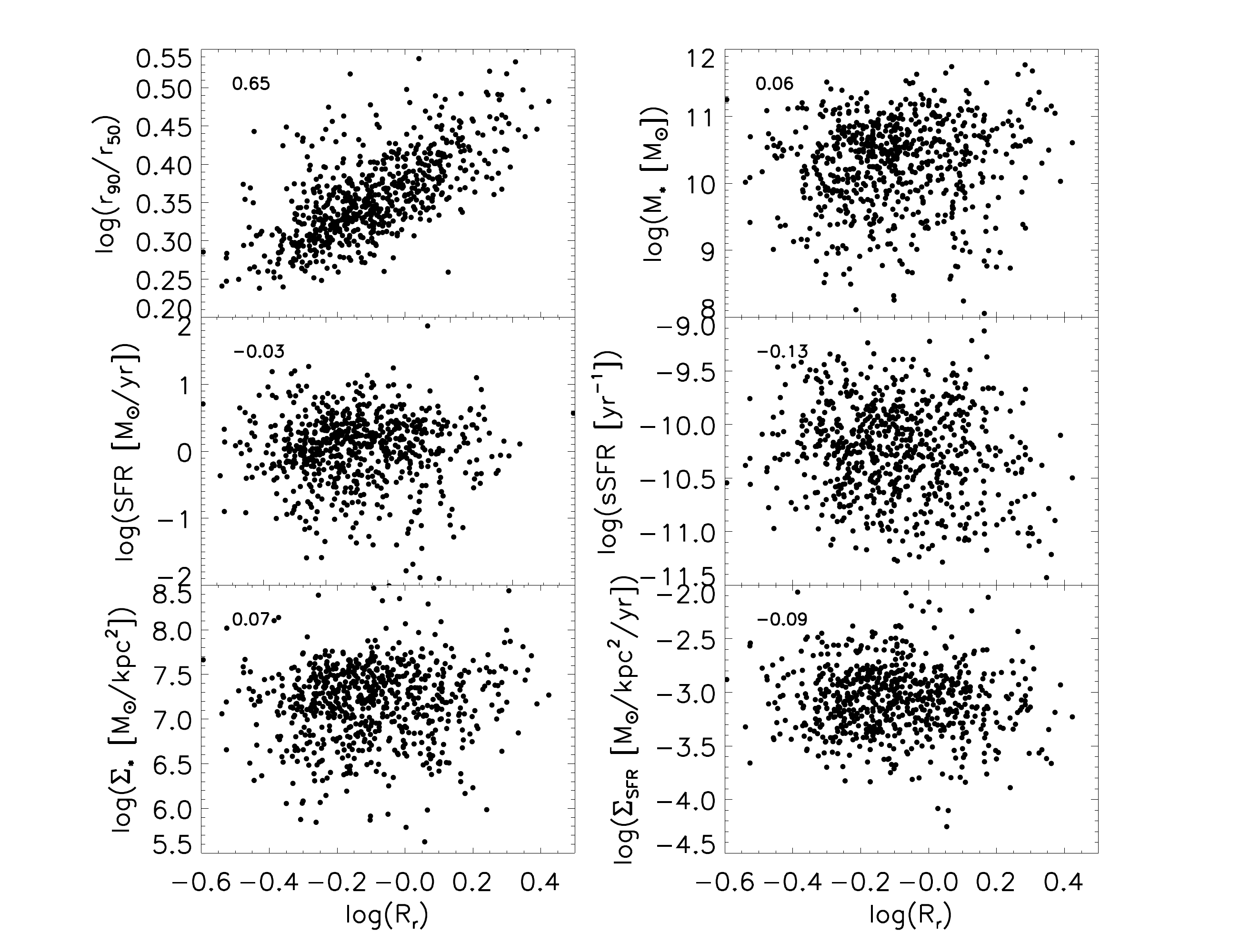}  

\caption{$r$-band slope ratio v.s. concentration parameter, stellar mass, total SFR, sSFR, $\Sigma_*$ and $\Sigma_{SFR}$. The only strong correlation is with concentration, with later type disk galaxies having more strongly down-bending radial SB profiles.}
\label{ps12_fig_krat_corre}
\end{center}
\end{figure}

We plot the slope ratios versus the concentration parameter ($C=r_{90}/r_{50}$), total stellar mass $M_*$, total star formation rate SFR, specific star formation rate (sSFR=${\rm SFR}/M_*$), average stellar surface mass density ($\Sigma_*=M_*/\pi r_{90}^2$) and average SFR surface density ($\Sigma_{\rm SFR}={\rm SFR}/\pi r_{90}^2$) in Fig. \ref{ps12_fig_krat_corre}. Here $r_{50}$ is the radius containing 50\% of the $r$-band Petrosian flux, the values for $M_*$ and SFR are derived in section \ref{stacked_1d}. 

There is a strong correlation between the $R_r$ and $C$ but no obvious correlation with the other parameters. The correlation of SB profile shape and $C$ (Hubble type) is consistent with PT06 who found the disk SB profile types strongly correlate with Hubble types, with down-bending disks typically found to be later type galaxies.

\begin{figure}
\begin{center}
\includegraphics[scale=0.6, angle=90]{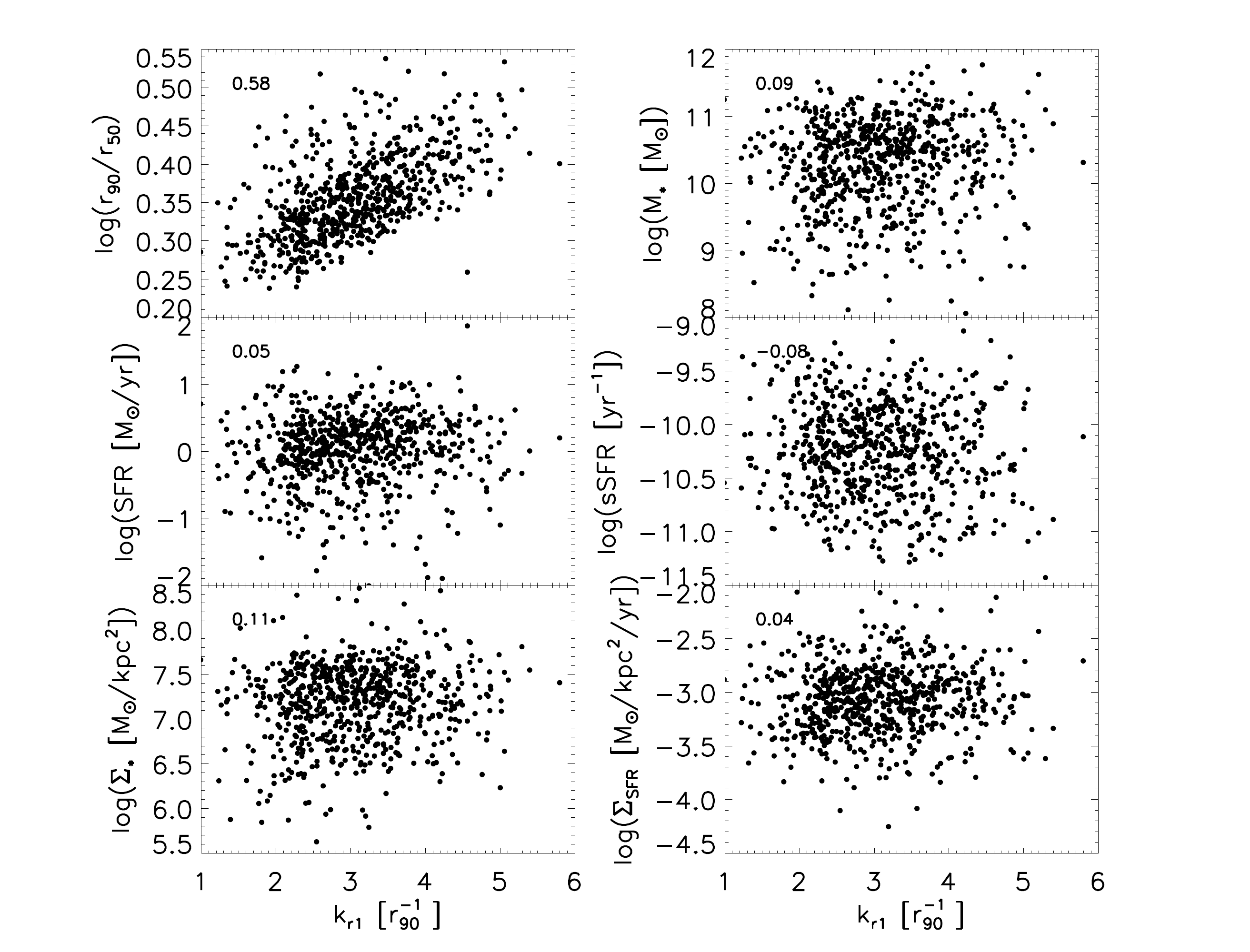}  
\caption{$r$-band $k_{r1}$ v.s. concentration parameter, stellar mass, total SFR, sSFR, $\Sigma_*$ and $\Sigma_{SFR}$. The only strong correlation is with concentration.}
\label{ps12_fig_k1_corre}
\end{center}
\end{figure}

We  plot $k_{r1}$ and $k_{r2}$ versus these physical parameters in Fig. \ref{ps12_fig_k1_corre}. It also shows obvious correlation with the $C$ but no obvious correlation with other parameters.  

\begin{figure}
\begin{center}
\includegraphics[scale=0.6, angle=90]{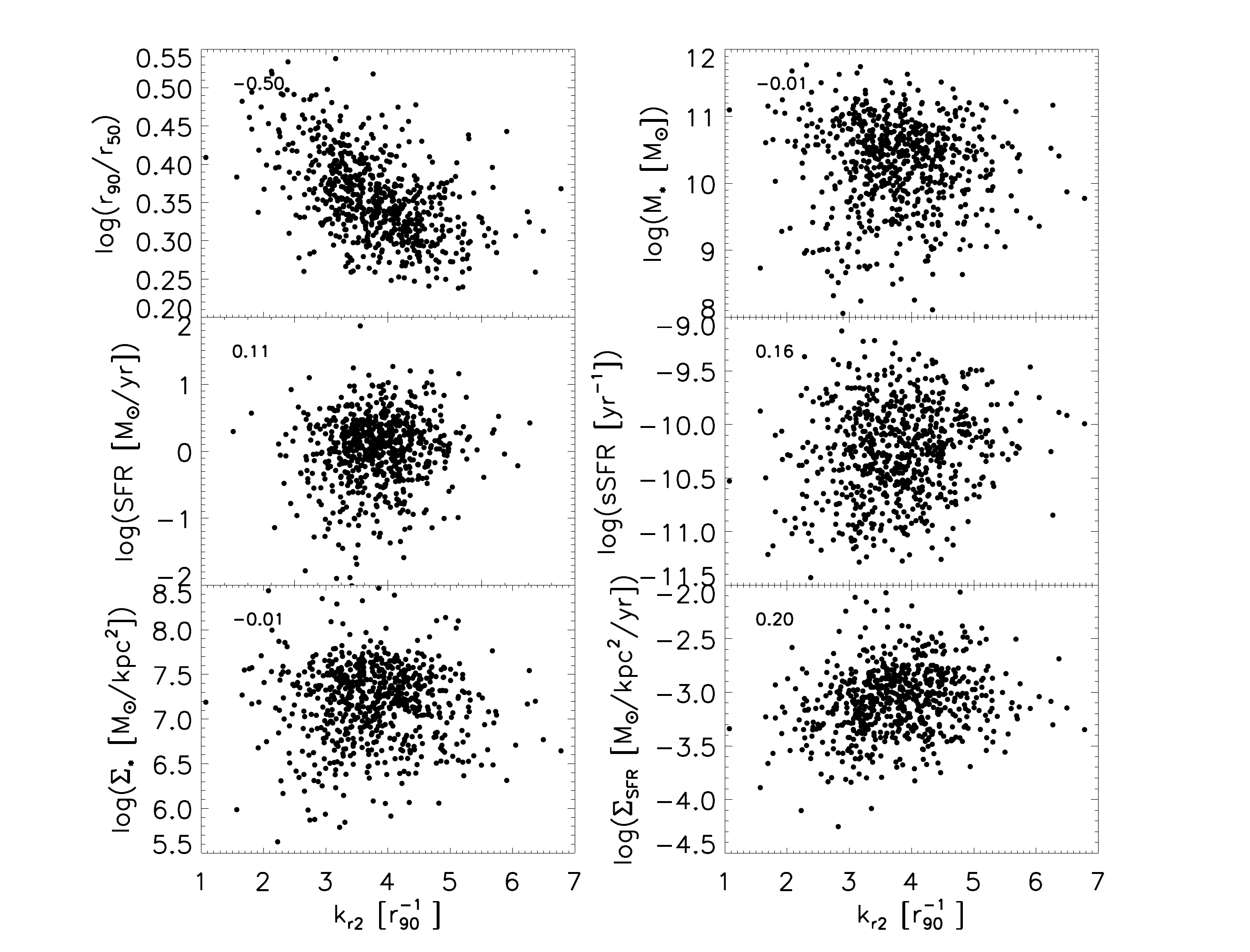}  
\caption{$r$-band $k_{r2}$ v.s. concentration parameter, stellar mass, total SFR, sSFR, $\Sigma_*$ and $\Sigma_{SFR}$. The only strong correlation is with concentration.}
\label{ps12_fig_k2_corre}
\end{center}
\end{figure}

\subsection{M/L radial profiles}
One of the key results in section \ref{stacked_1d} is that the stellar M/L radial profiles of most galaxies have a `U' shape with the minimum value located around $r_{90}$. However, \citet{bak08} found that the $g-r$ color profile (which should correlate with the M/L profile), varies with different SB profile types: Type II disks have a `U' shape color profile, Type I disks show a constant color in the outer disk, and Type III disk show a slight bluing after the break radius. It is therefore interesting to study the variation of the M/L radial profiles as a function of $R_r$.

For simplicity, we approximate the M/L profile using a skewed `V' shape (instead of the more complicated `U' shape) function similar to Eq. \ref{ps12_eq_k1} and \ref{ps12_eq_k2}:
\begin{equation}
\label{ps12_eq_km2l1}
\begin{array}{lr}
{\rm log}(\Upsilon(r))={\rm log}(\Upsilon_{10})+k_{\Upsilon 1}\,r  & ({\rm if}\, r<r_b),
\end{array}
\end{equation}
\begin{equation}
\label{ps12_eq_km2l2}
\begin{array}{lr}
{\rm log}(\Upsilon(r))={\rm log}(\Upsilon_{20})+k_{\Upsilon 2}\,r  & ({\rm if}\, r>r_b),
\end{array}
\end{equation}
where $\Upsilon$ is the stellar M/L radial profile in $\rm M_{\odot}/L_{\odot}$ and $\Upsilon_{10}$ and $\Upsilon_{10}$ are the central M/L interpolated using the inner and outer disk M/L profiles respectively.

\begin{figure}
\begin{center}
\includegraphics[scale=0.6, angle=90]{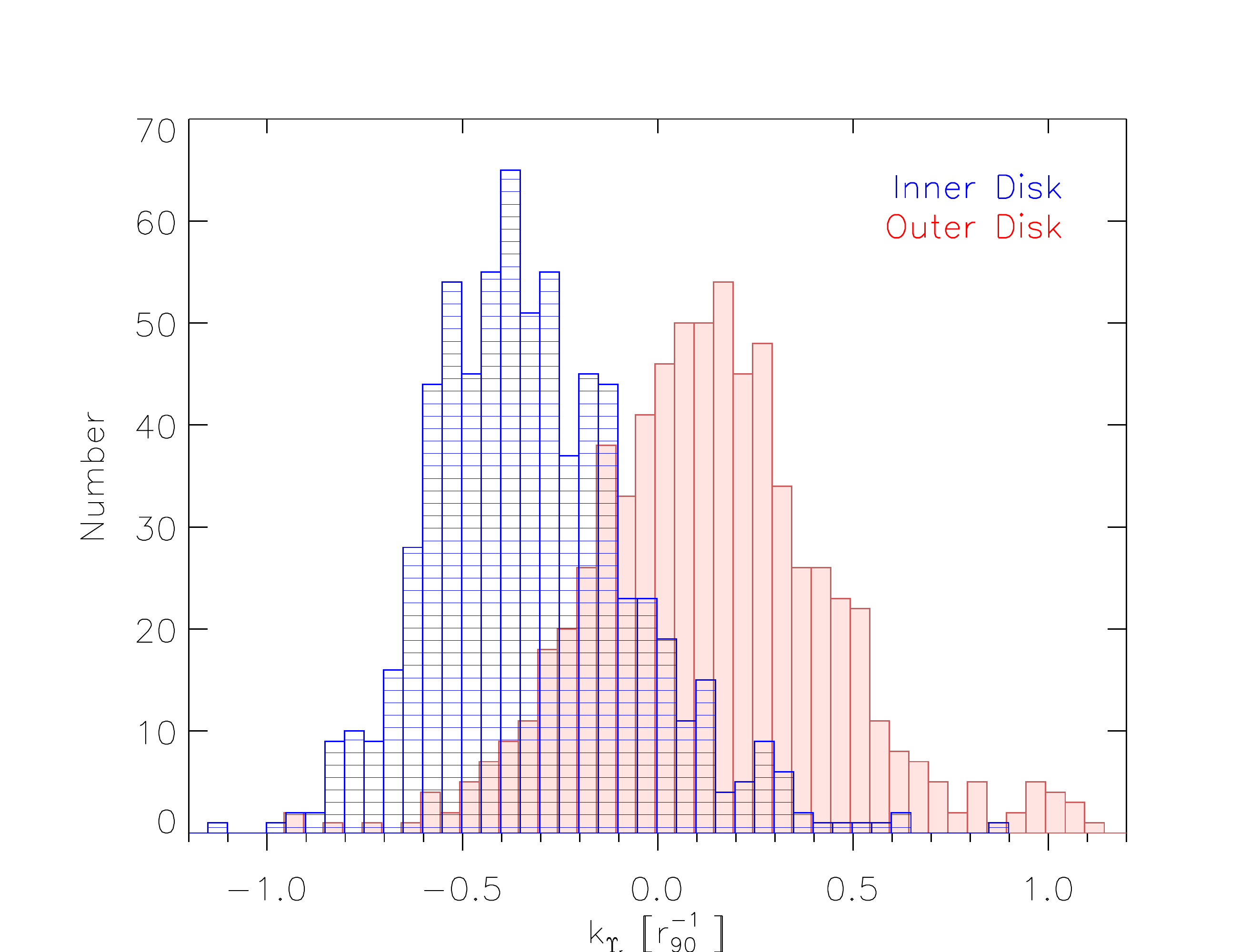}  
\caption{Histogram of the slopes of the inner and outer disk $r$-band M/L profile. Most galaxies show radial increases in M/L in the inner disk (blue histogram) and radial decreases in M/L in the outer disk (red histogram). The typical uncertainty in $k_{\Upsilon}$ is about 0.01.}
\label{ps12_fig_krm2l_histo}
\end{center}
\end{figure}

We  plot the histogram of the slopes of the inner ($k_{\Upsilon_r1}$) and outer ($k_{\Upsilon_r2}$) disk $r$-band M/L profile in Fig. \ref{ps12_fig_krm2l_histo}. The distribution of $k_{\Upsilon_r1}$ is a Gaussian centered around -0.35 $\rm log(L_{\odot}/M_{\odot})/r_{90}$, which means most galaxies show a decrease in M/L with increasing radius in the inner disk. The distribution of $k_{\Upsilon_r2}$ is also a Gaussian, but centered around 0.2 $\rm log(L_{\odot}/M_{\odot})/r_{90}$.  This means most galaxies show an increase in M/L with increasing radius in the outer disk (consistent with the results we summarized in section 5 earlier based on the composite profiles). However, Fig. \ref{ps12_fig_krm2l_histo} shows that there is a range of properties, and exceptions to the rules.

\begin{figure}
\begin{center}
\includegraphics[scale=0.6, angle=90]{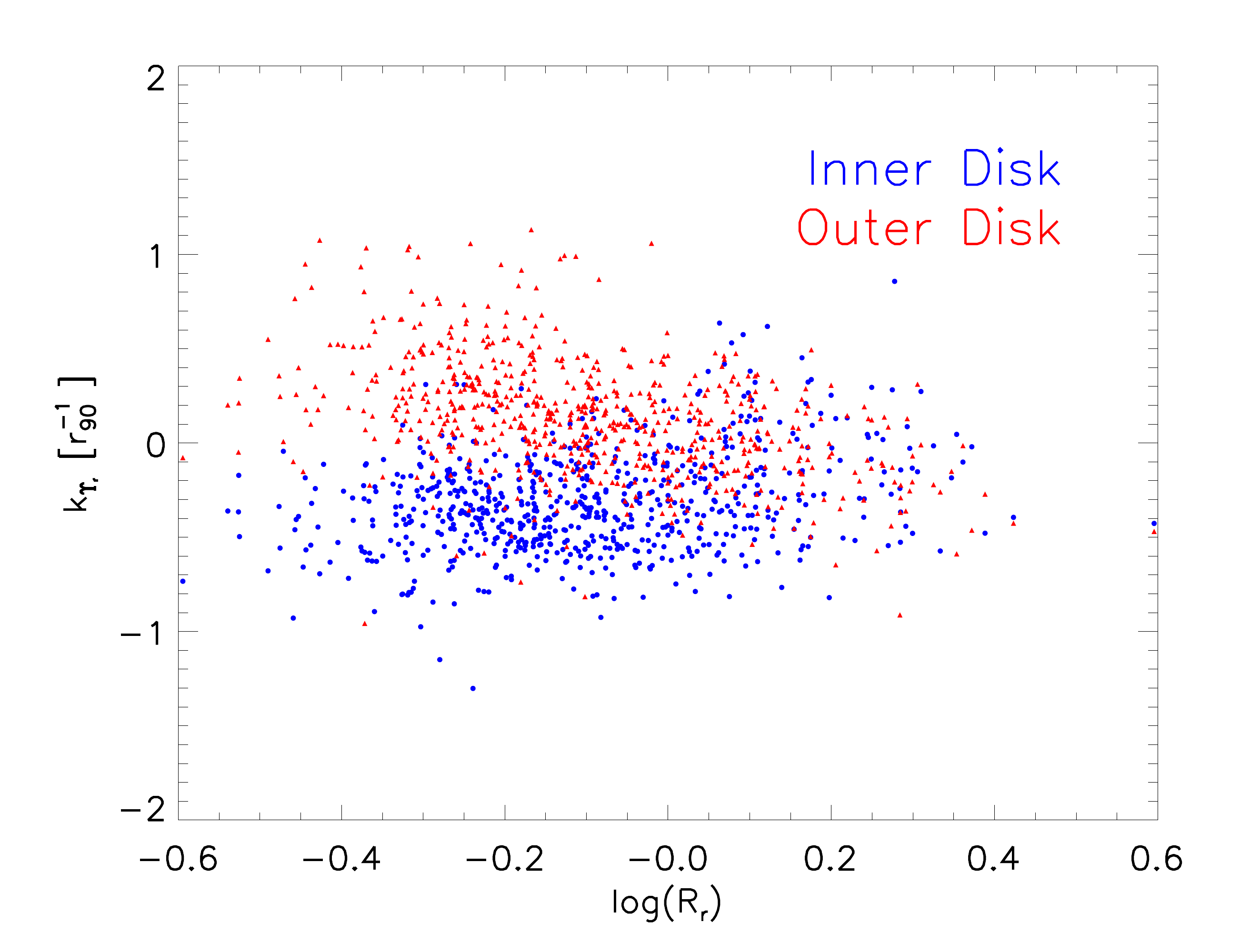}  
\caption{Correlation between $R_r$ and the slopes of the inner and outer disk M/L profile. Most down-bending disks (log($R_r <$ 0) show M/L values that decline with radius in the inner disk and increase with radius in the outer disk. The up-bending disks (log($R_r) >$ 0) typically show only weak radial gradients in M/L in either the inner or outer disk.}
\label{ps12_fig_krat_m2lslope}
\end{center}
\end{figure}

In order to see how the radial gradients in the M/L ratio relate to the surface brightness profiles, we plot the $R_r$ versus $k_{\Upsilon}$ in Fig. \ref{ps12_fig_krat_m2lslope}. This shows that the radial dependence of the M/L in the outer disk ($k_{\Upsilon_r2}$) correlates strongly with $R_r$. Several specific results are noteworthy. If we confine the analysis to galaxies with $R_r \leq$ 1 (down-bending surface brightness profiles), then galaxies with larger down-bending break amplitudes (smaller $R_r$) have stronger radial gradients in the M/L ratio in the outer disk. This is consistent with the result that even strongly down-bending r-band profiles can be produced by disks with single exponential surface mass density radial profiles. Moreover, for the down-bending galaxies, the inner disk slope $k_{\Upsilon_r1}$ is nearly a constant as a function of $R_r$, with nearly all such disks showing a decrease in M/L with increasing radius in the inner disk. Interestingly, the up-bending disks ($R_r > 1$) behave differently: any radial gradient in M/L is weak in both the inner and outer disk. This implies that the up-bending disks also have up-bending stellar surface mass-density profiles. 

\begin{figure}
\begin{center}
\includegraphics[scale=0.6, angle=90]{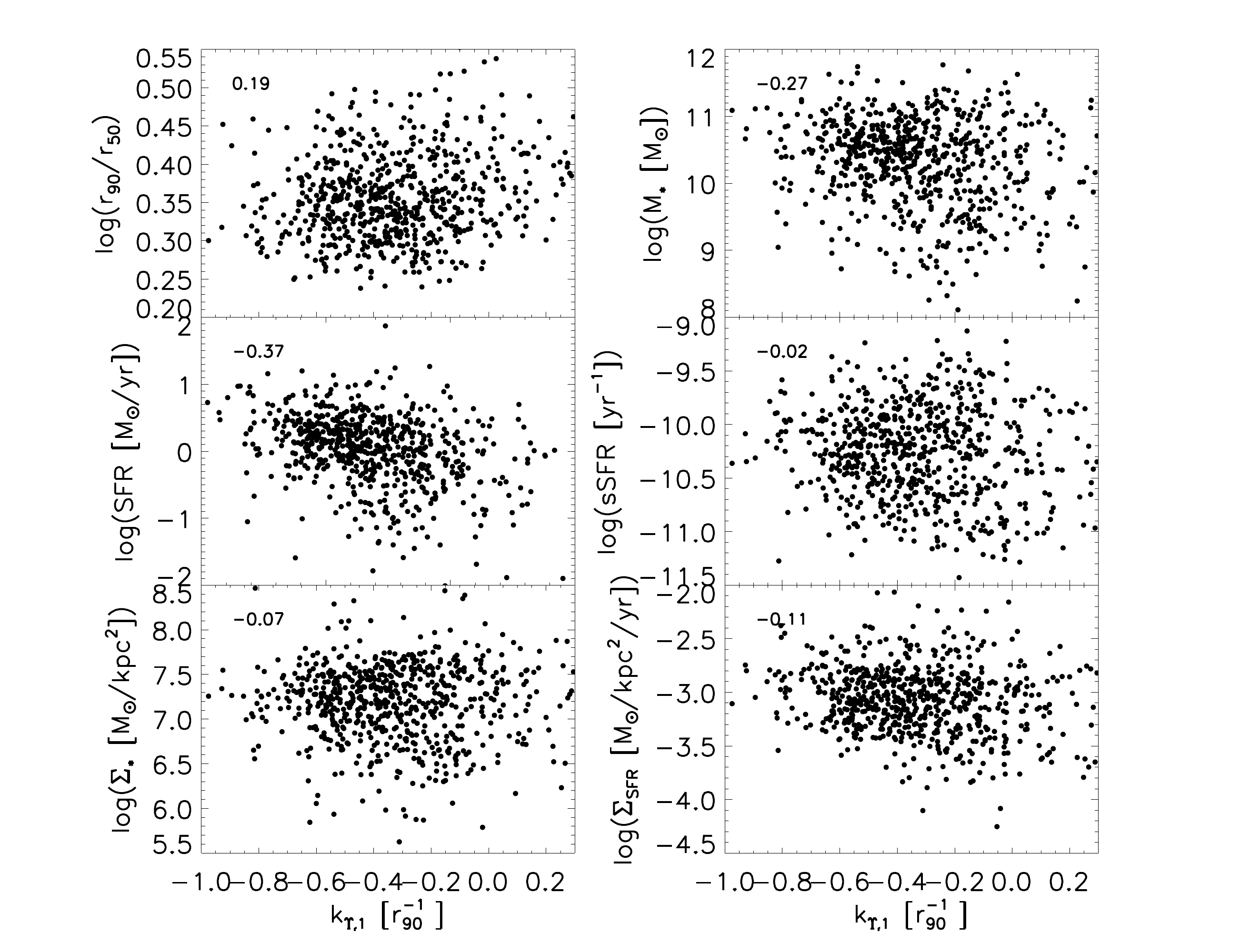}  
\caption{Slope of the inner disk $r$-band M/L profile, $k_{\Upsilon_r 1}$ v.s. concentration parameter, stellar mass, total SFR, sSFR, $\Sigma_*$ and $\Sigma_{SFR}$. There are no strong correlations.}
\label{ps12_fig_km2l1_plots}
\end{center}
\end{figure}

\begin{figure}
\begin{center}
\includegraphics[scale=0.6, angle=90]{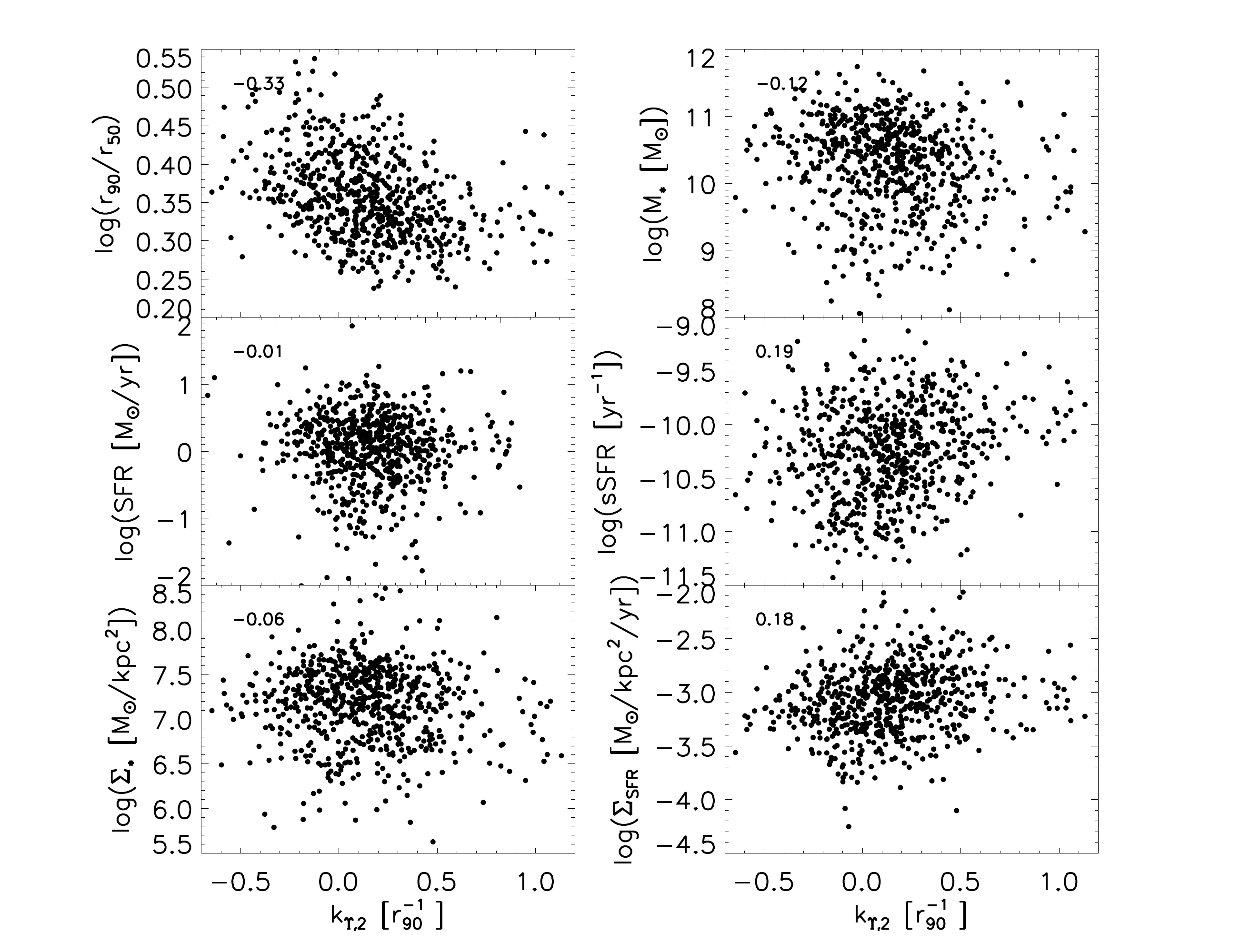}  
\caption{Slope of the outer disk $r$-band M/L profile, $k_{\Upsilon_r 2}$ v.s. concentration parameter, stellar mass, total SFR, sSFR, $\Sigma_*$ and $\Sigma_{SFR}$. The only strong correlation is with concentration: later type disk galaxies have stronger radial increases in M/L in the outer disk.}
\label{ps12_fig_km2l2_plots}
\end{center}
\end{figure}

We  plot both $k_{\Upsilon_r1}$ and $k_{\Upsilon_r2}$ versus different galaxy parameters in Fig. \ref{ps12_fig_km2l1_plots} and  \ref{ps12_fig_km2l2_plots}. The strongest correlation is between $C$ (bulge/disk ratio, Hubble type proxy) and the radial gradient of M/L in the outer disk ($k_{\Upsilon_r2}$. The early type disks are more likely to show M/L profiles in the outer disk that are flat. 

\section{Discussion}
\label{discussion}

\subsection{Comparison to Previous Studies}

Previous studies of radial profiles of the stellar component of disk galaxy profiles have usually been limited to the region interior to $R_{25}$ radius (or $r_{90}$). These typically have shown color profiles that become bluer with increasing radius \cite[e.g.][]{ma04}. It is only recently that systematic studies of the outer disks have shown the trend toward redder colors that is seen in our deep images.

While our results are based on integrated SEDs of the stellar population, similar results have been found using photometry of resolved stars to produce color magnitude diagrams (CMDs). For example, \citet{rs12} found a similar trend in NGC 7793 and \citet{thi14}  found the outer disk of M83 to be dominated by an old stellar population (although with the degree of dominance dependent on galactocentric radius).  However, \citet{gog10} also used CMDs of resolved stars, but found no up-turn in the radial age profile of NGC 300.  Some spectroscopic studies, such as \citet{del13, yoa10, yoa12} using IFU spectra, do show some examples of up-turning feature in the radial age profile at larger radii.

The general trend for nearby disk galaxies to have redder (older) outer disks was first reported by \citet{bak08} using the PT06 sample. 
They classified the radial color profiles into three different types corresponding to the SB profile types. 
They found that the Type II (down-bending) SB ($g$ and $r$-band) galaxies usually have a `U'-shaped color profile with the location of the minimum (bluest colors) corresponding to the SB profile break radius. The Type III (up-bending) SB galaxies also have a minimum in the radial color profile but its location does not correspond to the break radius. The Type I (single exponential) SB galaxies have a gradual decrease in the color profile until about 1.75 disk scale-lengths and the colors then stays constant in the outer disk.    \citet{azz08} found that Type II galaxies at intermediate redshifts have the `U'-shaped color profile as well. \citet{her13} found similar results in dwarf galaxies. \citet{dsou14} found  `U'-shaped color profiles in stacked images of late-type galaxies. 

\begin{figure}
\begin{center}
\includegraphics[scale=0.6, angle=90]{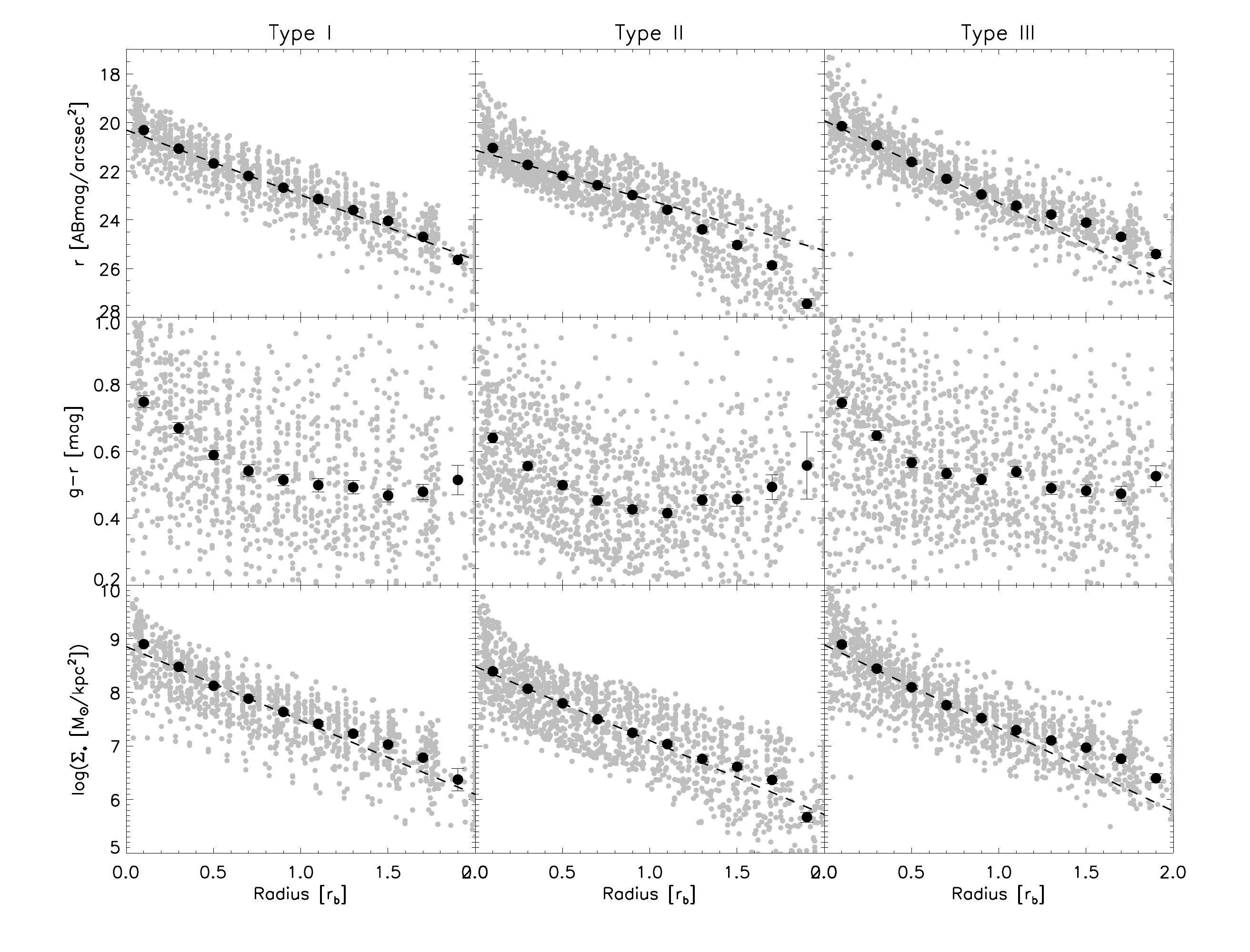} 

\caption{$r$-band SB (top row), $g-r$ color (middle row) and stellar surface mass density (bottom row) profiles of `Type I' (left column), `Type II' (middle column) and `Type III' (right column) disks. All three types have sample size of 100 galaxies. The plot is intended to have the same format as the Fig. 1 of \citet{bak08}: The gray dots are individual  galaxy profiles and the black dots show the median values. The vertical bar shows $\sigma/\sqrt{N}$ for each radial bin, where $N$ is the number of points in that radial bin. The dash lines are the linear fit to the black dots between $0.3-1\,r_{90}$. The radius is scaled in break radius $r_b$. (A high quality version of this figure is available online.)}
\label{ps12_fig_colorprofiles}
\end{center}
\end{figure}

To explicitly compare our results to those in \citet{bak08}, we selected the 100 most down-bending galaxies ($R_r<0.3$), the 100 galaxies with profiles closest to single exponentials ($0.89<R_r<1.106$) and the 100 most up-bending galaxies ($R_r>1.26$). 
We plot the radial profiles of $\mu(r)$, $g-r$ and $\Sigma_s$ fort these galaxies in Fig. \ref{ps12_fig_colorprofiles}, which reproduces Fig. 1 of \citet{bak08}.
It roughly matches some of the properties described by \citet{bak08}, however we do not see a clear minimum in the color profile in the Type III galaxies. In fact, the near constant color we see in the Type III galaxies beyond a radius of $\sim0.5\,r_b$, combined with the up-bending SB profile, results in a corresponding up-bending in the surface mass density radial profile. This is unlike the more common Type I and II disks, which have single exponential distributions of stellar mass density. 

\citet{bak12} used the deep SDSS Stripe82 imaging data to analyze SB and color radial profiles of 7 galaxies down to $30 mag/arcsec^2$ depth. The depth is similar to our data and it reaches the stellar halo region at around $27.5-30 mag/arcsec^2$. The only Type I galaxy in their sample also shows an up-turning feature in the radial color profile in the outer disk   although they claim it is due to the contribution of stellar halo light. They found that the stellar halo starts to affect the SB profile at the level of about 28 $\rm mag/arcsec^2$ and  claimed that the stellar halo could be responsible for the previous classification of Type III galaxies. Two of the examples, NGC 7716 and NGC 1087, are claimed to be affected by the tidal streams in the stellar halo. 

In this paper, we obtained SB profiles down to $28-30 \rm mag/arcsec^2$, similar to that of \citet{bak12}, thus should be able to detect the effect of the stellar halo and investigate whether galaxies with $R_r>0$ (Type III galaxies) are caused by the stellar halo. The results in Section \ref{stacked_2d} indicate that the ellipticity profile of the isophotes of the stacked images starts to decrease only slowly beyond $\sim 1.5 \, r_{90}$, and the stellar surface mass density profile starts to turn up after $\sim 2 \, r_{90}$ in generic disk galaxies. However, the SB profile of 100 most up-bending galaxies (Fig. \ref{ps12_fig_colorprofiles}) shows an upturn in the surface mass density around $r_b\, (\sim r_{90})$ where the halo light should be negligible. 
This indicates that although galaxies identified as Type III in previous studies might be caused by the stellar halo, there are a small portion of galaxies have $R_r>1$ (also cf. Fig. \ref{ps12_fig_slope_ratio}) and would be identified as Type III in the classification scheme of PT06.

\subsection{Implications for outer disk formation}

Let us conclude by discussing our results within the context of the mechanisms that determine the structure and stellar content of the outer disk. In the standard formation scenario \cite[e.g.][]{fal80, mmw98}, galaxies form in cold dark matter halos. Galaxies continue to accrete gas from the circum-galactic medium during their evolutionary history. Gas accreted in later times typically has higher angular momentum and will therefore remain in the outer disk because of the conservation of specific angular momentum during the infall process \cite[e.g.][]{guo11,fu13}. Since gas is the fuel of star formation, the outer disk should be more gas rich and composed of a younger population of stars compared to the inner disk. This is called the inside-out formation process. The inside-out picture is supported by numerous observations \cite[e.g.][]{wan11} and is widely accepted by the community. However, at first sight some of the results in this paper appear to be at odds with this simple picture, because the outer disks are frequently found to be older than the inner disk. 

There are two plausible ways that  could reconcile these results with the simple inside-out model. The first involves truncation of star formation in the outer disk.
One interesting feature of our results is that the minimum values of M/L and stellar age are generally located in the region in the galaxy disk where the stellar mass surface density has a value of $\sim 10^7M_{\odot}/kpc^2$ (lower panel of Fig. \ref{gm2l_stack}). This is the same value as the mean stellar mass surface density at the break radius of the PT06 type II galaxies, $10M_{\odot}/pc^2$ , \citep{bak08,bak12}.  Assuming the scale height of the disk to be $100pc$, we can convert this surface mass density into a volume number density of about 0.1 proton cm$^{-3}$. This is the general star formation threshold used in many simulations \cite[e.g.][]{ros08}, although the threshold is formulated as gas density. {\bf According to \citet{bb12}, the gas surface density is about $9\,M_{\odot}/pc^2$ at the break radius ($r_{90} \sim 0.7R_{25}$), which is very close to the stellar surface mass density at that point. We could expect the gas has a critical volume density at the break radius if we further assume that the star and gas have similar mean volume densities around $r_{90}$.}

\citet{ros08} argue that the `U'-shaped color/age radial profile is not caused by the star formation threshold but rather by a rapid drop of gas surface density. However, \citet{san09} argue that the gas surface density cannot drop so rapidly, based upon their fully cosmological simulation of disk galaxy formation. Instead, they suggest that the drop in star-formation in the outer disk is due to the warp/flare in gas disk which reduces the gas volume density more strongly than the gas surface density \cite[see also][]{fer98}. Our galaxies are so distant that we cannot easily detect gas and star formation occurring at the SFR surface densities typical for XUV disks, at least in radial profile form.  A similar study of local galaxies employing GALEX and HI data would be helpful to reveal the relation between the star formation threshold and the break. 
\citet{bb12} studied neutral gas profiles of 33 nearby spiral galaxies and  found that the combined  total gas radial profile is pure exponential that decreases slowly, with a scale-length  $\sim 0.6\,R_{25}$, out to about $2\,R_{25}$. The molecular-to-atomic Hydrogen transition radius in their study is about $0.7\,R_{25}$, which is comparable to $r_{90}$ in our paper and this supports the idea that the break radius might be caused by star formation threshold instead of rapid drop of total gas surface density.

A star formation threshold by itself cannot explain the pure exponential surface mass density profile that extends to the outer disk. A plausible explanation is to violate the implicit assumption that most of the stars currently in the outer disk were formed in-situ. There is in fact growing evidence that the radial migration of stars plays an important role in establishing the structure and stellar content of galaxy disks. \citet{sb02} proposed a dynamical churning mechanism that will cause radial stellar migration.  \citet{ros08} studied the effect of stellar radial migration in disk formation for an isolated model galaxy. They found that the stellar radial migration does have a large effect in redistributing stars. The migration distance can be very large (up to $\sim 10kpc$). 
They conclude that Type II disks can be formed as a consequence of the interplay between this outward transport of disk stars and a cutoff in the star formation in the outer disk. Their predicted age profile shows a `U'-shape, with a minimum at a stellar surface mass density of $10M_{\odot}/pc^2$. This fits our results very well. However, their models predict that the location of this region of minimum age corresponds to a downward break in the radial profile of the stellar mass surface density. This contrasts with our results that the stellar surface mass radial profile is a pure exponential with no obvious break. 

\citet{san09} extended this theme by simulating a disk galaxy in a full cosmological environment (in which the galaxy experiences cosmological gas infall and/or interactions with other galaxies). Gas infall and minor mergers not only expand the galactic disk but also induce density perturbations which enhance stellar radial migration. 
Their simulations yield age and sSFR radial profiles that appear similar to our observational results. Furthermore, they find that the breaks in the SB profiles only exist in bluer bands, and that the stellar mass surface density profile is close to exponential (again, in agreement with our results). They argue that break in the radial blue SB profile is due to a decrease in the star formation rate per unit area in the outer disk which is caused by a decrease in the volume density of gas at the break radius. They attribute this drop in density to the onset of a warp and flaring of the outer gas disk. Thus, while radial migration is important is producing an exponential stellar mass surface density profile, radial migration alone cannot explain the age gradient (the truncation of recent star formation in the outer disk is needed). \citet{ms09} reach qualitatively similar conclusions. More recently, \citet{bir13} have used numerical simulations to dissect in detail the build up of the stellar disk of a Milky way-like galaxy over cosmological time. This work again highlights the importance of the interplay between radial migration and the truncation of star formation in the outer disk.

Despite this emerging consensus, there is still some uncertainty about the detailed mechanism that drives radial migration. Previous theoretical studies \cite[][and references therein]{sel13} show that the churning effect (which was the mechanism originally proposed to derive stellar radial migration) does not heat up the disk. However, \citet{min12} simulated the evolution of an initially truncated galaxy in a full cosmological environment. They claim that the stellar migration is mainly due to the effects of the bar plus the spiral arms, instead of the churning mechanism (which mostly effects stars around the co-rotation radius). They found that the outer disk stars that migrated from inner disk should exhibit high velocity dispersions, especially in the radial direction. Thus observations of stellar velocity dispersions in the outer disk of the Milky Way might provide important constraints on the physics behind radial migration. 

One conclusion then would be that stellar migration has a significant influence on the outer disk formation and is mostly due to resonant effects induced by spiral arms and/or bars.  However, this the leads to a question as to  why   dwarf irregular galaxies, which do not have significant spiral arms or bars, also show `U'-shaped color/age profiles \citep{her13}.  A possible explanation is that radial migration in dwarf irregular galaxies is caused by resonant effects associated with bar-like density maxima. 

\begin{figure}[htbp]
\begin{center}
\includegraphics[scale=0.7]{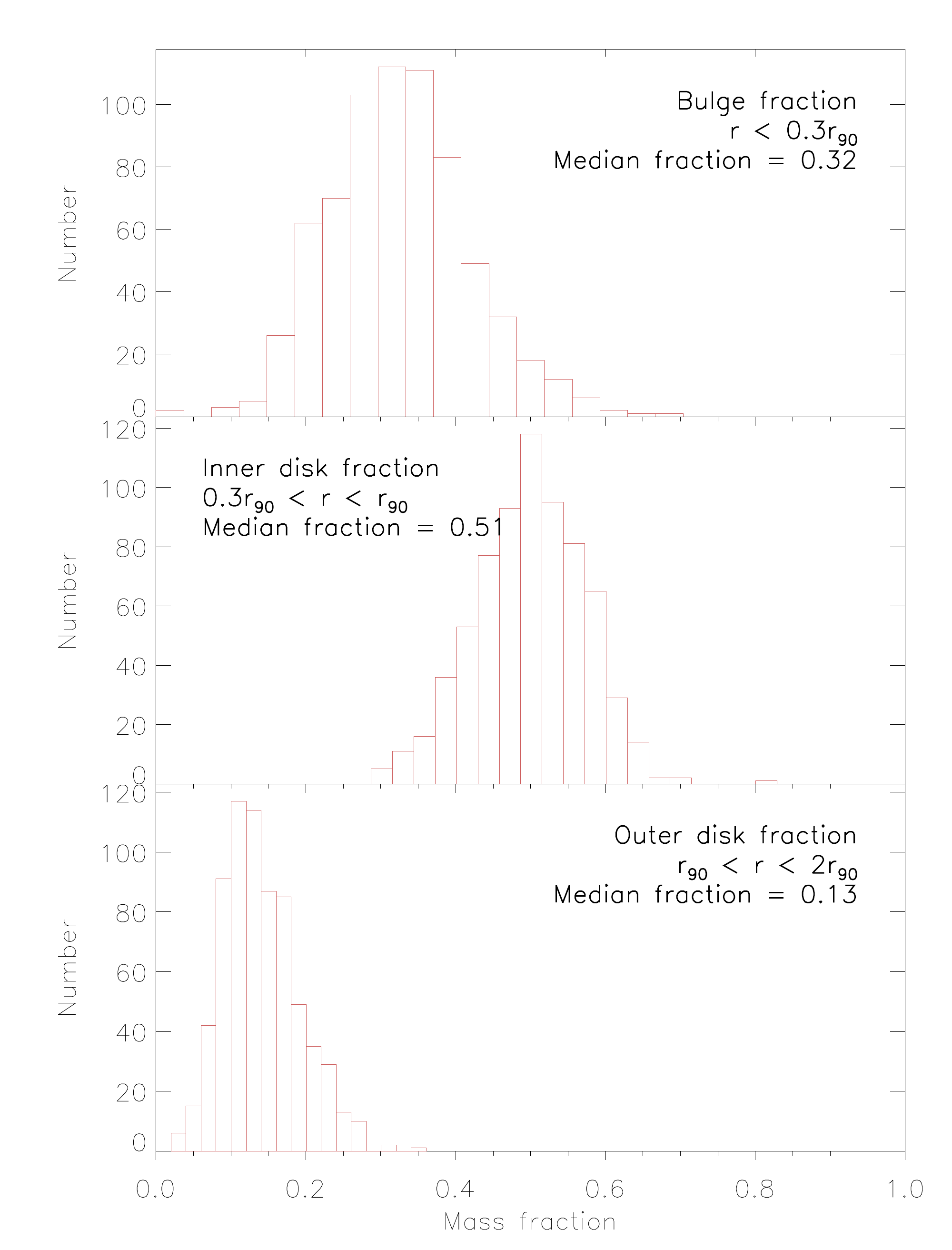}  
\caption{Observed stellar mass fraction of different parts of the galaxy. }
\label{mass_frac}
\end{center}
\end{figure}

Another question is  how much stellar mass can be transferred to the outer disk from the inner disk via radial migration? This is difficult  to answer precisely from our data but we can roughly estimate the percentage of migrated stellar mass needed in the outer disk.
Given a pure exponential disk with a scale-length of $0.3r_{90}$, a simple integration of the exponential function over the disk shows that the inner disk ($r<r_{90}$) contains about 85\% of the total stellar mass and the outer disk ($r>r_{90}$) contain less than 15\% of the total stellar mass. Here, using the derived stellar mass radial profiles for our sample galaxies, we calculate the percentage of stellar mass for different parts of each galaxy. The histograms of the percentage are shown in Fig. \ref{mass_frac}. The bulge here is defined as the part of the galaxy within 0.3\,$r_{90}$; the inner disk is defined as the part of the galaxy between 0.3\,$r_{90}$ and $r_{90}$; and the outer disk is defined as the part of the galaxy between $r_{90}$ and 2\,$r_{90}$. Suppose  57\% \citep{san09} of the outer disk stars were formed in the inner disk. Then the stellar mass that migrated from the inner to the outer disk is only $< 9\%$ of the total stellar mass of the galaxy. This is a small fraction, but nevertheless has a significant impact on the properties of the outer disk.

\section{Summary}
\label{summary}

Making use of deep stacked Pan-STARRS1 multi-band imaging observations, we  conducted an investigation of the structure and stellar content of 698 disk galaxies selected from SDSS to span a range of bulge/disk ratios and to be relatively face-on. The mass range we sample well extends from $\sim 10^9$ to 10$^{11}$ M$_{\odot}$. In particular, we modeled the five-band ($grizy$) images to determine the gradients in surface brightness (SB), color, stellar mass surface density, stellar M/L, specific star formation rate, and luminosity-weighted mean age over a range of radii typically reaching twice the characteristic size ($r_{90}$) of the galaxy. Compared to previous related studies our sample is much larger, is more representative of the full population of disk galaxis, and was analyzed using more robust and sophisticated comparison to stellar population synthesis models.
 
In summary:

\begin{itemize}

\item
We made ultra-deep stack images in 5 PS1 bands using all 698 galaxies to explore the properties of the faint stellar halo. We also made stack images using galaxies in three mass bins: low mass ($\rm M_*<10^{10}\,M_{\odot}$), medium mass ($\rm 10^{10}\,M_{\odot}< M_*<10^{10.5}\,M_{\odot}$) and high mass ($\rm M_*>10^{10.5}\,M_{\odot}$). The isophotes outside $2\,r_{90}$ are significantly rounder than those inside, indicating that the stellar halo may contaminate the light from the outer disk in the region outside $2\,r_{90}$. We found that the radial profile of the stellar surface mass density shows a distinct up-turn from a single exponential at around $2\,r_{90}$, further supporting the contribution of the halo light beyond this radius.  

\item

We used the combined data on the 698 individual galaxies to create normalized radial profiles in three bins of galaxy mass. The resulting characteristic radial SB profiles show downward bending breaks in the bluer bands (especially $g$-band), but become progressively smoother and more nearly single-exponentials as the bands move red-ward. The radial profiles of the stellar mass density are very close to single exponentials out to radii of twice $r_{90}$. This is true over the entire range in stellar mass sampled by our data. In all mass bins, the characteristic radial profiles of color, stellar M/L, and stellar age show a `U'-shape (they first decline with increasing radius, but then rise in the outer disk). The minima are located at radii of around 0.8 to 1.0 $r_{90}$ or at locations where the local stellar mass surface density is $\sim 10M_{\odot}/pc^2$.

\item

We created normalized radial profiles in three bins in the concentration parameter ($C = r_{90}/r_{50}$), which is a good proxy for bulge/disk ratio (Hubble Type). We find that the amplitude of the break in the SB profile decreases as $C$ increases (i.e. for earlier Hubble types). Moreover, the amplitude of the `U'-shaped M/L radial profile also declines as $C$ increases. In all cases, the radial stellar surface mass density remains a single exponential.   

\item

We developed an automatic algorithm to find the break radii in the $r$-band SB radial profiles of each galaxy. We then quantified the magnitude of the SB profile breaks using a slope ratio $R_r$, which is defined as the ratio of the slopes of the SB profiles in the inner ($k_1$) and outer ($k_2$) disk. We  measured an analogous ratio $R_m$ based on the slopes of the inner and outer surface mass density radial profiles. We found that $R_r$ has a log-normal distribution centered around $\log(R_r=-0.12)$ with a dispersion of 0.17, indicating that most $r$-band SB profiles are down-bending. On the other hand, $R_m$ has a log-normal distribution, but is centered around $\log(R_r=0.0)$ with a dispersion of 0.16. This means most of the stellar surface mass density profiles are close to pure (single) exponentials. Both results are consistent with our findings based on the binned, normalized profile shapes.

\item

We explored the correlation between the $r$-band SB profile slope ratio $R_r$ with the concentration parameter ($C$), total stellar mass ($M_*$), total SFR, specific SFR (sSFR), and average surface densities of stellar mass ($\Sigma_{M_*}$) and SFR ($\Sigma_{\rm SFR}$). There is a strong correlation between $C$ and $R_r$, consistent with the results described above based on the composite profiles. No obvious correlations were found between the $R_r$ and the other parameters. 

\item 
We approximated the radial M/L profile in individual galaxies using a skewed `V' shape function and characterized the results using the slopes of inner ($k_{\Upsilon_{r}1}$) and outer ($k_{\Upsilon_{r}2}$) disk $r$-band M/L profile. We found that $k_{\Upsilon_{r}1}$ has a log-normal distribution centered on $\sim -0.3$, implying that most galaxies have M/L ratios that decrease with radius in the inner disk. In contrast, $k_{\Upsilon_{r}2}$ has a log-normal distribution centered on $\sim$+0.2, implying that most galaxies have M/L ratios that increase with radius in the outer disk. We also examined $k_{\Upsilon_{r}1}$ and $k_{\Upsilon_{r}2}$ as functions of $R_r$ as well as other physical parameters. We found that $k_{\Upsilon_{r}2}$ strongly correlates with $R_r$ while the $k_{\Upsilon_{r}1}$ is almost a constant. The strongest other correlation is between $k_{\Upsilon_{r}2}$ and $C$. This is consistent with PT06 and \citet{bak08}, as well as with our inferences based on the composite profiles. 

\item 

We conclude that down-bending SB profiles, pure exponential surface mass density profiles, and older outer disks (beyond $r_{90}$) are typical for the majority of disk galaxies over a wide range in galaxy mass. However, these features weaken as the bulge/disk ratio increases. These results are consistent with recent numerical simulations of the evolution of disk galaxies. It appears that a combination of a radial migration of stars from the inner to the outer disk is required, but that a truncation of recent star-formation in the outer disk is also important. Together these two mechanisms may be able to create a smooth exponential radial mass profile while also creating an outer disk that is older than the inner disk.

\end{itemize}

\section*{Acknowledgments}
The Pan-STARRS1 Surveys (PS1) have been made possible through contributions of the Institute for Astronomy, the University of Hawaii, the Pan-STARRS Project Office, the Max-Planck Society and its participating institutes, the Max Planck Institute for Astronomy, Heidelberg and the Max Planck Institute for Extraterrestrial Physics, Garching, The Johns Hopkins University, Durham University, the University of Edinburgh, Queen's University Belfast, the Harvard-Smithsonian Center for Astrophysics, the Las Cumbres Observatory Global Telescope Network Incorporated, the National Central University of Taiwan, the Space Telescope Science Institute, the National Aeronautics and Space Administration under Grant No. NNX08AR22G issued through the Planetary Science Division of the NASA Science Mission Directorate, the National Science Foundation under Grant No. AST-1238877, the University of Maryland, and Eotvos Lorand University (ELTE). Part of this research was made during trips to JHU by DT and to JHU by GRM. These visits were in part funded by Research Collaboration Awards from the University of Western Australia.
This research made use of the ``K-corrections calculator'' service available at http://kcor.sai.msu.ru/


\appendix
\section{PSF}
\label{psf}

Deep stacked images can reveal the effects of the extended and faint wings of the image point spread function (PSF) \citep{dej08}. Thus, it is important to examine the contribution of the extended wings of the PSF to the outer disks and stellar halos. \citet{zib04} calculated an effective PSF for the SDSS using a stacked image of stars selected from their galaxy images following a similar galaxy stack procedure described in section \ref{stacked_2d}. Although their effective PSF has an extended wing they claim that the extended PSF wing has only a minor effect on the stellar halo detection, based on an examination of a PSF-convolved model galaxy. However, \citet{dej08} claimed that the mean or median stacked image is better than the mode image used by \citet{zib04} in deriving the PSF. He further claimed that the PSF has a major effect on the halo detection (20-80\% of the halo light) along the minor axis of edge-on galaxies. However, he also concluded that an extended halo component is still needed to fully fit the observations. 

\begin{figure}
\begin{center}
\includegraphics[scale=0.6, angle=90]{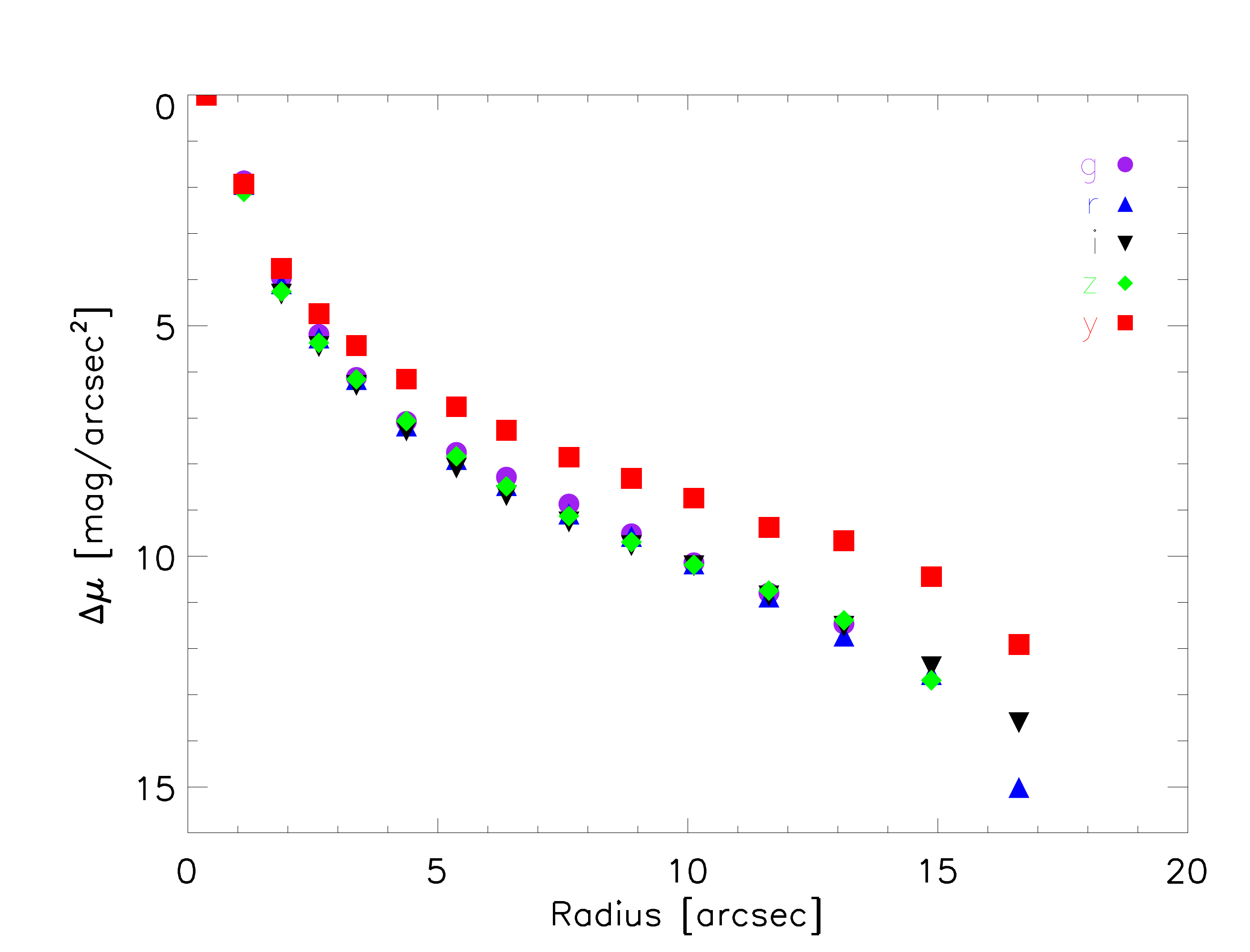}  
\caption{The PSF profiles.}
\label{fig_psf}
\end{center}
\end{figure}

Here we follow the procedure described by \citet{dej08} for calculating the effective PSF using a  stacking method: We first select a star, which is within 2 mag difference from the central brightness of the target galaxy but far away from other objects. We take these stars from each image and rotate and rescale the image centered on the star following the procedure described in section \ref{stacked_2d}. All  objects other than the selected star in the image are then masked. We then combine all the star cut-outs to calculate a mean image of the stars. The surface brightnesses of the stars are scaled to the central SB of the star and outerlier pixels are clipped in the calculation. The final effective PSFs are calculated and shown in Fig. \ref{fig_psf}. The resultant effective PSFs in different bands do have extended wings out to radii of 15''. The $g,\,r,\,i,\,z$-band PSFs are very close to each other but the $y$-band PSF has an up turn relative to the bluer bands beyond a radius of 3''. However, the PSFs of all 5 bands decline sharply by 7-9 mag within 7'', the $r_{90}$ of the stacked galaxy image. This is far below the SB around the break radius. We thus conclude that although the extended PSF wing might have some effects on the detection of the stellar halo, it should have very minimal effects on the composite stellar disk profiles.

\begin{figure}
\begin{center}
\includegraphics[scale=0.6]{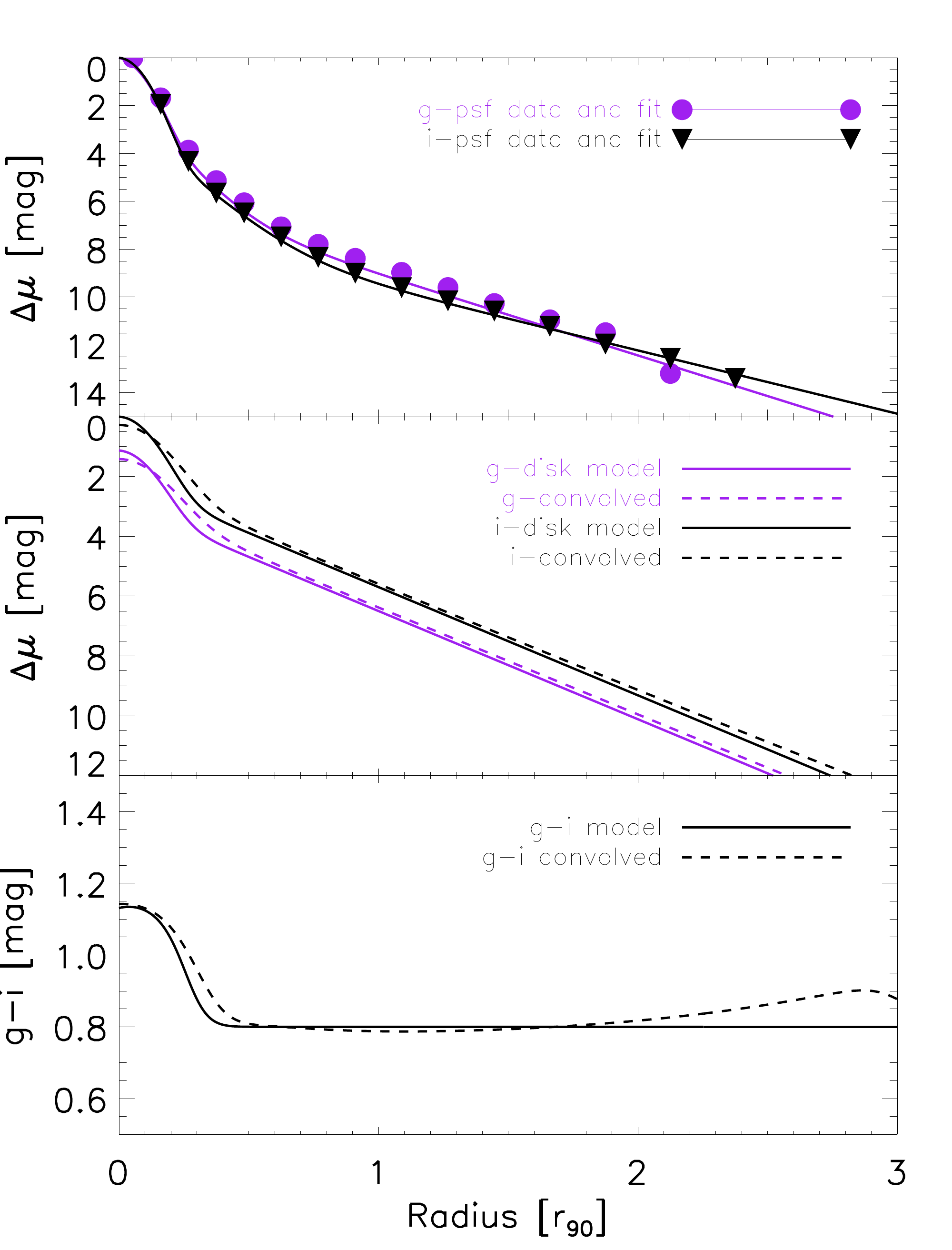}  
\caption{The 1-D galaxy model and the PSF convolved profiles. Upper-panel: $g$-band (purple) and $i$-band (black) PSF. The symbols are measured data and the lines are functional fit. The PSFs are fitted using a Gaussian plus a broken exponential profile. Middle-panel: $g$-band (purple) and $i$-band (black) SB profiles. Lower-panel: $g-i$ color profiles. Solid lines are the original model and dash lines are the model after convolution with the PSF. }
\label{fig_psf_effect}
\end{center}
\end{figure}

In order to quantify the effect of the PSF on the SB profiles, we made a 1-D galaxy model and convolved the model with the PSF. The model galaxy has an exponential disk and a point source in the center. The exponential disk has $g-i=0.8$ and the point source has $g-i=1.4$. The flux ratio of the point source and the exponential disk is 0.5. This is meant to be a simplified model of the galaxies in our sample with the larger bulge/disk ratio where the effects of the PSF wings on the measured colors of the outer disk will be most severe. The model galaxy and the PSF convolved SB and color profiles are plotted in Fig. \ref{fig_psf_effect}. The halo region ($\sim 2-3 r_{90}$) does have an spurious up-turn with the $g-i$ color by 0.1, but the color of the disk region ($\sim 0.5-2 r_{90}$) has almost no change from the input model. 

\section{Galaxy list}
Here we list our sample galaxies as well as some key parameters. This list has the same format as Table \ref{gal_list}.

\section{Radial profiles}
Here we list radial profiles of all our sample galaxies.

\section{Figures of each galaxy}
There are 698 galaxies and each galaxy has a 6-panel figure.
Upper-row from left to right:
5-band composite chi-squared detection image;
5-band surface brightness profiles; 
stellar surface mass density profile. 
Lower-row from left to right: 
3-color image with masks; 
g-r and r-i color profiles;
g, r, and i-band M/L profiles. 
The yellow ellipses in the images and the yellow vertical dash lines in the radial profile plots
show the location of $r_{90}$. The brown dash-dot line show the break radius. 
The two blue straight lines show the fit to the inner and outer disk 
r-band surface brightness profile. The black vertical dash lines show 
the inner ($0.3\,r_{90}$) and outer ($2\,r_{90}$) boundary of the stellar disk.


\begin{thebibliography}{}

\bibitem[Atkinson\,et\,al.\,(2013)]{atk13} Atkinson, A. M., Abraham, R. G., \& Ferguson, A. M. N. 2013, ApJ, 765, 28

\bibitem[Azzollini\,et\,al.\,(2008)]{azz08} Azzollini, R., Trujillo, I., \& Beckman, J. E. 2008, ApJ, L72

\bibitem[Bakos\,et\,al.\,(2008)]{bak08} Bakos, J., Trujillo, I., \& Pohlen, M. 2008, ApJ, L103

\bibitem[Bakos\,\&\,Trujillo\,(2012)]{bak12} Bakos, J., \& Trujillo, I. 2012, arXiv:1204.3082

\bibitem[Bell\,\&\,de Jong\,(2001)]{bdj01} Bell, E., \& de Jong, R. 2001, ApJ, 550, 212

\bibitem[Bell\,et\,al.\,(2003)]{bel03} Bell, E. F., et al. 2003, ApJS, 149, 289

\bibitem[Bigiel\,\&\,Blitz\,(2012)]{bb12} Bigiel, F., \& Blitz, L. 2012, ApJ, 756, 183

\bibitem[Bird\,et\,al.\,(2013)]{bir13} Bird, J. C. et al. 2013, ApJ, 773, 43

\bibitem[Brinchmann\,et\,al.\,(2004)]{bri04} Brinchmann, J., et al. 2004, MNRAS, 351, 1151

\bibitem[Bruzual\,\&\,Charlot\,(2003)]{bc03} Bruzual, G. \& Charlot, S. 2003, MNRAS, 344, 1000

\bibitem[Cardelli\,et\,al.\,(1989)]{cal89}Cardelli, J., Clayton, G., \& Mathis, J. 1989, ApJ.     345, 245

\bibitem[Charlot\,\&\,Fall\,(2000)]{cf00} Charlot, S. \& Fall, S. M. 2000, ApJ, 539, 718

\bibitem[Chilingarian\,et\,al.\,(2010)]{chi10} Chilingarian, I., Melchior, A.-L., Zolotukhin, I. 2010, MNRAS, 405, 1409

\bibitem[Chilingarian\,\&\,Zolotukhin\,(2012)]{chi12} Chilingarian, I., Zolotukhin, I. 2012, MNRAS, 419, 1727


\bibitem[da Cunha\,et\,al.\,(2008)]{dac08} da Cunha, E., et al. 2008, MNRAS, 388, 1595

\bibitem[de Jong\,et\,al.\,(2007)]{dej07} de Jong, R. S., et al. 2007, ApJ, 667, L49

\bibitem[de Jong\,(2008)]{dej08} de Jong, R. S. 2008, MNRAS, 388, 1521

\bibitem[Debattista\,et\,al.\,(2006)]{deb06} Debattista, V. P., et al. 2006, ApJ, 645, 209

\bibitem[Delgado\,et\,al.\,(2014)]{del13} Delgado, R. M. G., et al. 2014, A\&A, 562, A47

\bibitem[D'Souza\,et\,al.\,(2014)]{dsou14}D'Souza, R., Kauffmann, G., Wang, J., \& Vegetti, S. 2014, arXiv:1404.2123

\bibitem[Elmegreen\,\&\,Hunter\,(2006)]{elm06} Elmegreen, B. G. \& Hunter, D. A. 2006, ApJ, 636, 712

\bibitem[Erwin\,et\,al.\,(2005)]{erw05}Erwin, P., Beckman, J., Pohlen, M. 2005, ApJ, 626, L81

\bibitem[Erwin\,et\,al.\,(2008)]{erw08} Erwin, P., Pohlen, M., Beckman, J. 2008, AJ, 135, 20

\bibitem[Fall\,\&\,Efstathiou\,(1980)]{fal80} Fall, S. M. \& Efstathiou, G. 1980, MNRAS, 193, 189

\bibitem[Ferguson\,\&\,Clarke\,(2001)]{fer01} Ferguson, A. M. N. \& Clarke, C. J. 2001, MNRAS, 325, 781

\bibitem[Ferguson\,(1998)]{fer98} Ferguson, A. M. N. 1998, PHD Thesis

\bibitem[Freeman\,(1970)]{fre70} Freeman, K.C. 1970, ApJ, 160, 811

\bibitem[Fu\,et\,al.\,(2013)]{fu13} Fu, J., et al. 2013, MNRAS 434, 1531

\bibitem[Gadotti\,(2009)]{gad09} Gadotti, D.A.  2009, MNRAS, 393, 1531

\bibitem[Gogarten\,et\,al.\,(2010)]{gog10} Gogarten, S. M., et al. 2010, ApJ, 712, 858

\bibitem[Guo\,el\,al.\,(2011)]{guo11} Guo, Q., et al. 2011, MNRAS, 413, 101

\bibitem[Herrmann\,et\,al.\,(2013)]{her13} Herrmann, K., Hunter, D. \& Elmegreen, B. 2013, AJ, 146, 104

\bibitem[Kaiser\,et\,al.\,(2002)]{kai02} Kaiser, N., et al. 2002, SPIE, 4836, 154

\bibitem[Kauffmann\,et\,al.\,(2003)]{kau03} Kauffmann, G., et al. 2003, MNRAS, 341, 54

\bibitem[Kennicutt\,(1989)]{ken89} Kennicutt, R. C. 1989, ApJ, 344, 685

\bibitem[Lackner\,\&\,Gunn\,(2012)]{lg12} Lackner, C. N., and Gunn, J. E. 2012, MNRAS, 421, 2277

\bibitem[Li\,et\,al.\,(2011)]{li11} Li, Z., et al. 2011, ApJS, 197, 22

\bibitem[MacArthur\,et\,al.\,(2004)]{ma04} MacArthur, L., et al. 2004, ApJ, 152, 175

\bibitem[Majewski\,(1993)]{maj93} Majewski, S. R. 1993, ARA\&A, 31, 575

\bibitem[Masters\,et\,al.\,(2010)]{mas10} Masters, K. L., et al. 2010, MNRAS, 404, 792

\bibitem[Mat\'{i}nez-Serrano\,et\,al.\,(2009)]{ms09} Mat\'{i}nez-Serrano, F. J., et al. 2009, ApJ, 705, L133

\bibitem[Minchev\,et\,al.\,(2012)]{min12} Minchev, I., et al. 2012, A\&A, 548, A126

\bibitem[Mart\'{i}n-Navarro\,et\,al.\,(2012)]{mn12} Mart\'{i}n-Navarro, I. et al. 2012, MNRAS, 427, 1102

\bibitem[Mart\'{i}n-Navarro\,et\,al.\,(2014)]{mn14} Mart\'{i}n-Navarro, I. et al. 2014, MNRAS, 441, 2809

\bibitem[Mo\,et\,al.\,(1998)]{mmw98} Mo, H., Mao, S., \& White, S. 1998, MNRAS, 295, 319

\bibitem[Mu\~{n}oz-Mateos\,et\,al.\,(2009)]{mm09}Mu\~{n}oz-Mateos, J. C., et al. 2009, ApJ, 701, 1965

\bibitem[Pohlen\,et\,al.\,(2002))]{poh02} Pohlen, M., et al. 2002, A\&A, 392, 807

\bibitem[Pohlen\,\&\,Trujillo\,(2006)]{pt06} Pohlen, M.\,\&\,Trujillo, I. 2006, A\&A, 759, 772

\bibitem[Radburn-Smith\,et\,al.\,(2012)]{rs12} Radburn-Smith, D. J., et al. 2012, AJ, 753, 138

\bibitem[Rest\,et\,al.\,(2013)]{res13} Rest, A., et al. 2013, arXiv:1310.3828

\bibitem[Ro$\rm \check{s}$kar\,et\,al.\,(2008)]{ros08} Ro$\rm \check{s}$kar, R., et al. 2008, ApJ, 675, L65

\bibitem[S$\rm \acute{a}$nchez-Bl$\rm \acute{a}$zquez\,et\,al.\,(2009)]{san09} S$\rm \acute{a}$nchez-Bl$\rm \acute{a}$zquez, P., et al. 2009, MNRAS, 398, 591

\bibitem[Schaye\,(2004)]{sch04} Schaye, J. 2004, ApJ, 609, 667

\bibitem[Schiminovich\,et\,al.\,(2007)]{sch07} Schiminovich, D., et al. 2007, ApJS, 173, 315

\bibitem[Schlafly\,et\,al.\,(2012)]{sch12} Schlafly, E. F., et al. 2012, ApJ 756, 158

\bibitem[Schlegel\,et\,al.\,(1998)]{sfd98} Schlegel, D., Finkbeiner, D., \& Davis, M. 1998, ApJ, 500, 525

\bibitem[Sellwood\,\&\,Binney\,(2002)]{sb02} Sellwood, J. A. \& Binney, J. J. 2002, MNRAS, 336, 785

\bibitem[Sellwood\,(2013)]{sel13} Sellwood, J. A. 2013, arXiv: 1310.0403

\bibitem[Strauss\,(2002)]{str02} Strauss, M. A., et al. 2002, AJ, 124, 1810

\bibitem[Szalay\,et\,al.\,(1999)]{sza99}Szalay, A. S., et al. 1999, AJ, 117, 68

\bibitem[Thilker\,et\,al.\,(2007)]{t07} Thilker, D. A., et al. 2007, ApJS, 173, 538

\bibitem[Thilker\,et\,al.\,(in prep.)]{thi14} Thilker, D. A., et al. 2014, in preparation

\bibitem[Tonry\,et\,al.\,(2012)]{ton12} Tonry, J. L., et al. 2012, ApJ, 750, 99

\bibitem[van der Kruit\,(1979)]{vdk79} van der Kruit, P.C. 1979, A\&AS, 38, 15

\bibitem[van der Kruit\,(1987)]{vdk87} van der Kruit, P.C. 1987, A\&AS, 173, 59

\bibitem[Wang\,et\,al.(2011)]{wan11} Wang, J., et al. 2011, MNRAS, 412, 1081

\bibitem[Yoachim\,et\,al.\,(2010)]{yoa10} Yoachim, P., Ro$\rm \check{s}$kar, \& Debattista, V. 2010, ApJ 716, L4

\bibitem[Yoachim\,et\,al.\,(2012)]{yoa12} Yoachim, P., Ro$\rm \check{s}$kar, \& Debattista, V. 2012, ApJ 752, 97


\bibitem[Younger\,et\,al.\,(2007)]{you07} Younger, J. D., et al. 2007, ApJ, 670, 269


\bibitem[Zibetti\,et\,al.\,(2004)]{zib04} Zibetti, S., White, S., and Brinkmann, J. 2004, MNRAS 347, 556

\bibitem[Zibetti\,\&\,Ferguson\,(2004)]{zf04} Zibetti, S., \& Ferguson, A. 2004, MNRAS 352, L6

\end{thebibliography}
\end{document}